\documentclass{aa}  

\usepackage[english]{babel} 
\usepackage[varg]{txfonts} 
\usepackage{upgreek} 

\usepackage{balance}   

\usepackage{orcidlink} 

\usepackage{graphicx} 
\usepackage{float} 
\usepackage{caption} 
\graphicspath{{Figures/}} 

\usepackage{siunitx} 
\usepackage{amsmath,amsfonts,amssymb} 
\usepackage{xparse} 

\usepackage{subcaption} 
\usepackage{multirow} 
\usepackage{tablefootnote} 

\usepackage{natbib} 
\usepackage{aas_macros} 
\bibpunct{(}{)}{;}{a}{}{,} 

\usepackage{hyperref} 
\hypersetup{colorlinks=true, linkcolor=blue, urlcolor=blue, citecolor=blue}

\begin{document}


\newcommand{\columnbreak}{\balance\break}

\newcommand{\vsini}{v\sin{i}} 
\NewDocumentCommand{\vsin}{m}{%
  $v\sin{i} \!=\! #1\,\kmps$%
}
\newcommand{\kmps}{\si{\kilo\metre\per\second}} 
\newcommand{\mps}{\si{\metre\per\second}} 

\newcommand{\teff}{\mathrm{T_{eff}}} 
\newcommand{\geff}{\mathrm{g_{eff}}} 
\newcommand{\logg}{\mathrm{\log{g}}} 
\renewcommand{\K}{\si{\kelvin}} 

\newcommand{\Msol}{\mathrm{M_\odot}} 
\newcommand{\Rsol}{\mathrm{R_\odot}} 
\newcommand{\Zsol}{\mathrm{Z_\odot}} 

\providecommand{\abs}[1]{\lvert#1\rvert} 

\NewDocumentCommand{\specline}{m m o}{%
  $\mathrm{#1{\scriptstyle\,#2}}$%
  \IfValueT{#3}{$\,\lambdaup#3$}
}

\NewDocumentCommand{\multispecline}{m m m}{%
  $\mathrm{#1{\scriptstyle\,#2}\,\lambdaup\lambdaup#3}$
}
\NewDocumentCommand{\ratio}{m m o m m o}{%
  \specline{#1}{#2}[#3]\,/\,\specline{#4}{#5}[#6]
}
\NewDocumentCommand{\spectype}{m m o}{%
  #1#2%
  \IfValueT{#3}{\,#3}
}


\title{
    The lack of fast rotators in Cyg OB2
}

\subtitle{
    I. Insights from spectral reclassification of its \spectype{B}{0} population
}

\author{
    D. Galán-Diéguez \inst{1, 2}\,\orcidlink{0000-0001-6191-8251} \and
    S.R. Berlanas \inst{1,2}\,\orcidlink{0000-0002-2613-8564} \and
    A. Herrero \inst{1,2}\,\orcidlink{0000-0001-8768-2179} \and
    M. Abdul-Masih \inst{1,2}\,\orcidlink{0000-0001-6566-7568} \and
    D.J. Lennon \inst{1,2}\,\orcidlink{0000-0003-3063-4867} \and \\
    C. Martínez-Sebastián \inst{1,2}\,\orcidlink{0009-0004-2491-5384} \and
    F. M. Pérez-Toledo \inst{2,3}\,\orcidlink{0000-0001-6265-2248}
}

\institute{
    Instituto de Astrofísica de Canarias, c/ Vía Láctea, s/n, E-38205 La Laguna, Tenerife, Spain \\
    \email{dgalandieguez.astr@gmail.com}
    \and
    Departamento de Astrofísica, Universidad de La Laguna, E-38206 La Laguna, Tenerife, Spain
    \and
    Gran Telescopio Canarias (GRANTECAN), c/ Cuesta de San José, s/n, E-38712 Breña Baja, La Palma, Spain
}

\date{
        Received month day year; accepted month day year
}


\abstract
    {
    Cygnus~OB2, located within the Cygnus X complex -- one of the most active star-forming regions of the Galaxy -- hosts hundreds of O- and B-type stars at different evolutionary stages. This rich association offers a unique opportunity to study the evolution and dynamic interactions of massive stars. However, despite extensive studies, a notable absence of a fast-rotating group ($\vsini>200\,\kmps$) among the O-type population of Cygnus~OB2 challenges current models of massive star evolution.
    }
%
%
    {
    Stellar rotation strongly impacts spectral line shapes of O-type stars, with high rotational velocities potentially leading to misclassifications. This study investigates whether some stars in Cygnus~OB2, classified at low spectral resolution as \spectype{B}{0}, are actually rapidly rotating late-O types. Such cases could explain the observed lack of fast rotators in Cygnus~OB2.
    }
%
%
    {
    Considering the effects of rotation, we reclassified the known \spectype{B}{0} population in Cygnus~OB2, using the \texttt{MGB} tool and both the new and pre-existing optical spectroscopy. Finally, we computed the projected rotational velocities using \texttt{iacob-broad}.
    }
%
%
    {
    We find that approximately $19\,\%$ of the initial \spectype{B}{0} population in Cygnus~OB2 are, in fact, late-O types. Further analysis shows that only six stars in the entire dataset have projected rotational velocities above $200\,\kmps$, with just one new O-type star exceeding this threshold.
    }
%
%
    {
    In our study of Cygnus~OB2, we continue to find a notable lack of fast rotators among its O-type population. We propose a combination of three factors as the most likely explanation: (i) the young age of Cygnus~OB2 may imply that fast rotators have not been produced yet due to binary interactions; (ii) fast rotators may have been dynamically ejected from the core as runaway stars; and (iii) local star formation conditions may hinder binary formation (reducing spin-up interactions) or result in slower rotational velocities at birth.
    }
%
%

\maketitle

\section{Introduction} \label{S: Introduction}
    {
    Hot massive stars play a crucial role in the large-scale evolution of the cosmos. Throughout their complex life cycles, these stars are associated with the most powerful and luminous phenomena in the Universe \citep{2012ARA&A..50..107L}. These events significantly affect their surroundings and host galaxies through mechanical and radiative feedback \citep{2009msfp.book...74O, 2015MNRAS.448.3248G, 2020ApJ...891..113R}.

    Traditionally, stellar masses and metallicity have been considered the key parameters that determine stellar formation, structure, and evolution. However, the role of rotation has become a critical factor, particularly in the evolution of massive stars at low metallicities \citep{2000ARA&A..38..143M}. Given its importance, several studies have investigated the rotational properties of O-type stars, revealing a consistent bimodal distribution. This pattern, initially noted by \citet{1977ApJ...213..438C} and confirmed in various studies of galactic \citep{1996ApJ...463..737P, 2022AA...665A.150H} and extragalactic \citep{2013AA...560A..29R, 2015A&A...580A..92R} O stars, indicates that most of these stars (approximately $50\!-\!70\,\%$ of the total population) present slow rotation, ranging from $40\!-\!80\,\kmps$. Moreover, in the Galaxy, the tail of fast rotators is dominated by apparently single stars with masses $\mathrm{<\!32\,\Msol}$, extending from $\sim\!200$ up to $400\,\kmps$ \citep{2022AA...665A.150H}. Extragalactic studies reveal a similar pattern, with the high-velocity tail preferentially populated by low-mass O stars \citep{2013AA...560A..29R}.

    This wide range of rotational velocities observed in massive stars cannot be fully explained by current single stellar models or by simply assuming a broad range of initial spin rates at birth \citep[but see][for an alternative perspective]{2024A&A...689A.320N}. Certain features in these distributions suggest that some stars undergo spin-up processes during their evolution. The most widely accepted explanation for the bimodality in O-type rotational velocities is the effect and/or influence of binary interactions: 
    \begin{enumerate}[i.]
        \item Given the high intrinsic multiplicity of massive stars \citep{2009AJ....137.3358M, 2014ApJS..215...15S}, many are expected to interact with a companion during their lifetime \citep{2012Sci...337..444S}.
        \item For stars with masses below $32\,\Msol$, angular momentum losses due to stellar winds are far less significant than the effects of binary spin-up interactions \citep{2022AA...665A.150H, 2024A&A...684A..35B}. 
    \end{enumerate}
  
    The bimodal distribution found in different regions, both galactic and extragalactic, shows that, at least to first order, the bimodality should be independent of metallicity up to $\mathrm{Z\!=\!0.5\,\Zsol}$. However, the Cygnus~OB2 association, located within Cygnus-X complex, presents a remarkable lack of O-type stars rotating above $200\,\kmps$ \citep{2020AA...642A.168B}. Given that more than $70\,\%$ of O-type stars are expected to undergo binary interactions throughout their lives \citep{2012Sci...337..444S}, a substantial fast-rotating population is expected within such an active and massive star-forming region as Cygnus~OB2. The apparent absence of such population raises a significant challenge to current evolutionary models; they predict that fast rotators should emerge as a natural consequence of multiplicity and binary interactions.

    Stellar rotation can be the dominant process that shapes the spectral lines of O-type stars; thus, high rotational velocities may lead to spectral misclassifications \citep[][Galán-Diéguez et al., in prep.]{2011A&A...530A..11M}. In addition, the disparity of criteria in stellar classification and variations in data quality (signal-to-noise and/or spectral resolution) can further contribute to classification errors. Given that the high-velocity tail largely consists of stars with masses $\mathrm{<\!32\,\Msol}$ (typically late-O types), and the potential impact of rotation on classification (especially toward B-type stars, vanishing the $\mathrm{He{\scriptstyle\,II}~\lambdaup4552}$ line; Galán-Diéguez et al., in prep.), this work focuses on the study of the \spectype{B}{0} population in Cygnus~OB2. Our aim is to assess whether some early-B stars are actually fast-rotating late-O types, potentially restoring the expected fast-rotator tail in the rotational velocity distribution.

    This paper is organized in six main sections as follows. Section~\ref{S: Observational sample} describes the observational campaign and sample selection. In Sect.~\ref{S: Spectral reclassification} we introduce the methods and tools used for the spectral classification of our \spectype{B}{0} stars; we also examine the effects of rotation on late-O and B0 classification. In Sect.~\ref{S: Projected rotational velocities} we outline the approach to line-broadening characterization, and present the new projected rotational velocities from our new late-O and \spectype{B}{0} sample. Section~\ref{S: Discussion} presents some explanations that can potentially account for the absence of fast rotators in Cygnus~OB2. Finally, in Sect.~\ref{S: Conclusions} we summarize the main conclusions of this work.
    }

\section{Observational sample} \label{S: Observational sample}
    {
    \begin{table*}[th!]
        \caption{
            Telescopes, instruments, and settings used in this work.
        } 
        \label{T: Telescopes_Configurations}
        \centering
        \small
        \begin{tabular}{ccccccccc}
            \hline
            \hline \\ [-1.8ex]
            Source & Telescope                      & Instrument & Grating & Resolving power & $\lambdaup$ range [{\fontsize{6.7}{4.8}\selectfont \AA}] & Date                      & \# Stars \\ \hline \\ [-1.8ex]
            1      & $\mathrm{2.56\,m}$ NOT - ORM & ALFOSC     & \#17    & 5000            & $6330 - 6870$                                            & October 2024              & 31       \\
            2      & $\mathrm{10.4\,m}$ GTC - ORM & OSIRIS+    & R2000B  & 2165            & $3950 - 5700$                                            & September 2024            & 34       \\
            3a     & $\mathrm{10.4\,m}$ GTC - ORM & OSIRIS     & R2500U  & 2555            & $3440 - 4610$                                            & GOSSS survey              & 11       \\
            3b     & $\mathrm{10.4\,m}$ GTC - ORM & OSIRIS     & R2500V  & 2515            & $4500 - 6000$                                            & GOSSS survey              & 11       \\
            3c     & $\mathrm{3.5\,m}$ - CAHA     & TWIN       & 1200    & 3000            & $3900 - 5100$                                            & GOSSS survey              & 15       \\
            4a     & $\mathrm{4.2\,m}$ WHT - ORM  & ISIS       & H2400B  & 13600           & $3900 - 5100$                                            & \citet{2018AA...620A..56B} & 1        \\
            4b     & $\mathrm{4.2\,m}$ WHT - ORM  & ISIS       & R1200B  & 7500            & $3900 - 5100$                                            & \citet{2018AA...620A..56B} & 2        \\
            4c     & $\mathrm{4.2\,m}$ WHT - ORM  & WYFFOS     & H2400B  & 5000            & $3800 - 5200$                                            & \citet{2018AA...620A..56B} & 5        \\
            5      & $\mathrm{2.54\,m}$ INT - ORM & IDS        & R1200B  & 4600            & $3900 - 5100$                                            & \citet{2020AA...642A.168B} & 4        \\
            6      & $\mathrm{2.54\,m}$ INT - ORM & IDS        & R1200B  & 4600            & $3900 - 5100$                                            & October 2023              & 3        \\ \hline \\ [-1.8ex]
            \multicolumn{8}{c}{Total number of spectra $\rightarrow$ 120}                                                                                                                      \\
            \multicolumn{8}{c}{Total number of stars observed $\rightarrow 32 $}                                                                                                               \\ \hline
        \end{tabular}
        \tablefoot{
            Spectral data from sources 3a, 3b, and 3c are part of the GOSSS catalog; for further details on the observations, see \citet{2011ApJS..193...24S, 2014ApJS..211...10S} and \citet{2016ApJS..224....4M}. Some of the stars in our sample have multiple spectra available, obtained during our observing campaign and/or from the existing literature.
        }
    \end{table*}

    The star sample used in this work is comprised of known members of Cygnus~OB2, which are classified as \spectype{B}{0}-types in the literature within $\mathrm{1^\circ}$ centered on Galactic coordinates $l\!=\!79.8^\circ$ and $b\!=\!+\,0.8^\circ$. Two stars, namely 2MASS~J20324719+4117500 and J20324863+4114298, were excluded from the final sample due to their low B magnitudes, which made them unsuitable for blue-range observations and subsequent spectral classification. Ultimately, the selection process yielded a final sample of 32 stars, with coordinates and photometric magnitudes provided in Table~\ref{T: Bstars_Sample}.

    The observation campaign was conducted during multiple runs using different telescopes (see Table~\ref{T: Telescopes_Configurations}). We performed observations with the $\mathrm{2.56\,m}$~\textit{Nordic Optical Telescope} (NOT) during a single three-night run (October 9-11, 2024) with the Alhambra Faint Object Spectrograph and Camera (ALFOSC) and its grism~\#17, which provides a spectral resolution $\mathrm{R\!\sim\!5000}$ at $6580\,\AA$ (CCD14). For the $\mathrm{10.4\,m}$~\textit{Gran Telescopio Canarias} (GTC) observations, we used OSIRIS+ in filler mode and long-slit configuration with the R2000B grism, obtaining spectra of $\mathrm{R\!\sim\!2165}$ at $4755\,\AA$. The spectral properties of the OSIRIS+ data allowed us to reclassify the B0-type stars of Cygnus~OB2, using specific classification ratios in the blue optical range. Our configuration of the ALFOSC instrument also enabled the confident detection of stars exhibiting high rotational velocities ($\vsini\!>\!200\,\kmps$). While its spectral resolution is insufficient for precise measurements of low $\vsini$, it is fully appropriate for our primary goal, which is the identification of fast rotators within the observed sample. Furthermore, we conducted two additional observation nights in the $\mathrm{2.54\,m}$~\textit{Isaac Newton Telescope} (INT) with the Intermediate Dispersion Spectrograph (IDS) and its 1200B grating ($\mathrm{R\!\sim\!4600}$).

    To complete our data and confirm the results, we reviewed the existing literature for additional spectra of our sample stars. We used the Galactic O-Star Spectroscopic Survey (GOSSS), which provides extensive blue-violet and high S/N spectra of Galactic O-type stars at resolutions of $\mathrm{R\!\sim\!2500-3000}$ \citep{2011ApJS..193...24S, 2014ApJS..211...10S, 2016ApJS..224....4M}. The GOSSS spectra used in this work were acquired with two different instruments: the OSIRIS spectrograph at the GTC, and the TWIN spectrograph at the $\mathrm{3.5\,m}$~\textit{Calar Alto Observatory telescope} (CAHA, Centro Astronomico Hispano-Alemán). We also collected spectra from \citet{2018AA...620A..56B}, observed at the $\mathrm{4.2\,m}$~\textit{William Herschel Telescope} (WHT, ORM) using the Intermediate-dispersion Spectrograph and Imaging System (ISIS) and the Wide-field Fibre Optic Spectrograph (WYFFOS). We also included \citet{2020AA...642A.168B} data from IDS@INT.

    The details of the observational runs and data characteristics are summarized in Table~\ref{T: Telescopes_Configurations}. The final spectra from the NOT and GTC telescopes were processed with \texttt{PypeIt}, a Python package for semi-automated spectroscopic data reduction \citep{2020JOSS....5.2308P, 2020zndo...3743493P}. The data from the INT observations were reduced using \texttt{PyRAF} \citep{2012ascl.soft07011S}, following standard routines for bias and flat-field subtraction as well as wavelength calibration.
    }

\section{Spectral reclassification of the \spectype{B}{0} population in Cygnus~OB2} \label{S: Spectral reclassification}
    {
    \subsection{Effects of rotation in the late-O and early-B range for spectral classification} \label{SS: Effects of rotation}
        {
        Stellar rotation can strongly affect the spectral features of massive stars. As demonstrated by \citet{2011A&A...530A..11M} and Galán-Diéguez et al. (in prep.), rotational velocity, inclination, and signal-to-noise ratio can significantly alter the profiles and relative depths of key diagnostic lines. One of the most critical indicators in the late-O to early-B region is the depth ratio of \ratio{He}{II}[4542]{He}{I}[4388], which plays a key role in spectral classification \citep{2011ApJS..193...24S}. This ratio is particularly sensitive to rotational broadening.
        
        For late-O stars, increasing $\vsini$ progressively weakens the \specline{He}{I} line more rapidly than \specline{He}{II}, shifting the line depth ratio toward values typically associated with earlier spectral types (Galán-Diéguez et al., in prep.). In contrast, for early-B stars (\spectype{B}{0.2}–\spectype{B}{0.5}), the \specline{He}{II}[4542] line is already weak at these effective temperatures, and it becomes increasingly difficult to detect for $\vsini \!\gtrsim\! 200\,\kmps$. As a result, the line ratio trend is inverted, mimicking later spectral types.

        This effect of rotation becomes particularly problematic at the O9.7 subtype, which lies in the transition zone from O- to B-types. At high rotational velocities, \spectype{O}{9.7} stars can exhibit \ratio{He}{II}[4542]{He}{I}[4388] ratios that closely resemble those of \spectype{B}{0} stars. As a result, fast-rotating \spectype{O}{9.7} stars are especially prone to being misclassified as early-B types. This bias may be intensified by diverse classification criteria and heterogeneous data quality (moderate-resolution spectra with limited S/N, a common scenario in large-scale surveys).

        These results suggest that some stars cataloged in the literature as \spectype{B}{0} in Cygnus~OB2 might in fact be fast-rotating late-O stars misclassified due to rotational effects and observational constraints. This scenario has relevant implications. Since the rotational velocity distribution of O-type stars is known to be bimodal -- with fast rotators generally being late O-types below $32\,\Msol$ \citep{2022AA...665A.150H} -- such misclassifications might potentially account for the observed lack of high-$\vsini$ stars in Cygnus~OB2.
        }

    \subsection{Updated spectral types for the \spectype{B}{0} population of Cygnus~OB2} \label{SS: Updated spectral types}
        {
        An accurate spectral classification (including rotational effects) of the \spectype{B}{0} population within Cygnus~OB2 is crucial for understanding this region of intense massive star formation. However, discrepancies are present in the literature, likely related to differences in spectral resolutions, signal-to-noise ratios, and classification criteria. To ensure a consistent classification, we followed the criteria described in \citet{2011ApJS..193...24S} -- specifically in their Table~4 -- which lists the diagnostic line ratios for O- and early B-type stars. In particular, the relative strengths of \multispecline{He}{II}{4542, 4200}, \multispecline{He}{I}{4388, 4144}, and \specline{Si}{III}[4552] are used to distinguish between spectral subtypes. 

        To facilitate the classification we used the Marxist Ghost Buster code \citep[\texttt{MGB,}][]{2015hsa8.conf..603M}. This IDL-based script compares the observed spectrum against a two-dimensional grid of standard stars organized by spectral type and luminosity class. By overplotting the spectra, we identified the standard star that best matched the diagnostic line ratios of the sample star, thereby determining its spectral type and luminosity class. \texttt{MGB} incorporates O-type standards from the GOSSS catalog \citep{2016ApJS..224....4M}, and a private collection of B stars kindly provided by Dr. Maíz Apellániz. Furthermore, \texttt{MGB} allows the user to interactively broaden the standard spectra by applying rotational convolution, enabling a more accurate comparison when the observed star shows significant line broadening. This functionality is particularly helpful for rapidly rotating stars, as it facilitates the accurate classification when line broadening may lead to misclassifications if not properly accounted for. A detailed example of this reclassification procedure -- including the analysis of helium and silicon line ratios and the use of \texttt{MGB} -- is provided in Appendix~\ref{A: SpT_class}.
    
        Table~\ref{T: OBstars_SpT_vsini} lists the updated spectral types, together with previous types from the literature. Our results generally agree with previous classifications, with most differences within one spectral subtype or luminosity class. However, the central core of our analysis is deeply related to subtle variations of spectral subtypes within the \spectype{B}{0} population of Cygnus~OB2. Our updated classification reveals that approximately $19\,\%$ of the stars previously identified as \spectype{B}{0} are actually late-O types; indeed, all these reclassified O-type stars fall within the \spectype{O}{9.7} subtype, except 2MASS J20295701+4109538, which is classified as \spectype{O}{9.5}.
        }
    }

\section{Projected rotational velocities of O- and B0-type stars in Cygnus~OB2} \label{S: Projected rotational velocities}
    {
    \subsection{Line-broadening characterization} \label{SS: Line-broadening characterization}
        {
        Projected rotational velocities ($\vsini$, where $i$ is the inclination angle of the stellar rotation axis relative to the line of sight) are measured directly from spectral line broadening. Common techniques for measuring $\vsini$ include the full width at half maximum (FWHM) approach \citep{1975ApJS...29..137S, 1992A&A...261..209H, 2002RMxAC..14..111A, 2005AJ....129..809S}, cross-correlation with template spectra \citep{1996ApJ...463..737P, 1997MNRAS.284..265H}, and profile fitting using synthetic lines from model atmospheres \citep{2002MNRAS.336..577R, 2006A&A...456.1131M, 2008A&A...479..541H}. The Fourier transform \citep[FT,][and references therein]{1933MNRAS..93..478C, 1976PASP...88..809S, 1990A&A...237..137D, 2008oasp.book.....G} is a extremely useful method for early-type stars, as it effectively disentangles rotational broadening from other effects, such as macroturbulence. 

        For this study, we derived $\vsini$ values for the \spectype{B}{0} star sample with spectra from the $\mathrm{2.56\,m}$~NOT, $\mathrm{2.54\,m}$~INT and $\mathrm{4.2\,m}$~WHT telescopes, using the \texttt{iacob-broad} tool \citep{2014A&A...562A.135S}. This IDL-based script computes $\vsini$ and macroturbulence ($\mathrm{\textit{v}_{mac}}$) using both the FT and goodness-of-fit (GOF) methods on selected diagnostic lines. The FT approach is based on the Fourier transform of a specific line profile, using the first zero to determine $\vsini$ \citep{2007A&A...468.1063S, 2008oasp.book.....G}. The GOF technique involves matching an observed line profile with a synthetic profile convolved with varying $\vsini-\mathrm{\textit{v}_{mac}}$ values, optimizing the fit through $\chi^2$ minimization. By providing two independent $\vsini$ measurements, \texttt{iacob-broad} enhances consistency checks and helps resolve challenging cases in broadening characterization.

        Given the available spectra for our star sample (which vary significantly in spectral resolution),  we first prioritized the highest possible resolution when measuring the rotational velocities. As a secondary criterion, we favored metallic lines (Table~\ref{T: Lines iacob-broad}), since they are not affected by significant Stark broadening or by nebular contamination. However, when these lines were not available, we followed the selection criteria established in \citet{1992A&A...261..209H}, \citet{2014A&A...562A.135S}, and \citet{2020AA...642A.168B}: 
        \begin{enumerate}[(i)]
            \item Initially, we selected the \specline{O}{III}[5592], \specline{Si}{IV}[4116], \specline{Si}{III}[4552], and  \specline{Si}{II}[6347] lines. The \specline{O}{III}[5592] line is typically stronger and more isolated for O-type stars.

            \item If none of the previous metallic lines were available, we used nebular-free \specline{He}{I} lines, which are typically less affected by Stark broadening than \specline{He}{II} lines.
            
            \begin{itemize}
                 \item For the NOT spectra, we used \specline{He}{I}[6678], carefully excluding the red part of the line in O-type stars to avoid potential contamination from \specline{He}{II}[6683]. The bulk of our $\vsini$ measurements are derived from this line (see Table~\ref{T: Lines iacob-broad}), in order to take advantage of the superior signal-to-noise ratio and spectral resolution of the NOT spectra.

                \item For the rest of the spectra, we adopted an unweighted average of the lines \multispecline{He}{I}{4388, 4471, 4713, 4922} (\specline{He}{I} reference in Table~\ref{T: Lines iacob-broad}). In the case of \specline{He}{I}[4471] and \specline{He}{I}[4713], we excluded the blue component of the line if the forbidden component -- indicating strong Stark effects -- was present.
            \end{itemize}
        \end{enumerate}

        \begin{table}[th!]
            \caption{
                Diagnostic spectral lines used in \texttt{iacob-broad} to compute the projected rotational velocities ($\vsini$).
            } 
            \label{T: Lines iacob-broad}
            \centering   
            \small
            \begin{tabular}{cccccccc}
                \hline
                \hline \\ [-1.8ex]
                Line      & $\mathrm{O{\scriptstyle\,III}}$ & $\mathrm{Si{\scriptstyle\,III}}$ & $\mathrm{He{\scriptstyle\,I}\,6678}$ & $\mathrm{He{\scriptstyle\,I}}$ \\ \hline \\ [-1.8ex]
                \# Stars  & 1                               & 2                                & 25                                   & 4                              \\
                \% Sample & 3                               & 6                                & 78                                   & 13                             \\ \hline        
            \end{tabular}                    
        \end{table}

        To ensure the robustness of our results, we compared the $\vsini$ values obtained using the FT technique with those derived from the GOF approach, as shown in Fig.~\ref{Fig: ComparisonFTGOF}. Whenever possible, we prioritized using rotational velocities computed through the GOF procedure, provided that their FT counterparts fell within the adopted uncertainty range. This method reduces potential subjectivity when selecting the first zero in the Fourier transform. Based on the agreement observed between the two methods in our data, we adopted an uncertainty of $20\,\kmps$ or $20\,\%$, whichever was greater  (see Fig.~\ref{Fig: ComparisonFTGOF}); at lower rotational velocities, additional broadening from spectral resolution and microturbulence may increase the uncertainties.

        Even so, \citet{2014A&A...562A.135S} note that spectral sampling imposes a lower limit below which rotation no longer dominates the broadening of spectral lines -- although $\vsini$ can still be determined. This limit is defined as
        \begin{equation}
            \mathrm{
                \left(\textit{v} \sin{\textit{i}}\right)_{lim} = c \frac{\lambdaup}{\Delta\lambdaup} \sim\,c/R,
            }
            \label{Eq: vsini_limit}
        \end{equation}
        mainly resulting in $\mathrm{\left(\textit{v}\sin{\textit{i}}\right)_{lim}\!\sim\!60\,\kmps}$ for most of our computed velocities; the gray-shaded box in Fig.~\ref{Fig: ComparisonFTGOF} highlights the region below this limit. Within this domain, the spectral resolution is insufficient to accurately fit $\vsini-\mathrm{\textit{v}_{mac}}$ in the GOF procedure. In addition, when applying the FT technique, velocities below the $\mathrm{\left(\textit{v}\sin{\textit{i}}\right)_{lim}}$ threshold are close to the Nyquist frequency, leading to edge effects in the Fourier transform. Hence, we found in this region the only three cases where the FT-GOF results fell outside the reliable bounds, forcing us to adopt the $\vsini$ values computed from the $\mathrm{GOF(\textit{v}_{mac})\!=\!0}$ method (green squares in Fig.~\ref{Fig: ComparisonFTGOF}). As a result, in the low-rotation regime the precision of our $\vsini$ measurements is increasingly compromised for lower and lower velocities, as instrumental broadening tends to dominate over rotational effects. However, in our work, this limitation is not critical since our main goal is to identify and characterize fast rotators within Cygnus~OB2; stars with $\vsini$ values below the approximate threshold $\mathrm{\left(\textit{v}\sin{\textit{i}}\right)_{lim}}$ lie outside the primary scope of our analysis and do not affect our conclusions.

        \begin{figure}[]
            \centering
            \resizebox{\hsize}{!}{\includegraphics[scale=1]{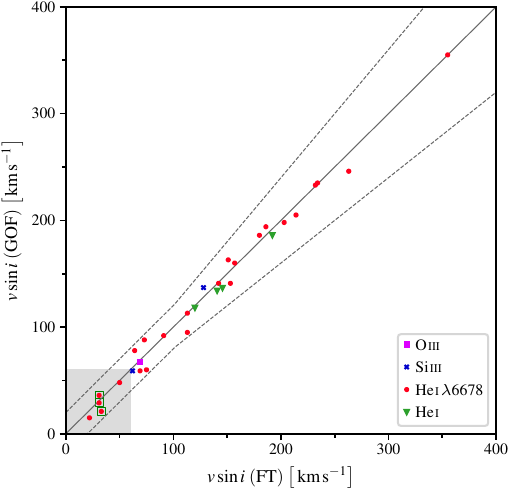}}
            \caption{
                Comparison of projected rotational velocities resulting from \texttt{iacob-broad}, using the Fourier transform (FT) and the goodness-of-fit (GOF) methods. The dashed lines indicate deviations of $20\,\kmps$ or $20\,\%$ (whichever is the largest) from the 1:1 correlation. The gray-shaded square denotes rotational velocities below the $\mathrm{c/R}$ threshold, estimated for a spectral resolution of $\mathrm{R\!=\!5000}$. The green squares show those $\vsini$ in which $\mathrm{GOF(\textit{v}_{mac})\!=\!0}$ is used. 
           }
            \label{Fig: ComparisonFTGOF}
        \end{figure}

        A similar argument applies to the impact of Stark broadening when using \specline{He}{I} lines for $\vsini$ computation. Although the agreement between our $\vsini\left(\mathrm{FT}\right) - \vsini\left(\mathrm{GOF}\right)$ results confirms their reliability, our main interest remains fast rotators, for which rotational broadening is the dominant mechanism.        

        In Table~\ref{T: OBstars_SpT_vsini} we present the projected rotational velocities of our sample of stars, which are further analyzed in Fig \ref{Fig: vsini_SpectralClass}. The spectral reclassification detailed in Sect.~\ref{S: Spectral reclassification} resulted in the identification of six new O stars, while the remaining 26 stars retain their classification as B-types. Notably, our sample includes only six stars with rotational velocities $\vsini\!>\!200\,\kmps$, of which only one O-type star exceeds this threshold.
        
        \begin{figure}[h!]
            \centering
            \resizebox{\hsize}{!}{\includegraphics[scale=1]{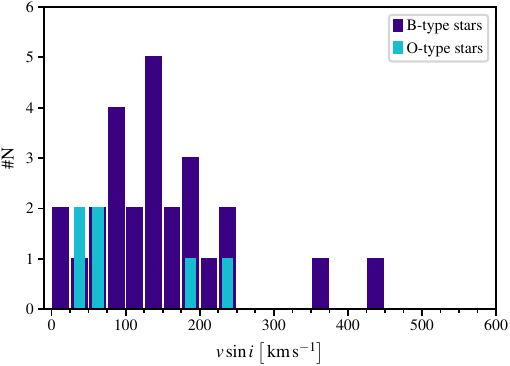}}
            \caption{
                Computed projected rotational velocities for the star sample in this work. Different shades of blue represent the corresponding spectral types (O- or B-type) after our spectral reclassification.
            }
            \label{Fig: vsini_SpectralClass}
        \end{figure}
        }

    \subsection{Cygnus~OB2 rotational velocity distribution from its O-type population} \label{S: Cygnus OB2 rotational velocity distribution from its O-type population}
        {
        \begin{figure*}[th!]
            \centering
            \resizebox{\hsize}{!}{\includegraphics[scale=1]{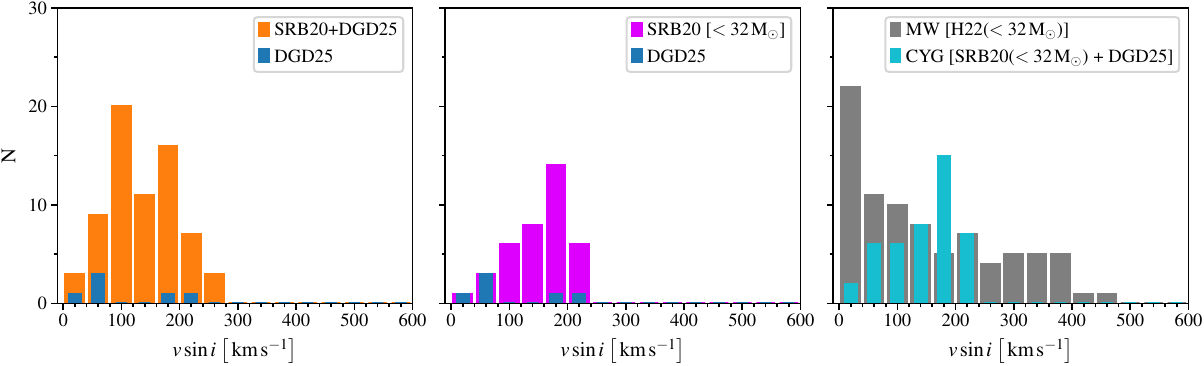}}
            \caption
                {
                Projected rotational velocities for the O-type population in Cygnus~OB2. \textit{Left}: histogram representing all $\vsini$ values for the Cygnus~OB2 O-type stars (in orange), combining results from this study with those from \citet{2020AA...642A.168B}. The rotational velocities computed in this work for O stars are plotted in blue. \textit{Middle}: $\vsini$ data for O-type stars with spectroscopic masses $\mathrm{M\!<\!32\,\Msol}$ \citep[in violet,][]{2020AA...642A.168B}. \textit{Right}: $\vsini$ results from \citet{2022AA...665A.150H} for Galactic O-type stars with $\mathrm{M\!<\!32\,\Msol}$ (gray histogram). The cyan overlay merges the blue and violet histograms from the middle panel.
                }
            \label{Fig: vsini_Distribution_1}
        \end{figure*}
        \begin{figure*}[th!]
            \centering
            \resizebox{\hsize}{!}{\includegraphics[scale=1]{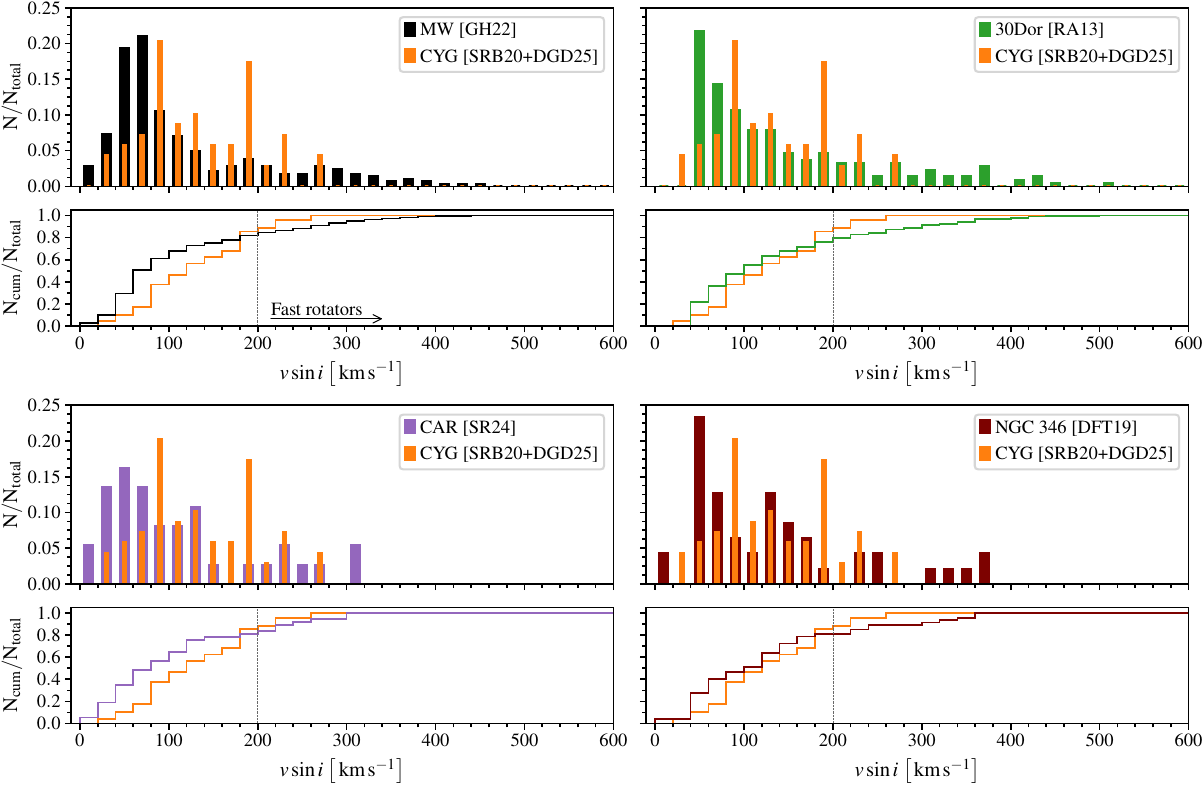}}
            \caption
                {
                    Distribution of projected rotational velocities for O-type stars $\vsini$ across different regions of the local Universe. The histograms in black, green, purple, and brown represent the $\vsini$ measurements for Galactic stars in the solar neighborhood \citep{2022AA...665A.150H}, 30~Doradus \citep{2013AA...560A..29R}, Carina~OB1 \citep{2025AA...695A.248B}, and NGC~346 \citep{2019AA...626A..50D}, respectively. The orange histogram displays the rotational velocity distribution for Cygnus~OB2, integrating O-type stars from \citet{2020AA...642A.168B} with those newly identified in our reclassification of the \spectype{B}{0} sample.
                }
            \label{Fig: vsini_Distribution_2}
        \end{figure*}

        In this section we compare our $\vsini$ measurements of Cygnus~OB2 O-type stars with those from similar studies in Fig.~\ref{Fig: vsini_Distribution_1} and \ref{Fig: vsini_Distribution_2}. The reference works comprise: (i) \citet{2020AA...642A.168B}, which concentrates on the O-type population of Cygnus~OB2; (ii) \citet{2022AA...665A.150H}, focusing on Galactic O stars from the solar neighborhood; (iii) \citet{2013AA...560A..29R}, which examines O-type stars in 30~Doradus; (iv) \citet{2025AA...695A.248B}, studying O stars in Carina~OB1; and (v) \citet{2019AA...626A..50D}, which measures O- and B-type stars in NGC~346. \footnote{For this work we only consider the O population of NGC~346.}
        
        Most of these studies used \texttt{iacob-broad} to characterize line broadening, but with slightly different approaches regarding the diagnostic lines. \citet{2013AA...560A..29R}, \citet{2020AA...642A.168B} (hereafter [SRB20]), and \citet{2025AA...695A.248B} followed criteria similar to ours, prioritizing \specline{O}{III}[5592] and \specline{Si}{III}[4552]. By comparison, \citet{2022AA...665A.150H} (henceforth [GH22]) primarily relied on \specline{O}{III}[5592], but also used \specline{Si}{III}[4552], \multispecline{N}{V}{4603, 4620} and \specline{N}{IV}[6380] when the \specline{O}{III} line was too weak. In contrast, \citet{2019AA...626A..50D} did not use \texttt{iacob-broad}; instead, they estimated $\vsini$ from independent FT and profile fitting applied to \specline{Mg}{II}[4481], \specline{Si}{III}[4552], or \specline{He}{I}[4026], depending on the spectral type and line quality.

        In Fig.~\ref{Fig: vsini_Distribution_1} (left and middle panels) we present in blue the projected rotational velocities of our newly classified O stars within Cygnus~OB2. The orange histogram illustrates the combination of those results with the $\vsini$ values of [SRB20], computed from the O-type population of Cygnus~OB2. These authors identified a distribution of rotational velocities with a peak of slow rotators at $80\!-\!120\,\kmps$, with no sign of stars rotating above $250\,\kmps$. In the middle panel, we specifically highlight O-type stars from [SRB20] with spectroscopic masses below $32\,\Msol$, shown in violet. The right panel merges the previous blue and violet histograms into a new one colored in cyan. This is then compared to the $\vsini$ histogram of O stars with stellar masses below $\mathrm{32\,\Msol}$ from [GH22] (in gray). We used $\mathrm{32\,\Msol}$ non-rotating evolutionary tracks from \citet{2012A&A...537A.146E} with solar metallicity $\mathrm{Z\!=\!0.14}$ to identify the $\mathrm{M\!<\!32\,\Msol}$ stars.

        Figure \ref{Fig: vsini_Distribution_1} (left panel) clearly indicates that Cygnus~OB2 rotational velocities (shown in orange) are skewed toward higher $\vsini$ values. This observed shift highlights the impact of spectral resolution on the accurate computation of stellar rotation: when rotational velocities fall below $\mathrm{\left(\textit{v}\sin{\textit{i}}\right)_{lim}}$, both the FT and GOF methods face significant limitations, leading to an overestimation of the computed $\vsini$ (see Sect.~\ref{SS: Line-broadening characterization}). The main peak of slow rotators identified by [SRB20] ($\vsini \!\leq\! 100\,\kmps$) mainly consists of stars observed at $\mathrm{R \!\sim\! 5000}$, causing a shift to higher velocities. This effect is similarly present in our study, yet it is amplified as we assumed $\mathrm{GOF(\textit{v}_{mac})\!=\!0}$ for most of the low-regime cases; in this range, our $\vsini$ values are even further overestimated as \texttt{iacob-broad} fails to effectively disentangle the rotational and macroturbulence velocities. 

        The middle and right panels of Fig.~\ref{Fig: vsini_Distribution_1} focus on galactic stars with spectroscopic masses $\mathrm{<\! 32\,\Msol}$. Simulations of binary populations \citep{2013ApJ...764..166D} and the analysis of O-type galactic stars \citep{2022AA...665A.150H} suggest that the fast-rotating tail in the rotational velocity distributions of massive stars is primarily populated by stars in this mass range. However, the middle panel highlights a significant pattern in Cygnus~OB2: its O-type population exhibits no evidence of stars rotating at $250\,\kmps$ or higher (violet histogram). This absence becomes even more clear in the right panel when compared with results from [GH22]: we do not find an extended tail of fast rotators in Cygnus~OB2, even after incorporating into [SRB20] our $\vsini$ results -- which involve newly identified O stars from our \spectype{B}{0} sample.

        We also compare the distribution of rotational velocities for O stars in Cygnus~OB2 with those from other regions in Fig.~\ref{Fig: vsini_Distribution_2}. Presented from left to right, top to bottom, the results include: the Galactic O stars from [GH22] in black; the O-type stars in 30~Doradus \citep{2013AA...560A..29R} in green; the O-star population of Carina~OB1 \citep{2025AA...695A.248B} with purple color; and the O stars from NGC~346 \citep{2019AA...626A..50D} in brown. The projected rotational velocities for the O-type stars of Cygnus~OB2, including the results from [SRB20] and our work, are shown in orange.

        The [GH22] data exhibit a bimodal distribution (top-left panel), characterized by a peak of slow rotators at $40-80\,\kmps$ and a high-velocity tail extending up to $450\,\kmps$. In contrast, the Cygnus~OB2 population lacks this extended tail, with rotational velocities reaching only $\sim\!270\,\kmps$; in addition, its shift in the low-velocity peak can be attributed to the constraints imposed by spectral resolution, as stated for Fig.~\ref{Fig: vsini_Distribution_1}. The top-right panel of Fig.~\ref{Fig: vsini_Distribution_2} highlights further contrasts. While Cygnus~OB2 and 30~Doradus are both young, massive star-forming regions, they exist in different environments: Cygnus~OB2 is located in the solar neighborhood, whereas 30~Doradus resides in the Large Magellanic Cloud, an extragalactic region with lower metallicity \citep[$\mathrm{Z_{LMC} \!=\! 0.43\,\Zsol}$,][]{2016MNRAS.455.1855C}. Even so, 30~Doradus continues to display a bimodal distribution, exhibiting a peak at $60 \!-\! 80\,\kmps$ and a fast-rotator tail that reaches $600\,\kmps$. Its apparent absence of very slow rotators is likely due to a higher detection threshold at low $\vsini$, a consequence of limited spectral resolution. Finally, O-type stars in NGC~346 also follow a broad distribution of rotational velocities, peaking around $80\!-\!100\,\kmps$ and with evidence of fast rotators above $300\,\kmps$. This young, massive, star-forming region -- despite its different environment within the Small Magellanic Cloud \citep[$\mathrm{Z_{SMC}\!=\!0.14,Z_\odot}$ ][]{2007A&A...471..625T, 2007A&A...466..277H} -- also hosts a significant number of rapid rotators, in contrast with Cygnus~OB2.

        Notably, Carina~OB1 shows a $\vsini$ distribution similar to Cygnus~OB2, characterized by a remarkable absence of fast rotators (bottom-left panel in Fig.~\ref{Fig: vsini_Distribution_2}). Moreover, its distribution shows a well-defined peak at low $\vsini$, closely resembling the pattern seen in the Galactic population. This low-velocity peak of Carina~OB1 results from the high-resolution data, which reduces the detection threshold and shifts the observed distribution toward lower velocities. \\

        In conclusion, resolution effects introduce an artificial shift in the low-velocity regime. Still, the lack of rapid rotators in Cygnus~OB2 remains evident, even after adding the newly classified late-O stars from the \spectype{B}{0} sample. The rotational velocity distribution shows no extended high-velocity tail, in contrast to what is observed in other environments such as the Milky Way and 30~Doradus. This persistent difference suggests that the population of massive stars in Cygnus~OB2 may have undergone a distinct evolutionary or environmental history that limits the presence of fast rotators.
        }

    \subsection{Rotational velocities for the Cygnus~OB2 OB-type population} \label{S: Rotational velocities for Cygnus~OB2 OB-type population}
        {        
        In this section we include the \spectype{B}{0} stars from Cygnus~OB2 on its rotational velocity distribution (see Fig.~\ref{Fig: vsini_Distribution_3}). In our sample, we identify four \spectype{B}{0} stars with $\vsini\!>\!200\,\kmps$, with only two exceeding $250\,\kmps$. These results slightly improved the representation of high rotational velocities within the distribution.
    
        Notably, this subtle improvement is in line with trends observed in stellar rotation studies. Large-scale surveys, such as that by \citet{2010ApJ...722..605H}, have shown that early B-type stars generally present higher rotational velocities compared to their more massive O-type counterparts, particularly at the zero-age main sequence. This correlation with mass may suggests that including \spectype{B}{0} stars in the Cygnus~OB2 sample will increase the number of rapid rotators. However, according to Fig.~\ref{Fig: vsini_Distribution_3}, this does not alter our main conclusion: Cygnus~OB2 lacks an extended tail of fast rotators for the more massive stars.

        \begin{figure}[H]
            \centering
            \resizebox{\hsize}{!}{\includegraphics[scale=1]{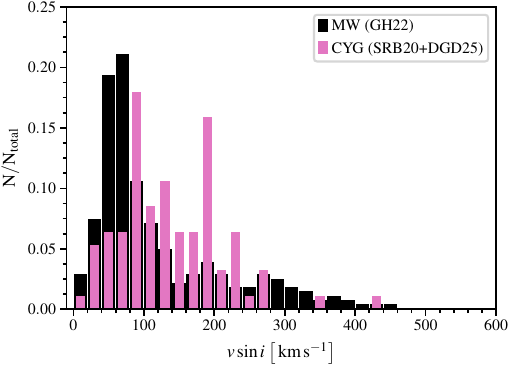}}
            \caption{
                Projected rotational velocities from the O+\spectype{B}{0} population of Cygnus~OB2 (shown in pink), compared with the $\vsini$ distribution of Galactic O stars \citep[in black, ][]{2022AA...665A.150H}. The Cygnus~OB2 distribution contains results obtained by \citet{2020AA...642A.168B} (O-type stars) and all the $\vsini$ computed in this work (new O and \spectype{B}{0} stars).
            }
            \label{Fig: vsini_Distribution_3}
        \end{figure}
    }

\section{Discussion} \label{S: Discussion}
    {
        {        
        The high multiplicity fraction of massive stars suggests that the bimodality observed in the rotational velocity distributions of O-type stars is likely driven by binary interactions \citep{2014ApJ...782....7D, 2022AA...665A.150H, 2024A&A...684A..35B}, although alternative explanations have also been proposed \citep[e.g.,][]{2024A&A...689A.320N}. In Sect.~\ref{S: Cygnus OB2 rotational velocity distribution from its O-type population}, we compare the projected rotational velocities in Cygnus~OB2 against distributions from other Galactic and extragalactic regions, and detect the absence of a fast-rotating tail in Cygnus~OB2. The probability density function, $f\left( x \right)$, for the distributions of \citet{2022AA...665A.150H} and \citet{2013AA...560A..29R}, calculated as    
        \begin{equation}
            \int_{0}^{\infty} f \left( x \right) \,dx = 1,
        \end{equation}
        reveals that the probability of finding O-type stars with $\vsini\!>\!200\,\kmps$ is approximately $0.18$ and $0.25$, respectively. Within the Cygnus~OB2 region, this would imply the presence of 13 to 18 O stars in this high-velocity range. Theoretical simulations by \citet{2013ApJ...764..166D} predict that roughly $19\,\%$ of hot massive stars will reach rotational speeds exceeding $200\,\kmps$, aligning closely with the empirical probabilities obtained by \citet{2022AA...665A.150H}. However, in Cygnus~OB2, we identified only nine O-type stars with $\vsini\!>\!200\,\kmps$ ($\sim\!14\,\%$), a fast-rotating population smaller than both theoretical and empirical predictions. The discrepancy is even worse for $\vsini\!>\!250\,\kmps$, where $10\!-\!12$ O stars are expected, yet only three are observed. 
        
        Following the initial study of the O-star population \citep{2020AA...642A.168B}, and our additional analysis of stars with $\mathrm{M\!<\!32\,\Msol}$ (newly identify late-O types), we confirmed the lack of an extended tail of fast rotators in Cygnus~OB2. In the following sections, we explore several potential explanations for this observed deficit of fast-rotating stars in the region: (i) the early stage of Cygnus~OB2, with insufficient time for spin-up via binary interactions; (ii) environmental conditions that may suppress binary formation or lead to slower initial rotations; (iii) the dynamical ejection of fast rotators as runaway stars; (iv) spin-axis alignment, which could lower projected velocities; and (v) magnetic braking as a mechanism for rotational slowdown.
        }

    \subsection{Evolutionary effects}
        {
        Massive stars often form in binary systems, where strong gravitational attraction leads to high multiplicity \citep{2012Sci...337..444S}. This multiplicity plays a pivotal role in shaping the evolution and fate of massive stars, which are heavily influenced by the interactions and evolutionary stages of their companions. Particularly, spin rates in binary systems are modified by tidal forces and mass transfer processes (see Fig.~\ref{Fig: BinarySystem_Evolution}). Such interactions are not present in isolated stars, but they significantly spin up one or both components of the system.

        During the initial pre-interaction phase, both stars remain within their Roche lobes in a detached configuration, with interactions limited to stellar winds and tidal forces \citep{2013ApJ...764..166D}. For $\mathrm{P\!<\!10\,d}$, tidal interactions dominate the spin rate change, working to synchronize the rotation of each star with their shared orbital motion. As this synchronization process occurs, it tends to equalize the angular velocities of both stars. This keeps the primary star in corotation with the orbit as it expands, and gradually increases its rotation until critical velocity is reached. Consequently, the primary can spin up to a rotational velocity exceeding $200\,\kmps$ before filling its Roche lobe -- with $300\,\kmps$ as an upper limit that tidal forces can achieve alone.

        The age of the O-type population in Cygnus~OB2 is estimated between 1 and $\mathrm{6\,Myr}$, with evidence of at least two main star formation episodes \citep{2020AA...642A.168B}. These findings are consistent with \citet{2008A&A...487..575N}, who proposed an age of $\mathrm{\sim 2.5\,Myr}$ for the association, while also detecting a slightly older population. Furthermore, \citet{2015MNRAS.449..741W} suggested a continuous star formation period spanning from 1 to $\mathrm{7\,Myr}$, supporting this age range. This maximum age of $\mathrm{6\!-\!7\,Myr}$ for the massive stellar population of Cygnus~OB2 aligns with the fact that only tides produce a significant spin-rate change during the pre-interaction phase; our $\vsini$ results together with those from \citet{2020AA...642A.168B} show that approximately $86\,\%$ of the O-type stars in Cygnus~OB2 have rotational velocities below $200\,\kmps$, while no star exceeds $300\,\kmps$.

        The lack of an extended fast-rotating tail in Cygnus~OB2 could be attributed to tidal forces alone driving spin rate changes as mass transfer events are typically expected after $\mathrm{8\!-\!10\,Myr}$. Indeed, the full potential of tidal acceleration may not yet have been reached, as reflected by the relatively modest population with velocities between $200\!-\!300\,\kmps$ (only nine stars in Cygnus~OB2). Despite a young evolutionary stage, it remains possible to find stars with velocities over $300\,\kmps$ in cases where mergers have occurred within the tightest binary systems \citep{2013ApJ...764..166D}, although this situation is unclear. Within Cygnus~OB2, we find only one star with $\vsini \!\sim\! 270\,\kmps$ \citep{2020AA...642A.168B}, potentially resulting from a merger according to the young age of the region.

        However, observations of other regions indicate that age alone is insufficient to explain the absence of a well-populated high-velocity tail in Cygnus~OB2. In 30~Doradus, \citet{2013AA...560A..29R} compared the rotational velocity distributions of the slightly older cluster NGC~2060 ($\mathrm{5\,Myr}$) and the younger NGC~2070 \citep[$\mathrm{1\!-\!2\,Myr}$,][]{2018A&A...618A..73S}, observing no significant difference: both clusters exhibit a pronounced peak of slow rotators and a high-velocity tail extending up to $\sim\!600\,\kmps$. NGC~346 also presents a fast-rotating population despite its young age \citep[$\mathrm{\gtrsim\!3\,Myr}$,][]{2019AA...626A..50D}. Crucially, the young evolutionary stage of their O-type populations does not seem to have strongly affected the $\vsini$ distributions. In contrast, \cite{2025AA...695A.248B} do not report the presence of an extended tail of fast rotators in Carina~OB1. This region is one of the youngest associations in the Galaxy and it contains Trumpler~14, which has an estimated age of $\mathrm{1\,Myr}$ \citep{2010A&A...515A..26S, 2025AA...695A.248B}.
        
        These differences in rotational velocities -- despite similar evolutionary statuses -- suggest that the lack of fast rotators in Cygnus~OB2 cannot be explained by age alone. Instead, it likely reflects a more complex interplay between the evolutionary stage and additional physical factors.

        \subsubsection*{The low-velocity population in Cygnus~OB2}
            {
            The presence of a high fraction of low-$\vsini$ stars in Cygnus~OB2, combined with the absence of a well-populated high-velocity tail, also suggests the importance of its evolutionary stage. Considering that slow and fast rotators might originate from binary interactions, then we would expect that both populations would be similarly affected by evolutionary processes. However, we observe slow rotators, but no fast rotators.
            
            Recent hydrodynamic simulations by \citet{2019Natur.574..211S} offer a potential explanation. Binary mergers increase angular momentum during the beginning of the merging phase, but rapidly lose it due to strong mass loss. As a result, merger products become magnetic slow rotators. Binary mergers are also expected to constitute a non-negligible fraction of pre-main sequence stars \citep[up to $\sim\!30\,\%$, see][]{2020MNRAS.491.5158T}, and this fraction may increase by $8^{+9}_{-4}\,\%$ before $\mathrm{\sim\!10\,Myr}$ when considering evolution of binary systems \citep{2013ApJ...764..166D, 2014ApJ...782....7D}. 

            However, the O-star population in Cygnus~OB2 ($\mathrm{1\!-\!6\,Myr}$) is not old enough to explain the lack of fast rotators as a result of a large number of mergers. While binary evolution and mergers may contribute to the slow-rotator population, they cannot fully account for the absence of fast-rotating stars. The high fraction of low-$\vsini$ stars in Cygnus~OB2 may instead indicate that some stars are simply born as slow rotators, reflecting an intrinsic distribution of angular momentum at birth. From this perspective, initial star formation conditions -- such as angular momentum transport and magnetic braking -- likely play a key role in shaping the observed low-rotation properties.
            }
        }

    \subsection{Properties of the molecular cloud}
        {
        \begin{figure*}[th!]
            \centering
            \resizebox{\hsize}{!}{\includegraphics[scale=1]{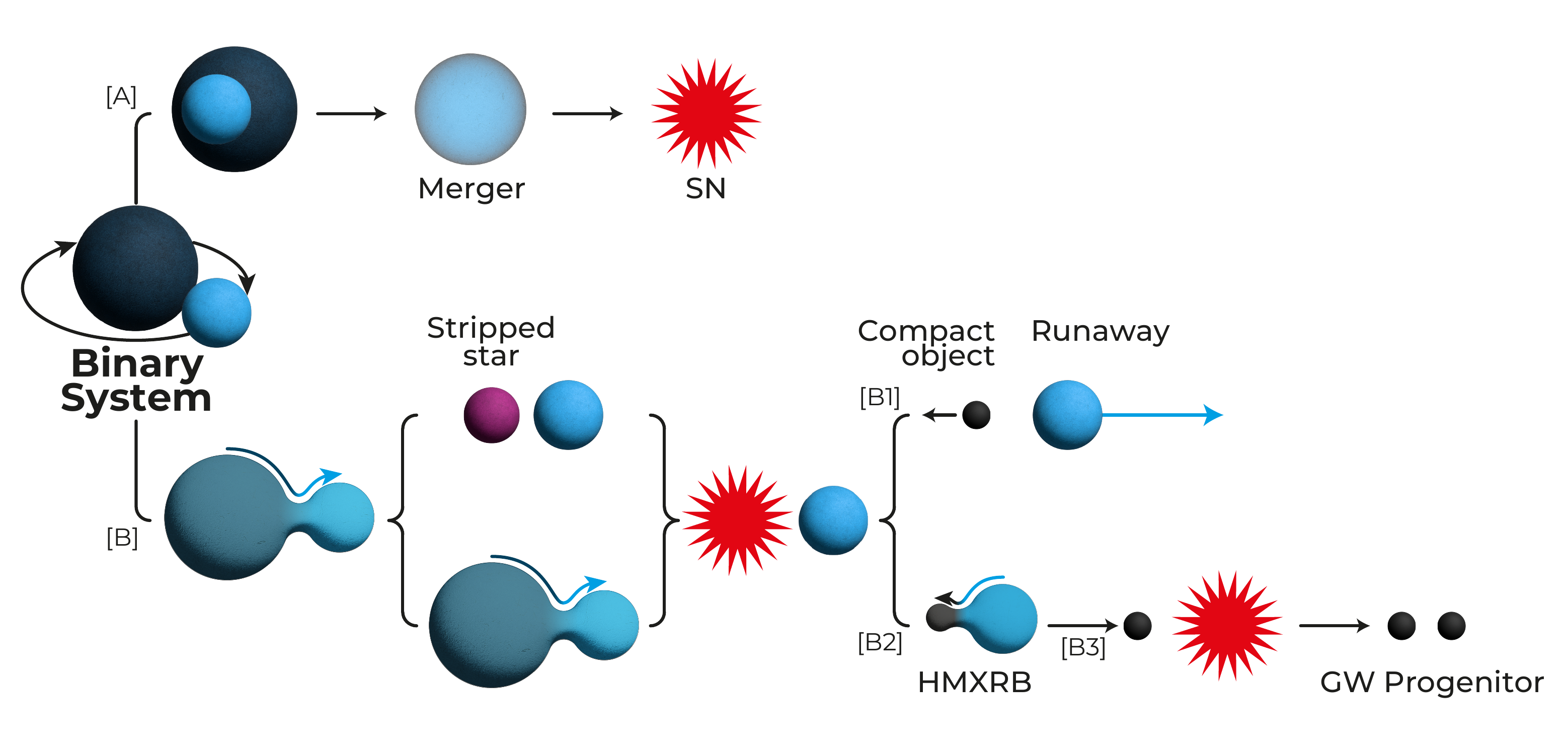}}
            \caption{
                Schematic evolutionary paths of a massive binary system. Initially, if the two stars orbit close enough to each other, they may eventually merge, forming a single, more massive star (top path). Alternatively, the more massive star may explode as a supernova (SN), potentially ejecting the companion as a high-velocity runaway star or resulting in a system with a compact object. The companion may later undergo its own SN event, creating a double compact-object system that could eventually merge and emit gravitational waves (bottom paths).
            }
                \label{Fig: BinarySystem_Evolution}
        \end{figure*}

        \subsubsection{Multiplicity fraction}
            {
            During the gravitational collapse of a molecular cloud, the conservation of momentum leads to a greater concentration of angular energy in the central region. This phenomenon should result in most forming stars rotating near critical velocities; however, since extreme rotators among massive O-type stars are rare, there must be mechanisms to transport away angular momentum. Several explanations have been proposed to resolve this discrepancy, including disk-mediated accretion \citep{2011MNRAS.416..580L} and magnetic coupling \citep{1998APS..APR..I801M}. Still, the impact of their spin-down effect would not be sufficient \citep[e.g.,][]{2012ApJ...748...97R}. The most accepted explanation, supported by observations, is that massive stars form in multiple systems, redistributing angular momentum into orbits and reducing the need for high initial spins. Indeed, multiplicity is intrinsic to massive stars \citep{2012Sci...337..444S}, with a close binary fraction of $\mathrm{\mathcal{F}_{bin} \!=\! 0.69 \!\pm\! 0.09}$.

            \subsubsection*{I. Dependence on the molecular cloud}
                {
                The exact formation mechanisms of multiple massive star systems remain poorly constrained \citep{2023ASPC..534..275O}. Current theories can be broadly grouped into three categories: (i) fragmentation of a molecular core or filament, (ii) gravitational instability within a massive accretion disk, or (iii) fragmentation through dynamical interactions.

                These pathways do not act in isolation. Feedback mechanisms (such as radiation, protostellar outflows, magnetic fields, and turbulence) can reshape the gas distribution and regulate star formation rates \citep{2023ASPC..534..275O}. In doing so, they influence formation outcomes and the resulting multiplicity fraction. If such mechanisms had strongly suppressed the formation of multiple systems in Cygnus OB2, the binary fraction would be lower. Fewer binaries then mean fewer interactions to spin up secondary stars, leaving a smaller population of rapid rotators.

                In addition, these feedback mechanisms may also influence the distribution of orbital separations. For instance, if binaries tend to form with wider orbits in Cygnus~OB2, tides and mass transfer episodes would be less efficient. This would reduce the number of high-velocity stars and help explain the observed scarcity of fast rotators.
                }

            \subsubsection*{II. The influence of metallicity}
                {
                Observational studies support that the formation of close binaries is largely independent of metallicity. \citet{Sana2025} report a slight anticorrelation between metallicity and the fraction of close massive binaries, with
                \begin{equation}
                    \mathrm{\mathcal{F}_{bin} = \left(0.59 \!\pm\! 0.06\right) + (-0.11 \!\pm\! 0.15)\log_{10}\left(Z/Z_\odot\right)}
                \end{equation} 
                for the O-type population. However, this trend is not statistically significant and the fraction of close massive binaries is similar across different metallicity environments. This suggests that metallicity has little effect on their formation.

                In summary, metallicity does not appear to significantly affect the formation of close massive binaries, where interactions such as tides and mass transfer can produce fast-rotating stars. This is supported by the presence of fast rotators in regions with different metallicities, such as 30~Doradus \citep[$\mathrm{Z_{LMC} \!=\!0.43\,\Zsol}$][]{2016MNRAS.455.1855C} and the solar neighborhood ($\mathrm{Z \!\sim\! \Zsol}$). Therefore, the scarcity of a fast-rotating tail in Cygnus~OB2 cannot be explained by metallicity altering the binary fraction.
                }
                 
            \subsubsection*{III. Dynamical interactions}
                {
                Survival of multiple systems depend on the density history of their birth region \citep{2023ASPC..534..129W}. After formation, these binary and multiple systems can be altered or destroyed by processes such as secular decay (particularly in high-order systems) and external dynamical encounters. Close encounters are more frequent and energetic in high-density regions, disrupting existing systems and reducing multiplicity. In contrast, low-density environments shield stellar systems from such interactions, allowing them to remain largely unaltered. 

                Evidence presented by \citet{2018MNRAS.476.2493G} shows that the presence of wide binaries in Cygnus~OB2 is inconsistent with a dense cluster origin, where such systems are typically disrupted. Instead, the survival of these systems within Cygnus~OB2 is indicative of its low-density nature. Thus, in associations like Cygnus~OB2, stars are less affected by external encounters or irradiation \citep{2014MNRAS.438..639W}. However, secular processes may still play a role in shaping multiplicity. Consequently, it is unlikely that dynamical interactions have significantly modified its primordial binary fraction. 

                Despite, a low multiplicity fraction in Cygnus~OB2 -- set by its formation conditions rather than by dynamical interactions -- may account for the absence of a fast-rotating tail. With fewer close binaries, key spin-up mechanisms such as tidal forces and mass transfer are less frequent, leading to a lower number of fast rotators compared to other regions.
                }
            }

        \subsubsection{Stellar formation as a regulator of initial rotation}
            {        
            Observed variations in rotational velocities among early-B stars in high- and low-density environments are likely driven by differences in their formation conditions \citep[see][]{2007AJ....133.1092W}. According to magnetically-regulated accretion models \citep{1991ApJ...370L..39K, 1994ApJ...429..781S, 2005ApJ...634.1214L, 2005MNRAS.356..167M}, the inital stellar rotation is given by
            \begin{equation}
                v_{\mathrm{init}} \sim M^{5/7} \, \dot{M}_{\mathrm{acc}}^{3/7} \, B^{-6/7} \, R^{-11/7},
            \end{equation}
            where $M$ is the stellar mass, $\dot{M}_{\mathrm{acc}}$ is the accretion rate, $B$ is the magnetic field, and $R$ is the radius of the star.
            
            In dense environments, higher core densities contribute to shorter collapse times and higher accretion rates during stellar assembly \citep{2003ApJ...585..850M}. Therefore, stars are expected to initially rotate faster. In addition, circumstellar disks in dense clusters are easily disrupted; photoevaporation from nearby O stars \citep{1998ApJ...499..758J, 2006MNRAS.365.1333W, 2006MNRAS.370L..85S} and gravitational interactions \citep{2006A&A...454..811P} shorten disk lifetimes. These processes limit the period over which magnetic coupling can remove angular momentum, allowing stars to retain higher rotation rates when forming.
            
            In contrast, O stars within Cygnus~OB2 likely formed in a low-density environment. Thus, longer-lived disks and lower accretion rates may have led to slower rotational velocities at birth. As a result, spin-up from tides and mass transfer might be less evident at the current age of Cygnus~OB2, thus explaining its observed lack of fast rotators.
            }
        }

     \subsection{Runaway scenario}
        {        
        As a massive binary system evolves, it can follow several evolutionary paths (see Fig.~\ref{Fig: BinarySystem_Evolution}). Typically, the more massive star will evolve faster, eventually reaching the supernova stage. The SN explosion may disrupt the system, ejecting the companion at high velocity as a runaway star. This runaway is left behind as a single star that moves across the sky with a relatively large proper motion.

        During their initial evolutionary stages, binary stars are in a pre-interaction phase, when only tidal forces affect their rotation. In this phase, stars may experience either acceleration to velocities between $200\!-\!300\,\kmps$ or deceleration when their initial velocities exceed $300\,\kmps$ \citep{2011BSRSL..80..543D, 2013ApJ...764..166D}. After this stage, binaries with orbits that are large enough to avoid merging enter in a mass-transfer state from the primary to the secondary (path [B] in Fig.~\ref{Fig: BinarySystem_Evolution}). This mass exchange occurs via Roche-lobe overflow, with angular momentum transferred either by direct impact onto the secondary’s surface (in short-period systems, $\mathrm{P\!\sim\!2\!-\!5\,d}$) or via an accretion disk (for longer periods, $\mathrm{P\!\sim\!5\!-\!100\,d}$). Following mass transfer, the primary may become a stripped star if it does not first explode as an SN, while the secondary turns into a massive fast-rotating star in the range of $300\!-\!600\,\kmps$.

        \begin{figure}[t]
            \centering
            \resizebox{\hsize}{!}{\includegraphics[scale=1]{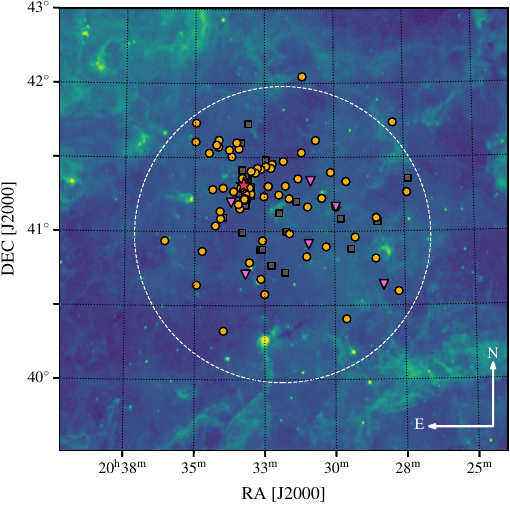}}
            \caption{
                \textit{MSX} $\mathrm{8\,\mu m}$ image showing the spatial distribution of the O-type stars from Cygnus~OB2 analyzed by \citet{2020AA...642A.168B} (orange circles) and those identified in this work (pink triangles). The B0 population is plotted with squares in gray. For reference, the red star locates CygOB2~\#8A, with J2016.0 coordinates $\mathrm{RA\!=\!20\!:\!33\!:\!15.078}$ and $\mathrm{DEC\!=\!+41\!:\!18\!:\!50.479}$. The dashed circle defines a $\mathrm{1^\circ}$ radius region, centered on Galactic coordinates $l\!=\!79.8^\circ$ and $b\!=\!+0.8^\circ$ \citep{2000A&A...360..539K}. North is up, and east is to the left. 
            }
            \label{Fig: Starfield_Ostars_Cygnus}
        \end{figure}
        
        Throughout the evolution of a binary system, the secondary star spins up through tidal forces and mass transfer. If the primary star undergoes a SN explosion and ejects the secondary (Fig.~\ref{Fig: BinarySystem_Evolution} path [B1]), the resulting runaway star is expected to retain a high rotational velocity. Thus, a potential explanation for the observed lack of fast rotators in Cygnus~OB2 is the spatial distribution of the observed population; the O-type stars analyzed by \citet{2020AA...642A.168B} and in this work are concentrated within a $1^\circ$ radius around its central region (see Fig.~\ref{Fig: Starfield_Ostars_Cygnus}). Given the distance to Cygnus~OB2 \citep[$\mathrm{1.74\,kpc}$, ][]{2019MNRAS.484.1838B}, a star ejected with a velocity of $30\,\kmps$ on the sky plane would only need $\mathrm{1\,Myr}$ to be outside the central area. Thus, fast rotators may now lie beyond the central region of Cygnus~OB2, having been ejected as high proper motion runaway stars.

        Indeed, the observed spatial distribution of fast-rotating O-type stars in 30~Doradus provides empirical support for the runaway scenario. \citet{2013AA...560A..29R} show that the high-velocity tail is significantly more populated among stars located outside the central clusters NGC~2070 and NGC~2060. Moreover, \citet{2022A&A...668L...5S} report an overabundance of rapid rotators ($\vsini \!>\! 200\,\kmps$) among the runaway population (see their Fig.~4). This trend reinforces the idea that runaway stars could account for the missing population of fast rotators in Cygnus~OB2.

        \subsubsection*{Star sample and distance effect}
            {
            It is also important to consider how distances to star-forming regions affect the spatial coverage and size of the observed samples. Cygnus~OB2 lies at a distance of  $\mathrm{1.74\,kpc}$ \citep{2019MNRAS.484.1838B}, so $2^\circ$ on the sky corresponds to a physical radius of about $\mathrm{30\,pc}$. In contrast, at the distance of 30~Doradus \citep[$\mathrm{51.7\,kpc}$,][]{2005coex.conf..585P}, the same angular size spans nearly $\mathrm{900\,pc}$ -- 30 times the physical scale compared to Cygnus~OB2. 
            
            These differences imply that observations of distant regions can naturally include stars over a much larger physical area, including a mix of cluster members and field stars. In contrast, our observations of Cygnus~OB2 are constrained to its core -- a smaller physical volume limited by both its proximity and our angular observation coverage. Consequently, our star sample is dominated by cluster members, with few field stars represented.

            Assuming roughly constant and comparable stellar densities, the distance also impacts the relative number of stars observable in different regions. The smaller physical volume covered in Cygnus~OB2 leads to a lower number of sampled stars compared to more distant regions, such as 30~Doradus (see Table~\ref{T: N_samples}).
            
            To properly account for sample size effects, we recomputed the $\vsini$ distributions shown in Fig.~\ref{Fig: vsini_Distribution_2} using the bootstrap resampling technique (see Appendix~\ref{A: Bootstrap} for details). The bootstrapped distributions remain consistent with the original histograms, confirming that the observed lack of fast rotators in Cygnus~OB2 is not a product of sampling or bin size. Instead, this new analysis reveals a statistically significant deficit, indicating that the absence of a high-velocity tail is intrinsic to the region. However, this conclusion remains potentially biased as a result of the restricted spatial coverage of the Cygnus~OB2 sample.
            }
        }

    \subsection{Spin orientation}
        {    
        Spectroscopic data only provide projected rotation, as rotational velocities are derived from spectral line broadening. In this context, the lack of fast rotators in Cygnus~OB2 could reflect a preferential alignment of stellar spin axes with our line of sight. 
        
        Simulations of star formation \citep[e.g.,][]{2003MNRAS.339..577B, 2009MNRAS.397..232B, 2009MNRAS.392.1363B, 2010MNRAS.401.1505B} indicate that chaotic accretion, disk fragmentation, and dynamical interactions effectively dissipate angular momentum and produce a random spin-axis distribution among forming stars. Empirical support for this scenario is provided by observations of Be stars, which exhibit circumstellar disks that appear randomly oriented \citep{1999A&A...352L..31H}.
        
        While an alignment of stellar spin axes has been observed in some clusters -- such as NGC~6791 and NGC~6819, \citep{2017NatAs...1E..64C} -- it is noteworthy that these systems are significantly older (on gigayear timescales) than Cygnus~OB2 ($\mathrm{1-6\,Myr}$). Given this age difference, and considering the expected turbulence and dynamical interactions during massive star formation, a global spin-axis alignment in Cygnus~OB2 is highly improbable. In any case, if such an alignment exists -- with spin axes preferentially oriented toward our line-of-sight -- it could lead to a systematic underestimation of rotational velocities in the observed O-type population of Cygnus~OB2.
        }

    \subsection{Magnetic fields}
        {
        An alternative origin for magnetic fields (other than mergers) is the fossil field hypothesis. In this scenario, the magnetic field is produced during the star formation process with a stable, large-scale configuration \citep{2010A&A...517A..58D}. Moreover, theoretical models propose that rotation may drive internal dynamos, which are capable of generating magnetic fields \citep{1999A&A...349..189S}. 
        
        In both scenarios, metallicity plays a key role in regulating angular momentum loss though magnetic fields. At higher metallicities, magnetic braking is more efficient \citep{2023ApJ...952...79L}, potentially accounting the lack of fast rotators in Cygnus~OB2. The higher metallicity environment of Cygnus~OB2 compared to 30~Doradus ($\mathrm{Z_{LMC}\!=\!0.5\,\Zsol}$) may enhance angular momentum loss through magnetic fields, contributing to the scarcity of very fast rotators.

        However, magnetic breaking appears to be an unlikely mechanism. Observations indicate that most O-type stars in the Milky Way show no detectable magnetic fields. Among the small number of confirmed magnetic O-type stars, surface field strengths are typically on the order of a few hundred gauss up to several $\mathrm{kG}$ \citep{2009ARA&A..47..333D, 2012MNRAS.426.2208G}. While such fields are generally considered too weak to cause significant spin-down, the precise field strength required for effective braking remains uncertain.
        
        Even so, theoretical models suggest that strong magnetic fields could form and remain confined to the stellar core, avoiding detection in the outer layers \citep{2019MNRAS.486.1652P}. If magnetic braking is indeed responsible for the lack of fast rotators within Cygnus~OB2, this leads to the question of why similar effects are not observed in other massive star-forming regions.
        }
    }

\section{Conclusions} \label{S: Conclusions}
    {    
    We revisited the distribution of projected rotational velocities in Cygnus~OB2, aiming to clarify the observed lack of fast rotators ($\vsini\!>\!200\,\kmps$) in its O-type population. This feature, initially identified by \citet{2020AA...642A.168B}, contrasts with theoretical predictions \citep{2013ApJ...764..166D} and observations from other Galactic and extragalactic populations \citep[e.g.,][]{2022AA...665A.150H, 2015A&A...580A..92R}.

    Our work contains a detailed spectral reanalysis of stars classified as \spectype{B}{0} in Cygnus~OB2. By combining new spectroscopic data with spectral classification tools (which incorporate rotational broadening), we identified significant misclassifications. We find that approximately $22\,\%$ of our sample are actually late-O stars, predominantly \spectype{O}{9.7}. This suggests that some fast rotators may have been misclassified due to inconsistent classification criteria, limited signal-to-noise ratios, moderate spectral resolutions, and the impact of rotation on diagnostic lines.

    Although we recovered some fast rotators after our reclassification, the number of stars with $\vsini\!>\!200\,\kmps$ remains notably limited: only ten O-type stars within Cygnus~OB2 satisfy this threshold. This is significantly lower than both theoretical predictions and empirical evidence, which estimate that $\sim19\%$ of massive stars should occupy this high-velocity tail \citep[see, for reference,][]{2013ApJ...764..166D, 2013AA...560A..29R, 2022AA...665A.150H}.

    We consider that the deficit of fast rotators in Cygnus~OB2 is most likely due to a combination of the following three factors:
    \begin{itemize}
        \item Evolutionary stage: Given the young age of Cygnus~OB2 \citep[$\mathrm{1\!-\!6\,Myr}$;][]{2020AA...642A.168B}, our results suggest that spin-up via mass transfer may not have occurred yet. Instead, tidal interactions -- limited to boosting rotation up to $300\,\kmps$ -- have likely dominated the angular momentum evolution so far. However, comparisons with older and younger stellar populations \citep[e.g., NGC~2060 and NGC~2070 in 30~Doradus;][]{2013AA...560A..29R} show that age alone cannot explain the lack of fast rotators in Cygnus~OB2. \\
               
        \item Environmental properties: Certain star formation conditions in Cygnus~OB2 may reduce binary formation and consequently reduce spin-up interactions. In addition, slow initial rotation -- from longer disk lifetimes \citep[e.g.,][]{2006A&A...454..811P, 2006MNRAS.370L..85S} and lower accretion rates \citep{2003ApJ...585..850M} during stellar formation in a low-density environment -- likely contributes to the lack of fast rotators. \\
        
        \item Runaway scenario: Fast rotators may have been dynamically ejected from the association core as surviving companions of supernova explosions. As a result, they could now reside beyond the central $1^\circ$ region of Cygnus~OB2, typically surveyed in most studies. \\
    \end{itemize}

    Our study highlights the need for continued investigation into the dynamics and rotational properties of the Cygnus~OB2 massive star population. Although we have addressed classification inconsistencies and reduced some observational biases, the deficit of fast rotators remains unresolved. Further progress will require high-resolution, high S/N, and multi-epoch spectroscopy, extended beyond the central region of the association. Complementary kinematic surveys to identify potential runaway stars, along with a deeper characterization of binary properties (e.g. binary fraction) are also essential to fully understand the rotational velocity distribution of Cygnus~OB2.
    }

    {
        \begin{acknowledgements} 
            {
            The authors acknowledge support funding from the Ministry of Science, Innovation and Universities (MCIU) through the Spanish State Research Agency (AEI) via grant PID2021-122397NB-C21 (cofunded by the European Regional Development Fund, FEDER) and the Severo Ochoa Program 2020-2023 (CEX2019-000920-S). SRB also thanks financial support from NextGeneration EU/PRTR and MIU (UNI/551/2021) through grant Margarita Salas-ULL. MAM project received support from the ``La Caixa'' Foundation (ID 100010434) under the fellowship code LCF/BQ/PI23/11970035. This work is partially based on observations made with the three following telescopes: (i) we thank the Nordic Optical Telescope, owned in collaboration by the University of Turku and Aarhus University, and operated jointly by Aarhus University, the University of Turku and the University of Oslo, representing Denmark, Finland and Norway, the University of Iceland and Stockholm University at the Observatorio del Roque de los Muchachos, La Palma, Spain, of the Instituto de Astrofisica de Canarias. The data presented in this work were obtained in part with ALFOSC, which is provided by the Instituto de Astrofisica de Andalucia (IAA) under a joint agreement with the University of Copenhagen and NOT; (ii) the Gran Telescopio Canarias is installed at the Spanish Observatorio del Roque de los Muchachos of the Instituto de Astrofísica de Canarias, on the island of La Palma. GTC data was obtained with the instrument OSIRIS, built by a Consortium led by the Instituto de Astrofísica de Canarias in collaboration with the Instituto de Astronomía of the Universidad Autónoma de México. OSIRIS was funded by GRANTECAN and the National Plan of Astronomy and Astrophysics of the Spanish Government; (iii) the Isaac Newton Telescope and its service mode are operated on the island of La Palma by the Isaac Newton Group of Telescopes in the Spanish Observatorio del Roque de los Muchachos of the Instituto de Astrofísica de Canarias. This work has made use of data from the European Space Agency (ESA) mission {\it Gaia} (\url{https://www.cosmos.esa.int/gaia}), processed by the {\it Gaia} Data Processing and Analysis Consortium (DPAC, \url{https://www.cosmos.esa.int/web/gaia/dpac/consortium}). Funding for the DPAC has been provided by national institutions, in particular the institutions participating in the {\it Gaia} Multilateral Agreement. During the spectroscopic data reduction we used \texttt{PyRAF}, a Python environment for \texttt{IRAF}. \texttt{IRAF} is distributed by the National Optical Astronomy Observatories, which is operated by the Association of Universities for Research in Astronomy, Inc. (AURA) under cooperative agreement with the National Science Foundation. In addition to the codes explicitly cited in the text, this paper makes use of the following packages: \textsc{matplotlib} \citep{2007CSE.....9...90H}, \textsc{numpy} \citep{2020Natur.585..357H}, and \textsc{scipy} \citep{2020NatMe..17..261V}. We are deeply grateful to J. Maíz Apellániz for generously providing access to his private \texttt{MGB} library, which was essential in our spectral classification. Finally, we thank the anonymous referee for the constructive, useful and positive feedback, which has significantly contributed to the improvement of this paper.
            }
        \end{acknowledgements}

        {
        \bibliographystyle{aa}   
        \bibliography{export-bibtex}

@article{Sana2025,
  author       = {Sana, H. and Shenar, T. and Bodensteiner, J. and Britavskiy, N. and Langer, N. and Lennon, D. J. and Mahy, L. and Mandel, I. and de Mink, S. E. and Patrick, L. R. and Villaseñor, J. I. and Dirickx, M. and Abdul-Masih, M. and Almeida, L. A. and Backs, F. and Berlanas, S. R. and Bernini-Peron, M. and Bowman, D. M. and Bronner, V. A. and Crowther, P. A. and Deshmukh, K. and Evans, C. J. and Fabry, M. and Gieles, M. and Gilkis, A. and González-Torà, G. and Gräfener, G. and Götberg, Y. and Hawcroft, C. and Hénault-Brunet, V. and Herrero, A. and Holgado, G. and Izzard, R. G. and de Koter, A. and Janssens, S. and Johnston, C. and Josiek, J. and Justham, S. and Kalari, V. M. and Klencki, J. and Kubát, J. and Kubátová, B. and Lefever, R. R. and van Loon, J. Th. and Ludwig, B. and Mackey, J. and Maíz Apellániz, J. and Maravelias, G. and Marchant, P. and Mazeh, T. and Menon, A. and Moe, M. and Najarro, F. and Oskinova, L. M. and Ovadia, R. and Pauli, D. and Pawlak, M. and Ramachandran, V. and Renzo, M. and Rocha, D. F. and Sander, A. A. C. and Schneider, F. R. N. and Schootemeijer, A. and Schösser, E. C. and Schürmann, C. and Sen, K. and Shahaf, S. and Simón-Díaz, S. and van Son, L. A. C. and Stoop, M. and Toonen, S. and Tramper, F. and Valli, R. and Vigna-Gómez, A. and Vink, J. S. and Wang, C. and Willcox, R.},
  title        = {A high fraction of close massive binary stars at low metallicity},
  journal      = {Nature Astronomy},
  year         = {2025},
  volume       = {},
  number       = {},
  pages        = {},
  doi          = {10.1038/s41550-025-02610-x},
  url          = {https://doi.org/10.1038/s41550-025-02610-x},
  issn         = {2397-3366}
}

@article{ADtest,
  title={A test of goodness of fit},
  author={Anderson, Theodore W and Darling, Donald A},
  journal={Journal of the American statistical association},
  volume={49},
  number={268},
  pages={765--769},
  year={1954},
  doi = {10.1080/01621459.1954.10501232},
  publisher={Taylor \& Francis}
}

@article{Scholz+87,
  title={K-sample Anderson--Darling tests},
  author={Scholz, Fritz W and Stephens, Michael A},
  journal={Journal of the American Statistical Association},
  volume={82},
  number={399},
  pages={918--924},
  year={1987},
  doi={10.2307/2288805},
  publisher={Taylor \& Francis}
}

@article{Kushary01052000,
    author = {Debashis Kushary},
    title = {Bootstrap Methods and Their Application},
    journal = {Technometrics},
    volume = {42},
    number = {2},
    pages = {216--217},
    year = {2000},
    publisher = {ASA Website},
    doi = {10.1080/00401706.2000.10486018},
    URL = {https://www.tandfonline.com/doi/abs/10.1080/00401706.2000.10486018},
    eprint = {https://www.tandfonline.com/doi/pdf/10.1080/00401706.2000.10486018}
}

@article{10.1214/16-SS113,
    author = {Zeinab Mashreghi and David Haziza and Christian L{\'e}ger},
    title = {{A survey of bootstrap methods in finite population sampling}},
    volume = {10},
    journal = {Statistics Surveys},
    number = {none},
    publisher = {Amer. Statist. Assoc., the Bernoulli Soc., the Inst. Math. Statist., and the Statist. Soc. Canada},
    pages = {1 -- 52},
    year = {2016},
    doi = {10.1214/16-SS113},
    URL = {https://doi.org/10.1214/16-SS113}
}

@ARTICLE{2025AA...695A.248B,
       author = {{Berlanas}, S.~R. and {Mahy}, L. and {Herrero}, A. and {Ma{\'\i}z Apell{\'a}niz}, J. and {Blomme}, R. and {Comer{\'o}n}, F. and {Negueruela}, I. and {Molina Lera}, J.~A. and {Pantaleoni Gonz{\'a}lez}, M. and {Daflon}, S. and {Santos}, W. and {Kalari}, V.~M.},
        title = "{Gaia-ESO survey: Massive stars in the Carina Nebula: II. The spectroscopic analysis of the O-star population}",
      journal = {\aap},
         year = 2025,
        month = mar,
       volume = {695},
          eid = {A248},
        pages = {A248},
          doi = {10.1051/0004-6361/202453269},
archivePrefix = {arXiv},
       eprint = {2501.16508},
 primaryClass = {astro-ph.SR},
       adsurl = {https://ui.adsabs.harvard.edu/abs/2025A&A...695A.248B}
}

@ARTICLE{2024A&A...689A.320N,
       author = {{Naz{\'e}}, Ya{\"e}l and {Britavskiy}, Nikolay and {Labadie-Bartz}, Jonathan},
        title = "{TESS observations of non-Be fast rotators}",
      journal = {\aap},
         year = 2024,
        month = sep,
       volume = {689},
          eid = {A320},
        pages = {A320},
          doi = {10.1051/0004-6361/202450966},
archivePrefix = {arXiv},
       eprint = {2407.08305},
 primaryClass = {astro-ph.SR},
       adsurl = {https://ui.adsabs.harvard.edu/abs/2024A&A...689A.320N}
}

@ARTICLE{2024A&A...684A..35B,
       author = {{Britavskiy}, N. and {Renzo}, M. and {Naz{\'e}}, Y. and {Rauw}, G. and {Vynatheya}, P.},
        title = "{Tracing the evolution of short-period binaries with super-synchronous fast rotators}",
      journal = {\aap},
         year = 2024,
        month = apr,
       volume = {684},
          eid = {A35},
        pages = {A35},
          doi = {10.1051/0004-6361/202348484},
archivePrefix = {arXiv},
       eprint = {2401.11304},
 primaryClass = {astro-ph.SR},
       adsurl = {https://ui.adsabs.harvard.edu/abs/2024A&A...684A..35B}
}

@ARTICLE{2023ApJ...952...79L,
       author = {{Li}, Lei and {Zhu}, Chunhua and {Guo}, Sufen and {Liu}, Helei and {L{\"u}}, Guoliang},
        title = "{The Effects of Rotation, Metallicity, and Magnetic Field on the Islands of Failed Supernovae}",
      journal = {\apj},
         year = 2023,
        month = jul,
       volume = {952},
       number = {1},
          eid = {79},
        pages = {79},
          doi = {10.3847/1538-4357/acd9ca},
archivePrefix = {arXiv},
       eprint = {2306.15879},
 primaryClass = {astro-ph.HE},
       adsurl = {https://ui.adsabs.harvard.edu/abs/2023ApJ...952...79L}
}

@INPROCEEDINGS{2023ASPC..534..129W,
       author = {{Wright}, N.~J. and {Kounkel}, M. and {Zari}, E. and {Goodwin}, S. and {Jeffries}, R.~D.},
        title = "{OB Associations}",
    booktitle = {Protostars and Planets VII},
         year = 2023,
       editor = {{Inutsuka}, S. and {Aikawa}, Y. and {Muto}, T. and {Tomida}, K. and {Tamura}, M.},
       series = {Astronomical Society of the Pacific Conference Series},
       volume = {534},
        month = jul,
        pages = {129},
       adsurl = {https://ui.adsabs.harvard.edu/abs/2023ASPC..534..129W}
}

@INPROCEEDINGS{2023ASPC..534..275O,
       author = {{Offner}, S.~S.~R. and {Moe}, M. and {Kratter}, K.~M. and {Sadavoy}, S.~I. and {Jensen}, E.~L.~N. and {Tobin}, J.~J.},
        title = "{The Origin and Evolution of Multiple Star Systems}",
    booktitle = {Protostars and Planets VII},
         year = 2023,
       editor = {{Inutsuka}, S. and {Aikawa}, Y. and {Muto}, T. and {Tomida}, K. and {Tamura}, M.},
       series = {Astronomical Society of the Pacific Conference Series},
       volume = {534},
        month = jul,
        pages = {275},
          doi = {10.48550/arXiv.2203.10066},
archivePrefix = {arXiv},
       eprint = {2203.10066},
 primaryClass = {astro-ph.SR},
       adsurl = {https://ui.adsabs.harvard.edu/abs/2023ASPC..534..275O}
}

@ARTICLE{2023A&A...674A...1G,
       author = {{Gaia Collaboration} and {Vallenari}, A. and {Brown}, A.~G.~A. and {Prusti}, T. and {de Bruijne}, J.~H.~J. and {Arenou}, F. and {Babusiaux}, C. and {Biermann}, M. and {Creevey}, O.~L. and {Ducourant}, C. and {Evans}, D.~W. and {Eyer}, L. and {Guerra}, R. and {Hutton}, A. and {Jordi}, C. and {Klioner}, S.~A. and {Lammers}, U.~L. and {Lindegren}, L. and {Luri}, X. and {Mignard}, F. and {Panem}, C. and {Pourbaix}, D. and {Randich}, S. and {Sartoretti}, P. and {Soubiran}, C. and {Tanga}, P. and {Walton}, N.~A. and {Bailer-Jones}, C.~A.~L. and {Bastian}, U. and {Drimmel}, R. and {Jansen}, F. and {Katz}, D. and {Lattanzi}, M.~G. and {van Leeuwen}, F. and {Bakker}, J. and {Cacciari}, C. and {Casta{\~n}eda}, J. and {De Angeli}, F. and {Fabricius}, C. and {Fouesneau}, M. and {Fr{\'e}mat}, Y. and {Galluccio}, L. and {Guerrier}, A. and {Heiter}, U. and {Masana}, E. and {Messineo}, R. and {Mowlavi}, N. and {Nicolas}, C. and {Nienartowicz}, K. and {Pailler}, F. and {Panuzzo}, P. and {Riclet}, F. and {Roux}, W. and {Seabroke}, G.~M. and {Sordo}, R. and {Th{\'e}venin}, F. and {Gracia-Abril}, G. and {Portell}, J. and {Teyssier}, D. and {Altmann}, M. and {Andrae}, R. and {Audard}, M. and {Bellas-Velidis}, I. and {Benson}, K. and {Berthier}, J. and {Blomme}, R. and {Burgess}, P.~W. and {Busonero}, D. and {Busso}, G. and {C{\'a}novas}, H. and {Carry}, B. and {Cellino}, A. and {Cheek}, N. and {Clementini}, G. and {Damerdji}, Y. and {Davidson}, M. and {de Teodoro}, P. and {Nu{\~n}ez Campos}, M. and {Delchambre}, L. and {Dell'Oro}, A. and {Esquej}, P. and {Fern{\'a}ndez-Hern{\'a}ndez}, J. and {Fraile}, E. and {Garabato}, D. and {Garc{\'\i}a-Lario}, P. and {Gosset}, E. and {Haigron}, R. and {Halbwachs}, J. -L. and {Hambly}, N.~C. and {Harrison}, D.~L. and {Hern{\'a}ndez}, J. and {Hestroffer}, D. and {Hodgkin}, S.~T. and {Holl}, B. and {Jan{\ss}en}, K. and {Jevardat de Fombelle}, G. and {Jordan}, S. and {Krone-Martins}, A. and {Lanzafame}, A.~C. and {L{\"o}ffler}, W. and {Marchal}, O. and {Marrese}, P.~M. and {Moitinho}, A. and {Muinonen}, K. and {Osborne}, P. and {Pancino}, E. and {Pauwels}, T. and {Recio-Blanco}, A. and {Reyl{\'e}}, C. and {Riello}, M. and {Rimoldini}, L. and {Roegiers}, T. and {Rybizki}, J. and {Sarro}, L.~M. and {Siopis}, C. and {Smith}, M. and {Sozzetti}, A. and {Utrilla}, E. and {van Leeuwen}, M. and {Abbas}, U. and {{\'A}brah{\'a}m}, P. and {Abreu Aramburu}, A. and {Aerts}, C. and {Aguado}, J.~J. and {Ajaj}, M. and {Aldea-Montero}, F. and {Altavilla}, G. and {{\'A}lvarez}, M.~A. and {Alves}, J. and {Anders}, F. and {Anderson}, R.~I. and {Anglada Varela}, E. and {Antoja}, T. and {Baines}, D. and {Baker}, S.~G. and {Balaguer-N{\'u}{\~n}ez}, L. and {Balbinot}, E. and {Balog}, Z. and {Barache}, C. and {Barbato}, D. and {Barros}, M. and {Barstow}, M.~A. and {Bartolom{\'e}}, S. and {Bassilana}, J. -L. and {Bauchet}, N. and {Becciani}, U. and {Bellazzini}, M. and {Berihuete}, A. and {Bernet}, M. and {Bertone}, S. and {Bianchi}, L. and {Binnenfeld}, A. and {Blanco-Cuaresma}, S. and {Blazere}, A. and {Boch}, T. and {Bombrun}, A. and {Bossini}, D. and {Bouquillon}, S. and {Bragaglia}, A. and {Bramante}, L. and {Breedt}, E. and {Bressan}, A. and {Brouillet}, N. and {Brugaletta}, E. and {Bucciarelli}, B. and {Burlacu}, A. and {Butkevich}, A.~G. and {Buzzi}, R. and {Caffau}, E. and {Cancelliere}, R. and {Cantat-Gaudin}, T. and {Carballo}, R. and {Carlucci}, T. and {Carnerero}, M.~I. and {Carrasco}, J.~M. and {Casamiquela}, L. and {Castellani}, M. and {Castro-Ginard}, A. and {Chaoul}, L. and {Charlot}, P. and {Chemin}, L. and {Chiaramida}, V. and {Chiavassa}, A. and {Chornay}, N. and {Comoretto}, G. and {Contursi}, G. and {Cooper}, W.~J. and {Cornez}, T. and {Cowell}, S. and {Crifo}, F. and {Cropper}, M. and {Crosta}, M. and {Crowley}, C. and {Dafonte}, C. and {Dapergolas}, A. and {David}, M. and {David}, P. and {de Laverny}, P. and {De Luise}, F. and {De March}, R.},
        title = "{Gaia Data Release 3. Summary of the content and survey properties}",
      journal = {\aap},
         year = 2023,
        month = jun,
       volume = {674},
          eid = {A1},
        pages = {A1},
          doi = {10.1051/0004-6361/202243940},
archivePrefix = {arXiv},
       eprint = {2208.00211},
 primaryClass = {astro-ph.GA},
       adsurl = {https://ui.adsabs.harvard.edu/abs/2023A&A...674A...1G}
}

@ARTICLE{2022A&A...668L...5S,
       author = {{Sana}, H. and {Ram{\'\i}rez-Agudelo}, O.~H. and {H{\'e}nault-Brunet}, V. and {Mahy}, L. and {Almeida}, L.~A. and {de Koter}, A. and {Bestenlehner}, J.~M. and {Evans}, C.~J. and {Langer}, N. and {Schneider}, F.~R.~N. and {Crowther}, P.~A. and {de Mink}, S.~E. and {Herrero}, A. and {Lennon}, D.~J. and {Gieles}, M. and {Ma{\'\i}z Apell{\'a}niz}, J. and {Renzo}, M. and {Sabbi}, E. and {van Loon}, J. Th. and {Vink}, J.~S.},
        title = "{The VLT-FLAMES Tarantula Survey. Observational evidence for two distinct populations of massive runaway stars in 30 Doradus}",
      journal = {\aap},
         year = 2022,
        month = dec,
       volume = {668},
          eid = {L5},
        pages = {L5},
          doi = {10.1051/0004-6361/202244677},
archivePrefix = {arXiv},
       eprint = {2211.13476},
 primaryClass = {astro-ph.SR},
       adsurl = {https://ui.adsabs.harvard.edu/abs/2022A&A...668L...5S}
}

@ARTICLE{2022AA...665A.150H,
       author = {{Holgado}, G. and {Sim{\'o}n-D{\'\i}az}, S. and {Herrero}, A. and {Barb{\'a}}, R.~H.},
        title = "{The IACOB project. VII. The rotational properties of Galactic massive O-type stars revisited}",
      journal = {\aap},
         year = 2022,
        month = sep,
       volume = {665},
          eid = {A150},
        pages = {A150},
          doi = {10.1051/0004-6361/202243851},
archivePrefix = {arXiv},
       eprint = {2207.12776},
 primaryClass = {astro-ph.SR},
       adsurl = {https://ui.adsabs.harvard.edu/abs/2022A&A...665A.150H}
}

@ARTICLE{2021A&A...649A...3R,
       author = {{Riello}, M. and {De Angeli}, F. and {Evans}, D.~W. and {Montegriffo}, P. and {Carrasco}, J.~M. and {Busso}, G. and {Palaversa}, L. and {Burgess}, P.~W. and {Diener}, C. and {Davidson}, M. and {Rowell}, N. and {Fabricius}, C. and {Jordi}, C. and {Bellazzini}, M. and {Pancino}, E. and {Harrison}, D.~L. and {Cacciari}, C. and {van Leeuwen}, F. and {Hambly}, N.~C. and {Hodgkin}, S.~T. and {Osborne}, P.~J. and {Altavilla}, G. and {Barstow}, M.~A. and {Brown}, A.~G.~A. and {Castellani}, M. and {Cowell}, S. and {De Luise}, F. and {Gilmore}, G. and {Giuffrida}, G. and {Hidalgo}, S. and {Holland}, G. and {Marinoni}, S. and {Pagani}, C. and {Piersimoni}, A.~M. and {Pulone}, L. and {Ragaini}, S. and {Rainer}, M. and {Richards}, P.~J. and {Sanna}, N. and {Walton}, N.~A. and {Weiler}, M. and {Yoldas}, A.},
        title = "{Gaia Early Data Release 3. Photometric content and validation}",
      journal = {\aap},
         year = 2021,
        month = may,
       volume = {649},
          eid = {A3},
        pages = {A3},
          doi = {10.1051/0004-6361/202039587},
archivePrefix = {arXiv},
       eprint = {2012.01916},
 primaryClass = {astro-ph.IM},
       adsurl = {https://ui.adsabs.harvard.edu/abs/2021A&A...649A...3R}
}

@ARTICLE{2020JOSS....5.2308P,
       author = {{Prochaska}, J. and {Hennawi}, Joseph and {Westfall}, Kyle and {Cooke}, Ryan and {Wang}, Feige and {Hsyu}, Tiffany and {Davies}, Frederick and {Farina}, Emanuele and {Pelliccia}, Debora},
        title = "{PypeIt: The Python Spectroscopic Data Reduction Pipeline}",
      journal = {The Journal of Open Source Software},
         year = 2020,
        month = dec,
       volume = {5},
       number = {56},
          eid = {2308},
        pages = {2308},
          doi = {10.21105/joss.02308},
archivePrefix = {arXiv},
       eprint = {2005.06505},
 primaryClass = {astro-ph.IM},
       adsurl = {https://ui.adsabs.harvard.edu/abs/2020JOSS....5.2308P}
}

@ARTICLE{2020AA...642A.168B,
       author = {{Berlanas}, S.~R. and {Herrero}, A. and {Comer{\'o}n}, F. and {Sim{\'o}n-D{\'\i}az}, S. and {Lennon}, D.~J. and {Pasquali}, A. and {Ma{\'\i}z Apell{\'a}niz}, J. and {Sota}, A. and {Peller{\'\i}n}, A.},
        title = "{Spectroscopic characterization of the known O-star population in Cygnus OB2. Evidence of multiple star-forming bursts}",
      journal = {\aap},
         year = 2020,
        month = oct,
       volume = {642},
          eid = {A168},
        pages = {A168},
          doi = {10.1051/0004-6361/202039015},
archivePrefix = {arXiv},
       eprint = {2008.09917},
 primaryClass = {astro-ph.SR},
       adsurl = {https://ui.adsabs.harvard.edu/abs/2020A&A...642A.168B}
}

@ARTICLE{2020Natur.585..357H,
       author = {{Harris}, Charles R. and {Millman}, K. Jarrod and {van der Walt}, St{\'e}fan J. and {Gommers}, Ralf and {Virtanen}, Pauli and {Cournapeau}, David and {Wieser}, Eric and {Taylor}, Julian and {Berg}, Sebastian and {Smith}, Nathaniel J. and {Kern}, Robert and {Picus}, Matti and {Hoyer}, Stephan and {van Kerkwijk}, Marten H. and {Brett}, Matthew and {Haldane}, Allan and {del R{\'\i}o}, Jaime Fern{\'a}ndez and {Wiebe}, Mark and {Peterson}, Pearu and {G{\'e}rard-Marchant}, Pierre and {Sheppard}, Kevin and {Reddy}, Tyler and {Weckesser}, Warren and {Abbasi}, Hameer and {Gohlke}, Christoph and {Oliphant}, Travis E.},
        title = "{Array programming with NumPy}",
      journal = {\nat},
         year = 2020,
        month = sep,
       volume = {585},
       number = {7825},
        pages = {357-362},
          doi = {10.1038/s41586-020-2649-2},
archivePrefix = {arXiv},
       eprint = {2006.10256},
 primaryClass = {cs.MS},
       adsurl = {https://ui.adsabs.harvard.edu/abs/2020Natur.585..357H}
}

@software{2020zndo...3743493P,
       author = {{Prochaska}, J. Xavier and {Hennawi}, Joseph and {Cooke}, Ryan and {Westfall}, Kyle and {Wang}, Feige and {EmAstro} and {Tiffanyhsyu} and {Wasserman}, Asher and {Villaume}, Alexa and {Marijana777} and {Schindler}, JT and {Young}, David and {Simha}, Sunil and {Wilde}, Matt and {Tejos}, Nicolas and {Isbell}, Jacob and {Fl{\"o}rs}, Andreas and {Sandford}, Nathan and {Vasovi{\'c}}, Zlatan and {Betts}, Edward and {Holden}, Brad},
        title = "{pypeit/PypeIt: Release 1.0.0}",
         year = 2020,
        month = apr,
          eid = {10.5281/zenodo.3743493},
          doi = {10.5281/zenodo.3743493},
      version = {v1.0.0},
    publisher = {Zenodo},
       adsurl = {https://ui.adsabs.harvard.edu/abs/2020zndo...3743493P}
}

@ARTICLE{2020ApJ...891..113R,
       author = {{Rebolledo}, David and {Guzm{\'a}n}, Andr{\'e}s E. and {Contreras}, Yanett and {Garay}, Guido and {Medina}, S. -N.~X. and {Sanhueza}, Patricio and {Green}, Anne J. and {Castro}, Camila and {Guzm{\'a}n}, Viviana and {Burton}, Michael G.},
        title = "{Effect of Feedback of Massive Stars in the Fragmentation, Distribution, and Kinematics of the Gas in Two Star-forming Regions in the Carina Nebula}",
      journal = {\apj},
         year = 2020,
        month = mar,
       volume = {891},
       number = {2},
          eid = {113},
        pages = {113},
          doi = {10.3847/1538-4357/ab6d76},
archivePrefix = {arXiv},
       eprint = {2001.06969},
 primaryClass = {astro-ph.GA},
       adsurl = {https://ui.adsabs.harvard.edu/abs/2020ApJ...891..113R}
}

@ARTICLE{2020NatMe..17..261V,
       author = {{Virtanen}, Pauli and {Gommers}, Ralf and {Oliphant}, Travis E. and {Haberland}, Matt and {Reddy}, Tyler and {Cournapeau}, David and {Burovski}, Evgeni and {Peterson}, Pearu and {Weckesser}, Warren and {Bright}, Jonathan and {van der Walt}, St{\'e}fan J. and {Brett}, Matthew and {Wilson}, Joshua and {Millman}, K. Jarrod and {Mayorov}, Nikolay and {Nelson}, Andrew R.~J. and {Jones}, Eric and {Kern}, Robert and {Larson}, Eric and {Carey}, C.~J. and {Polat}, {\.I}lhan and {Feng}, Yu and {Moore}, Eric W. and {VanderPlas}, Jake and {Laxalde}, Denis and {Perktold}, Josef and {Cimrman}, Robert and {Henriksen}, Ian and {Quintero}, E.~A. and {Harris}, Charles R. and {Archibald}, Anne M. and {Ribeiro}, Ant{\^o}nio H. and {Pedregosa}, Fabian and {van Mulbregt}, Paul and {SciPy 1. 0 Contributors}},
        title = "{SciPy 1.0: fundamental algorithms for scientific computing in Python}",
      journal = {Nature Medicine},
         year = 2020,
        month = feb,
       volume = {17},
        pages = {261-272},
          doi = {10.1038/s41592-019-0686-2},
archivePrefix = {arXiv},
       eprint = {1907.10121},
 primaryClass = {cs.MS},
       adsurl = {https://ui.adsabs.harvard.edu/abs/2020NatMe..17..261V}
}

@ARTICLE{2020MNRAS.491.5158T,
       author = {{Tokovinin}, Andrei and {Moe}, Maxwell},
        title = "{Formation of close binaries by disc fragmentation and migration, and its statistical modelling}",
      journal = {\mnras},
         year = 2020,
        month = feb,
       volume = {491},
       number = {4},
        pages = {5158-5171},
          doi = {10.1093/mnras/stz3299},
archivePrefix = {arXiv},
       eprint = {1910.01522},
 primaryClass = {astro-ph.SR},
       adsurl = {https://ui.adsabs.harvard.edu/abs/2020MNRAS.491.5158T}
}

@ARTICLE{2019Natur.574..211S,
       author = {{Schneider}, Fabian R.~N. and {Ohlmann}, Sebastian T. and {Podsiadlowski}, Philipp and {R{\"o}pke}, Friedrich K. and {Balbus}, Steven A. and {Pakmor}, R{\"u}diger and {Springel}, Volker},
        title = "{Stellar mergers as the origin of magnetic massive stars}",
      journal = {\nat},
         year = 2019,
        month = oct,
       volume = {574},
       number = {7777},
        pages = {211-214},
          doi = {10.1038/s41586-019-1621-5},
archivePrefix = {arXiv},
       eprint = {1910.14058},
 primaryClass = {astro-ph.SR},
       adsurl = {https://ui.adsabs.harvard.edu/abs/2019Natur.574..211S}
}

@ARTICLE{2019MNRAS.486.1652P,
       author = {{Peres}, Inbal and {Sabach}, Efrat and {Soker}, Noam},
        title = "{Storing magnetic fields in pre-collapse cores of massive stars}",
      journal = {\mnras},
         year = 2019,
        month = jun,
       volume = {486},
       number = {2},
        pages = {1652-1657},
          doi = {10.1093/mnras/stz954},
archivePrefix = {arXiv},
       eprint = {1812.08518},
 primaryClass = {astro-ph.HE},
       adsurl = {https://ui.adsabs.harvard.edu/abs/2019MNRAS.486.1652P}
}

@ARTICLE{2019AA...626A..50D,
       author = {{Dufton}, P.~L. and {Evans}, C.~J. and {Hunter}, I. and {Lennon}, D.~J. and {Schneider}, F.~R.~N.},
        title = "{A census of massive stars in NGC 346. Stellar parameters and rotational velocities}",
      journal = {\aap},
         year = 2019,
        month = jun,
       volume = {626},
          eid = {A50},
        pages = {A50},
          doi = {10.1051/0004-6361/201935415},
archivePrefix = {arXiv},
       eprint = {1905.03359},
 primaryClass = {astro-ph.SR},
       adsurl = {https://ui.adsabs.harvard.edu/abs/2019A&A...626A..50D}
}

@ARTICLE{2019MNRAS.484.1838B,
       author = {{Berlanas}, S.~R. and {Wright}, N.~J. and {Herrero}, A. and {Drew}, J.~E. and {Lennon}, D.~J.},
        title = "{Disentangling the spatial substructure of Cygnus OB2 from Gaia DR2}",
      journal = {\mnras},
         year = 2019,
        month = apr,
       volume = {484},
       number = {2},
        pages = {1838-1842},
          doi = {10.1093/mnras/stz117},
archivePrefix = {arXiv},
       eprint = {1901.02959},
 primaryClass = {astro-ph.SR},
       adsurl = {https://ui.adsabs.harvard.edu/abs/2019MNRAS.484.1838B}
}

@ARTICLE{2018AA...620A..56B,
       author = {{Berlanas}, S.~R. and {Herrero}, A. and {Comer{\'o}n}, F. and {Sim{\'o}n-D{\'\i}az}, S. and {Cervi{\~n}o}, M. and {Pasquali}, A.},
        title = "{Oxygen and silicon abundances in Cygnus OB2. Chemical homogeneity in a sample of OB slow rotators}",
      journal = {\aap},
         year = 2018,
        month = nov,
       volume = {620},
          eid = {A56},
        pages = {A56},
          doi = {10.1051/0004-6361/201833989},
archivePrefix = {arXiv},
       eprint = {1809.06644},
 primaryClass = {astro-ph.SR},
       adsurl = {https://ui.adsabs.harvard.edu/abs/2018A&A...620A..56B}
}

@ARTICLE{2018A&A...618A..73S,
       author = {{Schneider}, F.~R.~N. and {Ram{\'\i}rez-Agudelo}, O.~H. and {Tramper}, F. and {Bestenlehner}, J.~M. and {Castro}, N. and {Sana}, H. and {Evans}, C.~J. and {Sab{\'\i}n-Sanjuli{\'a}n}, C. and {Sim{\'o}n-D{\'\i}az}, S. and {Langer}, N. and {Fossati}, L. and {Gr{\"a}fener}, G. and {Crowther}, P.~A. and {de Mink}, S.~E. and {de Koter}, A. and {Gieles}, M. and {Herrero}, A. and {Izzard}, R.~G. and {Kalari}, V. and {Klessen}, R.~S. and {Lennon}, D.~J. and {Mahy}, L. and {Ma{\'\i}z Apell{\'a}niz}, J. and {Markova}, N. and {van Loon}, J. Th. and {Vink}, J.~S. and {Walborn}, N.~R.},
        title = "{The VLT-FLAMES Tarantula Survey. XXIX. Massive star formation in the local 30 Doradus starburst}",
      journal = {\aap},
         year = 2018,
        month = oct,
       volume = {618},
          eid = {A73},
        pages = {A73},
          doi = {10.1051/0004-6361/201833433},
archivePrefix = {arXiv},
       eprint = {1807.03821},
 primaryClass = {astro-ph.SR},
       adsurl = {https://ui.adsabs.harvard.edu/abs/2018A&A...618A..73S}
}

@ARTICLE{2018MNRAS.476.2493G,
       author = {{Griffiths}, Daniel W. and {Goodwin}, Simon P. and {Caballero-Nieves}, Saida M.},
        title = "{Massive, wide binaries as tracers of massive star formation}",
      journal = {\mnras},
         year = 2018,
        month = may,
       volume = {476},
       number = {2},
        pages = {2493-2500},
          doi = {10.1093/mnras/sty412},
archivePrefix = {arXiv},
       eprint = {1802.04560},
 primaryClass = {astro-ph.GA},
       adsurl = {https://ui.adsabs.harvard.edu/abs/2018MNRAS.476.2493G}
}

@ARTICLE{2018A&A...612A..50B,
       author = {{Berlanas}, S.~R. and {Herrero}, A. and {Comer{\'o}n}, F. and {Pasquali}, A. and {Bertelli Motta}, C. and {Sota}, A.},
        title = "{New massive members of Cygnus OB2}",
      journal = {\aap},
         year = 2018,
        month = apr,
       volume = {612},
          eid = {A50},
        pages = {A50},
          doi = {10.1051/0004-6361/201731856},
archivePrefix = {arXiv},
       eprint = {1711.06945},
 primaryClass = {astro-ph.SR},
       adsurl = {https://ui.adsabs.harvard.edu/abs/2018A&A...612A..50B}
}

@ARTICLE{2017NatAs...1E..64C,
       author = {{Corsaro}, Enrico and {Lee}, Yueh-Ning and {Garc{\'\i}a}, Rafael A. and {Hennebelle}, Patrick and {Mathur}, Savita and {Beck}, Paul G. and {Mathis}, Stephane and {Stello}, Dennis and {Bouvier}, J{\'e}r{\^o}me},
        title = "{Spin alignment of stars in old open clusters}",
      journal = {Nature Astronomy},
         year = 2017,
        month = mar,
       volume = {1},
          eid = {0064},
        pages = {0064},
          doi = {10.1038/s41550-017-0064},
archivePrefix = {arXiv},
       eprint = {1703.05588},
 primaryClass = {astro-ph.SR},
       adsurl = {https://ui.adsabs.harvard.edu/abs/2017NatAs...1E..64C}
}

@ARTICLE{2016A&A...595A...1G,
       author = {{Gaia Collaboration} and {Prusti}, T. and {de Bruijne}, J.~H.~J. and {Brown}, A.~G.~A. and {Vallenari}, A. and {Babusiaux}, C. and {Bailer-Jones}, C.~A.~L. and {Bastian}, U. and {Biermann}, M. and {Evans}, D.~W. and {Eyer}, L. and {Jansen}, F. and {Jordi}, C. and {Klioner}, S.~A. and {Lammers}, U. and {Lindegren}, L. and {Luri}, X. and {Mignard}, F. and {Milligan}, D.~J. and {Panem}, C. and {Poinsignon}, V. and {Pourbaix}, D. and {Randich}, S. and {Sarri}, G. and {Sartoretti}, P. and {Siddiqui}, H.~I. and {Soubiran}, C. and {Valette}, V. and {van Leeuwen}, F. and {Walton}, N.~A. and {Aerts}, C. and {Arenou}, F. and {Cropper}, M. and {Drimmel}, R. and {H{\o}g}, E. and {Katz}, D. and {Lattanzi}, M.~G. and {O'Mullane}, W. and {Grebel}, E.~K. and {Holland}, A.~D. and {Huc}, C. and {Passot}, X. and {Bramante}, L. and {Cacciari}, C. and {Casta{\~n}eda}, J. and {Chaoul}, L. and {Cheek}, N. and {De Angeli}, F. and {Fabricius}, C. and {Guerra}, R. and {Hern{\'a}ndez}, J. and {Jean-Antoine-Piccolo}, A. and {Masana}, E. and {Messineo}, R. and {Mowlavi}, N. and {Nienartowicz}, K. and {Ord{\'o}{\~n}ez-Blanco}, D. and {Panuzzo}, P. and {Portell}, J. and {Richards}, P.~J. and {Riello}, M. and {Seabroke}, G.~M. and {Tanga}, P. and {Th{\'e}venin}, F. and {Torra}, J. and {Els}, S.~G. and {Gracia-Abril}, G. and {Comoretto}, G. and {Garcia-Reinaldos}, M. and {Lock}, T. and {Mercier}, E. and {Altmann}, M. and {Andrae}, R. and {Astraatmadja}, T.~L. and {Bellas-Velidis}, I. and {Benson}, K. and {Berthier}, J. and {Blomme}, R. and {Busso}, G. and {Carry}, B. and {Cellino}, A. and {Clementini}, G. and {Cowell}, S. and {Creevey}, O. and {Cuypers}, J. and {Davidson}, M. and {De Ridder}, J. and {de Torres}, A. and {Delchambre}, L. and {Dell'Oro}, A. and {Ducourant}, C. and {Fr{\'e}mat}, Y. and {Garc{\'\i}a-Torres}, M. and {Gosset}, E. and {Halbwachs}, J. -L. and {Hambly}, N.~C. and {Harrison}, D.~L. and {Hauser}, M. and {Hestroffer}, D. and {Hodgkin}, S.~T. and {Huckle}, H.~E. and {Hutton}, A. and {Jasniewicz}, G. and {Jordan}, S. and {Kontizas}, M. and {Korn}, A.~J. and {Lanzafame}, A.~C. and {Manteiga}, M. and {Moitinho}, A. and {Muinonen}, K. and {Osinde}, J. and {Pancino}, E. and {Pauwels}, T. and {Petit}, J. -M. and {Recio-Blanco}, A. and {Robin}, A.~C. and {Sarro}, L.~M. and {Siopis}, C. and {Smith}, M. and {Smith}, K.~W. and {Sozzetti}, A. and {Thuillot}, W. and {van Reeven}, W. and {Viala}, Y. and {Abbas}, U. and {Abreu Aramburu}, A. and {Accart}, S. and {Aguado}, J.~J. and {Allan}, P.~M. and {Allasia}, W. and {Altavilla}, G. and {{\'A}lvarez}, M.~A. and {Alves}, J. and {Anderson}, R.~I. and {Andrei}, A.~H. and {Anglada Varela}, E. and {Antiche}, E. and {Antoja}, T. and {Ant{\'o}n}, S. and {Arcay}, B. and {Atzei}, A. and {Ayache}, L. and {Bach}, N. and {Baker}, S.~G. and {Balaguer-N{\'u}{\~n}ez}, L. and {Barache}, C. and {Barata}, C. and {Barbier}, A. and {Barblan}, F. and {Baroni}, M. and {Barrado y Navascu{\'e}s}, D. and {Barros}, M. and {Barstow}, M.~A. and {Becciani}, U. and {Bellazzini}, M. and {Bellei}, G. and {Bello Garc{\'\i}a}, A. and {Belokurov}, V. and {Bendjoya}, P. and {Berihuete}, A. and {Bianchi}, L. and {Bienaym{\'e}}, O. and {Billebaud}, F. and {Blagorodnova}, N. and {Blanco-Cuaresma}, S. and {Boch}, T. and {Bombrun}, A. and {Borrachero}, R. and {Bouquillon}, S. and {Bourda}, G. and {Bouy}, H. and {Bragaglia}, A. and {Breddels}, M.~A. and {Brouillet}, N. and {Br{\"u}semeister}, T. and {Bucciarelli}, B. and {Budnik}, F. and {Burgess}, P. and {Burgon}, R. and {Burlacu}, A. and {Busonero}, D. and {Buzzi}, R. and {Caffau}, E. and {Cambras}, J. and {Campbell}, H. and {Cancelliere}, R. and {Cantat-Gaudin}, T. and {Carlucci}, T. and {Carrasco}, J.~M. and {Castellani}, M. and {Charlot}, P. and {Charnas}, J. and {Charvet}, P. and {Chassat}, F. and {Chiavassa}, A. and {Clotet}, M. and {Cocozza}, G. and {Collins}, R.~S. and {Collins}, P. and {Costigan}, G.},
        title = "{The Gaia mission}",
      journal = {\aap},
         year = 2016,
        month = nov,
       volume = {595},
          eid = {A1},
        pages = {A1},
          doi = {10.1051/0004-6361/201629272},
archivePrefix = {arXiv},
       eprint = {1609.04153},
 primaryClass = {astro-ph.IM},
       adsurl = {https://ui.adsabs.harvard.edu/abs/2016A&A...595A...1G}
}

@ARTICLE{2016ApJS..224....4M,
       author = {{Ma{\'\i}z Apell{\'a}niz}, J. and {Sota}, A. and {Arias}, J.~I. and {Barb{\'a}}, R.~H. and {Walborn}, N.~R. and {Sim{\'o}n-D{\'\i}az}, S. and {Negueruela}, I. and {Marco}, A. and {Le{\~a}o}, J.~R.~S. and {Herrero}, A. and {Gamen}, R.~C. and {Alfaro}, E.~J.},
        title = "{The Galactic O-Star Spectroscopic Survey (GOSSS). III. 142 Additional O-type Systems.}",
      journal = {\apjs},
         year = 2016,
        month = may,
       volume = {224},
       number = {1},
          eid = {4},
        pages = {4},
          doi = {10.3847/0067-0049/224/1/4},
archivePrefix = {arXiv},
       eprint = {1602.01336},
 primaryClass = {astro-ph.SR},
       adsurl = {https://ui.adsabs.harvard.edu/abs/2016ApJS..224....4M}
}

@ARTICLE{2016MNRAS.455.1855C,
       author = {{Choudhury}, Samyaday and {Subramaniam}, Annapurni and {Cole}, Andrew A.},
        title = "{Photometric metallicity map of the Large Magellanic Cloud}",
      journal = {\mnras},
         year = 2016,
        month = jan,
       volume = {455},
       number = {2},
        pages = {1855-1880},
          doi = {10.1093/mnras/stv2414},
archivePrefix = {arXiv},
       eprint = {1510.05769},
 primaryClass = {astro-ph.GA},
       adsurl = {https://ui.adsabs.harvard.edu/abs/2016MNRAS.455.1855C}
}

@ARTICLE{2015A&A...580A..92R,
       author = {{Ram{\'\i}rez-Agudelo}, O.~H. and {Sana}, H. and {de Mink}, S.~E. and {H{\'e}nault-Brunet}, V. and {de Koter}, A. and {Langer}, N. and {Tramper}, F. and {Gr{\"a}fener}, G. and {Evans}, C.~J. and {Vink}, J.~S. and {Dufton}, P.~L. and {Taylor}, W.~D.},
        title = "{The VLT-FLAMES Tarantula Survey. XXI. Stellar spin rates of O-type spectroscopic binaries}",
      journal = {\aap},
         year = 2015,
        month = aug,
       volume = {580},
          eid = {A92},
        pages = {A92},
          doi = {10.1051/0004-6361/201425424},
archivePrefix = {arXiv},
       eprint = {1507.02286},
 primaryClass = {astro-ph.SR},
       adsurl = {https://ui.adsabs.harvard.edu/abs/2015A&A...580A..92R}
}

@ARTICLE{2015MNRAS.449..741W,
       author = {{Wright}, Nicholas J. and {Drew}, Janet E. and {Mohr-Smith}, Michael},
        title = "{The massive star population of Cygnus OB2}",
      journal = {\mnras},
         year = 2015,
        month = may,
       volume = {449},
       number = {1},
        pages = {741-760},
          doi = {10.1093/mnras/stv323},
archivePrefix = {arXiv},
       eprint = {1502.05718},
 primaryClass = {astro-ph.SR},
       adsurl = {https://ui.adsabs.harvard.edu/abs/2015MNRAS.449..741W}
}

@INPROCEEDINGS{2015hsa8.conf..603M,
       author = {{Ma{\'\i}z Apell{\'a}niz}, J. and {Alfaro}, E.~J. and {Arias}, J.~I. and {Barb{\'a}}, R.~H. and {Gamen}, R.~C. and {Herrero}, A. and {Le{\~a}o}, J.~R.~S. and {Marco}, A. and {Negueruela}, I. and {Sim{\'o}n-D{\'\i}az}, S. and {Sota}, A. and {Walborn}, N.~R.},
        title = "{MGB and the new Galactic O-Star Spectroscopic Survey spectral classification standard grid}",
    booktitle = {Highlights of Spanish Astrophysics VIII},
         year = 2015,
       editor = {{Cenarro}, A.~J. and {Figueras}, F. and {Hern{\'a}ndez-Monteagudo}, C. and {Trujillo Bueno}, J. and {Valdivielso}, L.},
        month = may,
        pages = {603-603},
          doi = {10.48550/arXiv.1410.7615},
archivePrefix = {arXiv},
       eprint = {1410.7615},
 primaryClass = {astro-ph.SR},
       adsurl = {https://ui.adsabs.harvard.edu/abs/2015hsa8.conf..603M}
}

@ARTICLE{2015MNRAS.448.3248G,
       author = {{Geen}, Sam and {Rosdahl}, Joakim and {Blaizot}, Jeremy and {Devriendt}, Julien and {Slyz}, Adrianne},
        title = "{A detailed study of feedback from a massive star}",
      journal = {\mnras},
         year = 2015,
        month = apr,
       volume = {448},
       number = {4},
        pages = {3248-3264},
          doi = {10.1093/mnras/stv251},
archivePrefix = {arXiv},
       eprint = {1412.0484},
 primaryClass = {astro-ph.GA},
       adsurl = {https://ui.adsabs.harvard.edu/abs/2015MNRAS.448.3248G}
}

@ARTICLE{2014ApJS..215...15S,
       author = {{Sana}, H. and {Le Bouquin}, J. -B. and {Lacour}, S. and {Berger}, J. -P. and {Duvert}, G. and {Gauchet}, L. and {Norris}, B. and {Olofsson}, J. and {Pickel}, D. and {Zins}, G. and {Absil}, O. and {de Koter}, A. and {Kratter}, K. and {Schnurr}, O. and {Zinnecker}, H.},
        title = "{Southern Massive Stars at High Angular Resolution: Observational Campaign and Companion Detection}",
      journal = {\apjs},
         year = 2014,
        month = nov,
       volume = {215},
       number = {1},
          eid = {15},
        pages = {15},
          doi = {10.1088/0067-0049/215/1/15},
archivePrefix = {arXiv},
       eprint = {1409.6304},
 primaryClass = {astro-ph.SR},
       adsurl = {https://ui.adsabs.harvard.edu/abs/2014ApJS..215...15S}
}

@ARTICLE{2014ApJS..211...10S,
       author = {{Sota}, A. and {Ma{\'\i}z Apell{\'a}niz}, J. and {Morrell}, N.~I. and {Barb{\'a}}, R.~H. and {Walborn}, N.~R. and {Gamen}, R.~C. and {Arias}, J.~I. and {Alfaro}, E.~J.},
        title = "{The Galactic O-Star Spectroscopic Survey (GOSSS). II. Bright Southern Stars}",
      journal = {\apjs},
         year = 2014,
        month = mar,
       volume = {211},
       number = {1},
          eid = {10},
        pages = {10},
          doi = {10.1088/0067-0049/211/1/10},
archivePrefix = {arXiv},
       eprint = {1312.6222},
 primaryClass = {astro-ph.GA},
       adsurl = {https://ui.adsabs.harvard.edu/abs/2014ApJS..211...10S}
}

@ARTICLE{2014ApJ...782....7D,
       author = {{de Mink}, S.~E. and {Sana}, H. and {Langer}, N. and {Izzard}, R.~G. and {Schneider}, F.~R.~N.},
        title = "{The Incidence of Stellar Mergers and Mass Gainers among Massive Stars}",
      journal = {\apj},
         year = 2014,
        month = feb,
       volume = {782},
       number = {1},
          eid = {7},
        pages = {7},
          doi = {10.1088/0004-637X/782/1/7},
archivePrefix = {arXiv},
       eprint = {1312.3650},
 primaryClass = {astro-ph.SR},
       adsurl = {https://ui.adsabs.harvard.edu/abs/2014ApJ...782....7D}
}

@ARTICLE{2014MNRAS.438..639W,
       author = {{Wright}, Nicholas J. and {Parker}, Richard J. and {Goodwin}, Simon P. and {Drake}, Jeremy J.},
        title = "{Constraints on massive star formation: Cygnus OB2 was always an association}",
      journal = {\mnras},
         year = 2014,
        month = feb,
       volume = {438},
       number = {1},
        pages = {639-646},
          doi = {10.1093/mnras/stt2232},
archivePrefix = {arXiv},
       eprint = {1311.4537},
 primaryClass = {astro-ph.SR},
       adsurl = {https://ui.adsabs.harvard.edu/abs/2014MNRAS.438..639W}
}

@ARTICLE{2014A&A...562A.135S,
       author = {{Sim{\'o}n-D{\'\i}az}, S. and {Herrero}, A.},
        title = "{The IACOB project. I. Rotational velocities in northern Galactic O- and early B-type stars revisited. The impact of other sources of line-broadening}",
      journal = {\aap},
         year = 2014,
        month = feb,
       volume = {562},
          eid = {A135},
        pages = {A135},
          doi = {10.1051/0004-6361/201322758},
archivePrefix = {arXiv},
       eprint = {1311.3360},
 primaryClass = {astro-ph.SR},
       adsurl = {https://ui.adsabs.harvard.edu/abs/2014A&A...562A.135S}
}

@ARTICLE{2013AA...560A..29R,
       author = {{Ram{\'\i}rez-Agudelo}, O.~H. and {Sim{\'o}n-D{\'\i}az}, S. and {Sana}, H. and {de Koter}, A. and {Sab{\'\i}n-Sanjul{\'\i}an}, C. and {de Mink}, S.~E. and {Dufton}, P.~L. and {Gr{\"a}fener}, G. and {Evans}, C.~J. and {Herrero}, A. and {Langer}, N. and {Lennon}, D.~J. and {Ma{\'\i}z Apell{\'a}niz}, J. and {Markova}, N. and {Najarro}, F. and {Puls}, J. and {Taylor}, W.~D. and {Vink}, J.~S.},
        title = "{The VLT-FLAMES Tarantula Survey. XII. Rotational velocities of the single O-type stars}",
      journal = {\aap},
         year = 2013,
        month = dec,
       volume = {560},
          eid = {A29},
        pages = {A29},
          doi = {10.1051/0004-6361/201321986},
archivePrefix = {arXiv},
       eprint = {1309.2929},
 primaryClass = {astro-ph.SR},
       adsurl = {https://ui.adsabs.harvard.edu/abs/2013A&A...560A..29R}
}

@ARTICLE{2013ApJ...764..166D,
       author = {{de Mink}, S.~E. and {Langer}, N. and {Izzard}, R.~G. and {Sana}, H. and {de Koter}, A.},
        title = "{The Rotation Rates of Massive Stars: The Role of Binary Interaction through Tides, Mass Transfer, and Mergers}",
      journal = {\apj},
         year = 2013,
        month = feb,
       volume = {764},
       number = {2},
          eid = {166},
        pages = {166},
          doi = {10.1088/0004-637X/764/2/166},
archivePrefix = {arXiv},
       eprint = {1211.3742},
 primaryClass = {astro-ph.SR},
       adsurl = {https://ui.adsabs.harvard.edu/abs/2013ApJ...764..166D}
}

@ARTICLE{2012MNRAS.426.2208G,
       author = {{Grunhut}, J.~H. and {Wade}, G.~A. and {Sundqvist}, J.~O. and {ud-Doula}, A. and {Neiner}, C. and {Ignace}, R. and {Marcolino}, W.~L.~F. and {Rivinius}, Th. and {Fullerton}, A. and {Kaper}, L. and {Mauclaire}, B. and {Buil}, C. and {Garrel}, T. and {Ribeiro}, J. and {Ubaud}, S.},
        title = "{Investigating the spectroscopic, magnetic and circumstellar variability of the O9 subgiant star HD 57682}",
      journal = {\mnras},
         year = 2012,
        month = nov,
       volume = {426},
       number = {3},
        pages = {2208-2227},
          doi = {10.1111/j.1365-2966.2012.21799.x},
archivePrefix = {arXiv},
       eprint = {1207.6988},
 primaryClass = {astro-ph.SR},
       adsurl = {https://ui.adsabs.harvard.edu/abs/2012MNRAS.426.2208G}
}

@ARTICLE{2012ARA&A..50..107L,
       author = {{Langer}, N.},
        title = "{Presupernova Evolution of Massive Single and Binary Stars}",
      journal = {\araa},
         year = 2012,
        month = sep,
       volume = {50},
        pages = {107-164},
          doi = {10.1146/annurev-astro-081811-125534},
archivePrefix = {arXiv},
       eprint = {1206.5443},
 primaryClass = {astro-ph.SR},
       adsurl = {https://ui.adsabs.harvard.edu/abs/2012ARA&A..50..107L}
}

@ARTICLE{2012A&A...543A.101C,
       author = {{Comer{\'o}n}, F. and {Pasquali}, A.},
        title = "{New members of the massive stellar population in Cygnus}",
      journal = {\aap},
         year = 2012,
        month = jul,
       volume = {543},
          eid = {A101},
        pages = {A101},
          doi = {10.1051/0004-6361/201219022},
       adsurl = {https://ui.adsabs.harvard.edu/abs/2012A&A...543A.101C}
}

@ARTICLE{2012Sci...337..444S,
       author = {{Sana}, H. and {de Mink}, S.~E. and {de Koter}, A. and {Langer}, N. and {Evans}, C.~J. and {Gieles}, M. and {Gosset}, E. and {Izzard}, R.~G. and {Le Bouquin}, J. -B. and {Schneider}, F.~R.~N.},
        title = "{Binary Interaction Dominates the Evolution of Massive Stars}",
      journal = {Science},
         year = 2012,
        month = jul,
       volume = {337},
       number = {6093},
        pages = {444},
          doi = {10.1126/science.1223344},
archivePrefix = {arXiv},
       eprint = {1207.6397},
 primaryClass = {astro-ph.SR},
       adsurl = {https://ui.adsabs.harvard.edu/abs/2012Sci...337..444S}
}

@software{2012ascl.soft07011S,
       author = {{Science Software Branch at STScI}},
        title = "{PyRAF: Python alternative for IRAF}",
 howpublished = {Astrophysics Source Code Library, record ascl:1207.011},
         year = 2012,
        month = jul,
          eid = {ascl:1207.011},
archivePrefix = {ascl},
       eprint = {1207.011},
       adsurl = {https://ui.adsabs.harvard.edu/abs/2012ascl.soft07011S}
}

@dataset{2012yCat.1322....0Z,
       author = {{Zacharias}, N. and {Finch}, C.~T. and {Girard}, T.~M. and {Henden}, A. and {Bartlett}, J.~L. and {Monet}, D.~G. and {Zacharias}, M.~I.},
        title = "{VizieR Online Data Catalog: UCAC4 Catalogue (Zacharias+, 2012)}",
 howpublished = {VizieR On-line Data Catalog: I/322A.  Originally published in: 2013AJ....145...44Z},
         year = 2012,
        month = jul,
          eid = {I/322A},
       adsurl = {https://ui.adsabs.harvard.edu/abs/2012yCat.1322....0Z}
}

@ARTICLE{2012ApJ...748...97R,
       author = {{Rosen}, Anna L. and {Krumholz}, Mark R. and {Ramirez-Ruiz}, Enrico},
        title = "{What Sets the Initial Rotation Rates of Massive Stars?}",
      journal = {\apj},
         year = 2012,
        month = apr,
       volume = {748},
       number = {2},
          eid = {97},
        pages = {97},
          doi = {10.1088/0004-637X/748/2/97},
archivePrefix = {arXiv},
       eprint = {1201.4186},
 primaryClass = {astro-ph.SR},
       adsurl = {https://ui.adsabs.harvard.edu/abs/2012ApJ...748...97R}
}

@ARTICLE{2012A&A...537A.146E,
       author = {{Ekstr{\"o}m}, S. and {Georgy}, C. and {Eggenberger}, P. and {Meynet}, G. and {Mowlavi}, N. and {Wyttenbach}, A. and {Granada}, A. and {Decressin}, T. and {Hirschi}, R. and {Frischknecht}, U. and {Charbonnel}, C. and {Maeder}, A.},
        title = "{Grids of stellar models with rotation. I. Models from 0.8 to 120 M$_{{\ensuremath{\odot}}}$ at solar metallicity (Z = 0.014)}",
      journal = {\aap},
         year = 2012,
        month = jan,
       volume = {537},
          eid = {A146},
        pages = {A146},
          doi = {10.1051/0004-6361/201117751},
archivePrefix = {arXiv},
       eprint = {1110.5049},
 primaryClass = {astro-ph.SR},
       adsurl = {https://ui.adsabs.harvard.edu/abs/2012A&A...537A.146E}
}

@ARTICLE{2011MNRAS.416..580L,
       author = {{Lin}, Min-Kai and {Krumholz}, Mark R. and {Kratter}, Kaitlin M.},
        title = "{Spin-down of protostars through gravitational torques}",
      journal = {\mnras},
         year = 2011,
        month = sep,
       volume = {416},
       number = {1},
        pages = {580-590},
          doi = {10.1111/j.1365-2966.2011.19074.x},
archivePrefix = {arXiv},
       eprint = {1105.3205},
 primaryClass = {astro-ph.SR},
       adsurl = {https://ui.adsabs.harvard.edu/abs/2011MNRAS.416..580L}
}

@ARTICLE{2011A&A...530A..11M,
       author = {{Markova}, N. and {Puls}, J. and {Scuderi}, S. and {Sim{\'o}n-D{\'\i}az}, S. and {Herrero}, A.},
        title = "{Spectroscopic and physical parameters of Galactic O-type stars. I. Effects of rotation and spectral resolving power in the spectral classification of dwarfs and giants}",
      journal = {\aap},
         year = 2011,
        month = jun,
       volume = {530},
          eid = {A11},
        pages = {A11},
          doi = {10.1051/0004-6361/201015956},
archivePrefix = {arXiv},
       eprint = {1103.3357},
 primaryClass = {astro-ph.GA},
       adsurl = {https://ui.adsabs.harvard.edu/abs/2011A&A...530A..11M}
}

@ARTICLE{2011ApJS..193...24S,
       author = {{Sota}, A. and {Ma{\'\i}z Apell{\'a}niz}, J. and {Walborn}, N.~R. and {Alfaro}, E.~J. and {Barb{\'a}}, R.~H. and {Morrell}, N.~I. and {Gamen}, R.~C. and {Arias}, J.~I.},
        title = "{The Galactic O-Star Spectroscopic Survey. I. Classification System and Bright Northern Stars in the Blue-violet at R \raisebox{-0.5ex}\textasciitilde 2500}",
      journal = {\apjs},
         year = 2011,
        month = apr,
       volume = {193},
       number = {2},
          eid = {24},
        pages = {24},
          doi = {10.1088/0067-0049/193/2/24},
archivePrefix = {arXiv},
       eprint = {1101.4002},
 primaryClass = {astro-ph.GA},
       adsurl = {https://ui.adsabs.harvard.edu/abs/2011ApJS..193...24S}
}

@ARTICLE{2011BSRSL..80..543D,
       author = {{de Mink}, S.~E. and {Langer}, N. and {Izzard}, R.~G.},
        title = "{Binaries are the best single stars}",
      journal = {Bulletin de la Societe Royale des Sciences de Liege},
         year = 2011,
        month = jan,
       volume = {80},
        pages = {543-548},
          doi = {10.48550/arXiv.1010.2200},
archivePrefix = {arXiv},
       eprint = {1010.2200},
 primaryClass = {astro-ph.SR},
       adsurl = {https://ui.adsabs.harvard.edu/abs/2011BSRSL..80..543D}
}

@ARTICLE{2010ApJ...722..605H,
       author = {{Huang}, Wenjin and {Gies}, D.~R. and {McSwain}, M.~V.},
        title = "{A Stellar Rotation Census of B Stars: From ZAMS to TAMS}",
      journal = {\apj},
         year = 2010,
        month = oct,
       volume = {722},
       number = {1},
        pages = {605-619},
          doi = {10.1088/0004-637X/722/1/605},
archivePrefix = {arXiv},
       eprint = {1008.1761},
 primaryClass = {astro-ph.SR},
       adsurl = {https://ui.adsabs.harvard.edu/abs/2010ApJ...722..605H}
}

@ARTICLE{2010A&A...517A..58D,
       author = {{Duez}, V. and {Mathis}, S.},
        title = "{Relaxed equilibrium configurations to model fossil fields . I. A first family}",
      journal = {\aap},
         year = 2010,
        month = jul,
       volume = {517},
          eid = {A58},
        pages = {A58},
          doi = {10.1051/0004-6361/200913496},
       adsurl = {https://ui.adsabs.harvard.edu/abs/2010A&A...517A..58D}
}

@ARTICLE{2010A&A...515A..26S,
       author = {{Sana}, H. and {Momany}, Y. and {Gieles}, M. and {Carraro}, G. and {Beletsky}, Y. and {Ivanov}, V.~D. and {de Silva}, G. and {James}, G.},
        title = "{A MAD view of Trumpler 14}",
      journal = {\aap},
         year = 2010,
        month = jun,
       volume = {515},
          eid = {A26},
        pages = {A26},
          doi = {10.1051/0004-6361/200913688},
archivePrefix = {arXiv},
       eprint = {1003.2208},
 primaryClass = {astro-ph.SR},
       adsurl = {https://ui.adsabs.harvard.edu/abs/2010A&A...515A..26S}
}

@ARTICLE{2010MNRAS.401.1505B,
       author = {{Bate}, M.~R. and {Lodato}, G. and {Pringle}, J.~E.},
        title = "{Chaotic star formation and the alignment of stellar rotation with disc and planetary orbital axes}",
      journal = {\mnras},
         year = 2010,
        month = jan,
       volume = {401},
       number = {3},
        pages = {1505-1513},
          doi = {10.1111/j.1365-2966.2009.15773.x},
archivePrefix = {arXiv},
       eprint = {0909.4255},
 primaryClass = {astro-ph.SR},
       adsurl = {https://ui.adsabs.harvard.edu/abs/2010MNRAS.401.1505B}
}

@ARTICLE{2009ARA&A..47..333D,
       author = {{Donati}, J. -F. and {Landstreet}, J.~D.},
        title = "{Magnetic Fields of Nondegenerate Stars}",
      journal = {\araa},
         year = 2009,
        month = sep,
       volume = {47},
       number = {1},
        pages = {333-370},
          doi = {10.1146/annurev-astro-082708-101833},
archivePrefix = {arXiv},
       eprint = {0904.1938},
 primaryClass = {astro-ph.SR},
       adsurl = {https://ui.adsabs.harvard.edu/abs/2009ARA&A..47..333D}
}

@ARTICLE{2009MNRAS.397..232B,
       author = {{Bate}, Matthew R.},
        title = "{The dependence of star formation on initial conditions and molecular cloud structure}",
      journal = {\mnras},
         year = 2009,
        month = jul,
       volume = {397},
       number = {1},
        pages = {232-248},
          doi = {10.1111/j.1365-2966.2009.14970.x},
archivePrefix = {arXiv},
       eprint = {0905.3562},
 primaryClass = {astro-ph.SR},
       adsurl = {https://ui.adsabs.harvard.edu/abs/2009MNRAS.397..232B}
}

@ARTICLE{2009AJ....137.3358M,
       author = {{Mason}, Brian D. and {Hartkopf}, William I. and {Gies}, Douglas R. and {Henry}, Todd J. and {Helsel}, John W.},
        title = "{The High Angular Resolution Multiplicity of Massive Stars}",
      journal = {\aj},
         year = 2009,
        month = feb,
       volume = {137},
       number = {2},
        pages = {3358-3377},
          doi = {10.1088/0004-6256/137/2/3358},
archivePrefix = {arXiv},
       eprint = {0811.0492},
 primaryClass = {astro-ph},
       adsurl = {https://ui.adsabs.harvard.edu/abs/2009AJ....137.3358M}
}

@ARTICLE{2009MNRAS.392.1363B,
       author = {{Bate}, Matthew R.},
        title = "{The importance of radiative feedback for the stellar initial mass function}",
      journal = {\mnras},
         year = 2009,
        month = feb,
       volume = {392},
       number = {4},
        pages = {1363-1380},
          doi = {10.1111/j.1365-2966.2008.14165.x},
archivePrefix = {arXiv},
       eprint = {0811.1035},
 primaryClass = {astro-ph},
       adsurl = {https://ui.adsabs.harvard.edu/abs/2009MNRAS.392.1363B}
}

@INCOLLECTION{2009msfp.book...74O,
       author = {{Oey}, M.~S. and {Clarke}, C.~J.},
        title = "{Massive stars: Feedback effects in the local universe:}",
    booktitle = {Massive Stars: From Pop III and GRBs to the Milky Way. Space Telescope Science Institute Symposium Series No. 20. Edited by Mario Livio and Eva Villaver. Cambridge University Press},
         year = 2009,
       editor = {{Livio}, Mario and {Villaver}, Eva},
        pages = {74-92},
          doi = {10.1017/CBO9780511770593.006},
       adsurl = {https://ui.adsabs.harvard.edu/abs/2009msfp.book...74O}
}

@ARTICLE{2008A&A...487..575N,
       author = {{Negueruela}, I. and {Marco}, A. and {Herrero}, A. and {Clark}, J.~S.},
        title = "{New very massive stars in Cygnus OB2}",
      journal = {\aap},
         year = 2008,
        month = aug,
       volume = {487},
       number = {2},
        pages = {575-581},
          doi = {10.1051/0004-6361:200810094},
archivePrefix = {arXiv},
       eprint = {0806.2879},
 primaryClass = {astro-ph},
       adsurl = {https://ui.adsabs.harvard.edu/abs/2008A&A...487..575N}
}

@BOOK{2008oasp.book.....G,
       author = {{Gray}, David F.},
        title = "{The Observation and Analysis of Stellar Photospheres}",
         year = 2008,
       adsurl = {https://ui.adsabs.harvard.edu/abs/2008oasp.book.....G}
}

@ARTICLE{2008A&A...479..541H,
       author = {{Hunter}, I. and {Lennon}, D.~J. and {Dufton}, P.~L. and {Trundle}, C. and {Sim{\'o}n-D{\'\i}az}, S. and {Smartt}, S.~J. and {Ryans}, R.~S.~I. and {Evans}, C.~J.},
        title = "{The VLT-FLAMES survey of massive stars: atmospheric parameters and rotational velocity distributions for B-type stars in the Magellanic Clouds}",
      journal = {\aap},
         year = 2008,
        month = feb,
       volume = {479},
       number = {2},
        pages = {541-555},
          doi = {10.1051/0004-6361:20078511},
archivePrefix = {arXiv},
       eprint = {0711.2264},
 primaryClass = {astro-ph},
       adsurl = {https://ui.adsabs.harvard.edu/abs/2008A&A...479..541H}
}

@ARTICLE{2007ApJ...664.1102K,
       author = {{Kiminki}, Daniel C. and {Kobulnicky}, Henry A. and {Kinemuchi}, K. and {Irwin}, Jennifer S. and {Fryer}, Christopher L. and {Berrington}, R.~C. and {Uzpen}, B. and {Monson}, Andy J. and {Pierce}, Michael J. and {Woosley}, S.~E.},
        title = "{A Radial Velocity Survey of the Cyg OB2 Association}",
      journal = {\apj},
         year = 2007,
        month = aug,
       volume = {664},
       number = {2},
        pages = {1102-1120},
          doi = {10.1086/513709},
archivePrefix = {arXiv},
       eprint = {astro-ph/0609772},
 primaryClass = {astro-ph},
       adsurl = {https://ui.adsabs.harvard.edu/abs/2007ApJ...664.1102K}
}

@MISC{2007A&A...471..625T,
       author = {{Trundle}, C. and {Dufton}, P.~L. and {Hunter}, I. and {Evans}, C.~J. and {Lennon}, D.~J. and {Smartt}, S.~J. and {Ryans}, R.~S.~I.},
        title = "{The VLT-FLAMES survey of massive stars: evolution of surface N abundances and effective temperature scales in the Galaxy and Magellanic Clouds}",
         year = 2007,
        month = aug,
        pages = {625-643},
          doi = {10.1051/0004-6361:20077838},
archivePrefix = {arXiv},
       eprint = {0706.1731},
 primaryClass = {astro-ph},
    publisher = {EDP},
       adsurl = {https://ui.adsabs.harvard.edu/abs/2007A&A...471..625T}
}

@ARTICLE{2007A&A...468.1063S,
       author = {{Sim{\'o}n-D{\'\i}az}, S. and {Herrero}, A.},
        title = "{Fourier method of determining the rotational velocities in OB stars}",
      journal = {\aap},
         year = 2007,
        month = jun,
       volume = {468},
       number = {3},
        pages = {1063-1073},
          doi = {10.1051/0004-6361:20066060},
archivePrefix = {arXiv},
       eprint = {astro-ph/0703216},
 primaryClass = {astro-ph},
       adsurl = {https://ui.adsabs.harvard.edu/abs/2007A&A...468.1063S}
}

@ARTICLE{2007A&A...466..277H,
       author = {{Hunter}, I. and {Dufton}, P.~L. and {Smartt}, S.~J. and {Ryans}, R.~S.~I. and {Evans}, C.~J. and {Lennon}, D.~J. and {Trundle}, C. and {Hubeny}, I. and {Lanz}, T.},
        title = "{The VLT-FLAMES survey of massive stars: surface chemical compositions of B-type stars in the Magellanic Clouds}",
      journal = {\aap},
         year = 2007,
        month = apr,
       volume = {466},
       number = {1},
        pages = {277-300},
          doi = {10.1051/0004-6361:20066148},
archivePrefix = {arXiv},
       eprint = {astro-ph/0609710},
 primaryClass = {astro-ph},
       adsurl = {https://ui.adsabs.harvard.edu/abs/2007A&A...466..277H}
}

@ARTICLE{2007AJ....133.1092W,
       author = {{Wolff}, S.~C. and {Strom}, S.~E. and {Dror}, D. and {Venn}, K.},
        title = "{Rotational Velocities for B0-B3 Stars in Seven Young Clusters: Further Study of the Relationship between Rotation Speed and Density in Star-Forming Regions}",
      journal = {\aj},
         year = 2007,
        month = mar,
       volume = {133},
       number = {3},
        pages = {1092-1103},
          doi = {10.1086/511002},
archivePrefix = {arXiv},
       eprint = {astro-ph/0702133},
 primaryClass = {astro-ph},
       adsurl = {https://ui.adsabs.harvard.edu/abs/2007AJ....133.1092W}
}

@ARTICLE{2007CSE.....9...90H,
       author = {{Hunter}, John D.},
        title = "{Matplotlib: A 2D Graphics Environment}",
      journal = {Computing in Science and Engineering},
         year = 2007,
        month = jan,
       volume = {9},
       number = {3},
        pages = {90-95},
          doi = {10.1109/MCSE.2007.55},
       adsurl = {https://ui.adsabs.harvard.edu/abs/2007CSE.....9...90H}
}

@ARTICLE{2006A&A...456.1131M,
       author = {{Mokiem}, M.~R. and {de Koter}, A. and {Evans}, C.~J. and {Puls}, J. and {Smartt}, S.~J. and {Crowther}, P.~A. and {Herrero}, A. and {Langer}, N. and {Lennon}, D.~J. and {Najarro}, F. and {Villamariz}, M.~R. and {Yoon}, S. -C.},
        title = "{The VLT-FLAMES survey of massive stars: mass loss and rotation of early-type stars in the SMC}",
      journal = {\aap},
         year = 2006,
        month = sep,
       volume = {456},
       number = {3},
        pages = {1131-1151},
          doi = {10.1051/0004-6361:20064995},
archivePrefix = {arXiv},
       eprint = {astro-ph/0606403},
 primaryClass = {astro-ph},
       adsurl = {https://ui.adsabs.harvard.edu/abs/2006A&A...456.1131M}
}

@ARTICLE{2006A&A...454..811P,
       author = {{Pfalzner}, S. and {Olczak}, C. and {Eckart}, A.},
        title = "{The fate of discs around massive stars in young clusters}",
      journal = {\aap},
         year = 2006,
        month = aug,
       volume = {454},
       number = {3},
        pages = {811-814},
          doi = {10.1051/0004-6361:20064905},
archivePrefix = {arXiv},
       eprint = {astro-ph/0604018},
 primaryClass = {astro-ph},
       adsurl = {https://ui.adsabs.harvard.edu/abs/2006A&A...454..811P}
}

@ARTICLE{2006MNRAS.370L..85S,
       author = {{Shen}, Yue and {Lou}, Yu-Qing},
        title = "{Dispersal of gaseous circumstellar discs around high-mass stars}",
      journal = {\mnras},
         year = 2006,
        month = jul,
       volume = {370},
       number = {1},
        pages = {L85-L89},
          doi = {10.1111/j.1745-3933.2006.00194.x},
archivePrefix = {arXiv},
       eprint = {astro-ph/0605505},
 primaryClass = {astro-ph},
       adsurl = {https://ui.adsabs.harvard.edu/abs/2006MNRAS.370L..85S}
}

@ARTICLE{2006MNRAS.365.1333W,
       author = {{Weidner}, Carsten and {Kroupa}, Pavel},
        title = "{The maximum stellar mass, star-cluster formation and composite stellar populations}",
      journal = {\mnras},
         year = 2006,
        month = feb,
       volume = {365},
       number = {4},
        pages = {1333-1347},
          doi = {10.1111/j.1365-2966.2005.09824.x},
archivePrefix = {arXiv},
       eprint = {astro-ph/0511331},
 primaryClass = {astro-ph},
       adsurl = {https://ui.adsabs.harvard.edu/abs/2006MNRAS.365.1333W}
}

@ARTICLE{2005ApJ...634.1214L,
       author = {{Long}, M. and {Romanova}, M.~M. and {Lovelace}, R.~V.~E.},
        title = "{Locking of the Rotation of Disk-Accreting Magnetized Stars}",
      journal = {\apj},
         year = 2005,
        month = dec,
       volume = {634},
       number = {2},
        pages = {1214-1222},
          doi = {10.1086/497000},
archivePrefix = {arXiv},
       eprint = {astro-ph/0510659},
 primaryClass = {astro-ph},
       adsurl = {https://ui.adsabs.harvard.edu/abs/2005ApJ...634.1214L}
}

@ARTICLE{2005AJ....129..809S,
       author = {{Strom}, Stephen E. and {Wolff}, Sidney C. and {Dror}, David H.~A.},
        title = "{B Star Rotational Velocities in h and {\ensuremath{\chi}} Persei: A Probe of Initial Conditions during the Star Formation Epoch?}",
      journal = {\aj},
         year = 2005,
        month = feb,
       volume = {129},
       number = {2},
        pages = {809-828},
          doi = {10.1086/426748},
archivePrefix = {arXiv},
       eprint = {astro-ph/0410337},
 primaryClass = {astro-ph},
       adsurl = {https://ui.adsabs.harvard.edu/abs/2005AJ....129..809S}
}

@INPROCEEDINGS{2005coex.conf..585P,
       author = {{Panagia}, Nino},
        title = "{A Geometric Determination of the Distance to SN 1987A and the LMC}",
    booktitle = {IAU Colloquium 192: Cosmic Explosions, On the 10th Anniversary of SN1993J},
         year = 2005,
       editor = {{Marcaide}, Juan-Maria and {Weiler}, Kurt W.},
       volume = {99},
        month = jan,
        pages = {585},
          doi = {10.1007/3-540-26633-X_78},
       adsurl = {https://ui.adsabs.harvard.edu/abs/2005coex.conf..585P}
}

@ARTICLE{2005MNRAS.356..167M,
       author = {{Matt}, Sean and {Pudritz}, Ralph E.},
        title = "{The spin of accreting stars: dependence on magnetic coupling to the disc}",
      journal = {\mnras},
         year = 2005,
        month = jan,
       volume = {356},
       number = {1},
        pages = {167-182},
          doi = {10.1111/j.1365-2966.2004.08431.x},
archivePrefix = {arXiv},
       eprint = {astro-ph/0409701},
 primaryClass = {astro-ph},
       adsurl = {https://ui.adsabs.harvard.edu/abs/2005MNRAS.356..167M}
}

@ARTICLE{2003AJ....125.2531R,
       author = {{Reed}, B. Cameron},
        title = "{Catalog of Galactic OB Stars}",
      journal = {\aj},
         year = 2003,
        month = may,
       volume = {125},
       number = {5},
        pages = {2531-2533},
          doi = {10.1086/374771},
       adsurl = {https://ui.adsabs.harvard.edu/abs/2003AJ....125.2531R}
}

@ARTICLE{2003ApJ...585..850M,
       author = {{McKee}, Christopher F. and {Tan}, Jonathan C.},
        title = "{The Formation of Massive Stars from Turbulent Cores}",
      journal = {\apj},
         year = 2003,
        month = mar,
       volume = {585},
       number = {2},
        pages = {850-871},
          doi = {10.1086/346149},
archivePrefix = {arXiv},
       eprint = {astro-ph/0206037},
 primaryClass = {astro-ph},
       adsurl = {https://ui.adsabs.harvard.edu/abs/2003ApJ...585..850M}
}

@ARTICLE{2003MNRAS.339..577B,
       author = {{Bate}, Matthew R. and {Bonnell}, Ian A. and {Bromm}, Volker},
        title = "{The formation of a star cluster: predicting the properties of stars and brown dwarfs}",
      journal = {\mnras},
         year = 2003,
        month = mar,
       volume = {339},
       number = {3},
        pages = {577-599},
          doi = {10.1046/j.1365-8711.2003.06210.x},
archivePrefix = {arXiv},
       eprint = {astro-ph/0212380},
 primaryClass = {astro-ph},
       adsurl = {https://ui.adsabs.harvard.edu/abs/2003MNRAS.339..577B}
}

@ARTICLE{2003AJ....125..984M,
       author = {{Monet}, David G. and {Levine}, Stephen E. and {Canzian}, Blaise and {Ables}, Harold D. and {Bird}, Alan R. and {Dahn}, Conard C. and {Guetter}, Harry H. and {Harris}, Hugh C. and {Henden}, Arne A. and {Leggett}, Sandy K. and {Levison}, Harold F. and {Luginbuhl}, Christian B. and {Martini}, Joan and {Monet}, Alice K.~B. and {Munn}, Jeffrey A. and {Pier}, Jeffrey R. and {Rhodes}, Albert R. and {Riepe}, Betty and {Sell}, Stephen and {Stone}, Ronald C. and {Vrba}, Frederick J. and {Walker}, Richard L. and {Westerhout}, Gart and {Brucato}, Robert J. and {Reid}, I. Neill and {Schoening}, William and {Hartley}, M. and {Read}, M.~A. and {Tritton}, S.~B.},
        title = "{The USNO-B Catalog}",
      journal = {\aj},
         year = 2003,
        month = feb,
       volume = {125},
       number = {2},
        pages = {984-993},
          doi = {10.1086/345888},
archivePrefix = {arXiv},
       eprint = {astro-ph/0210694},
 primaryClass = {astro-ph},
       adsurl = {https://ui.adsabs.harvard.edu/abs/2003AJ....125..984M}
}

@ARTICLE{2002MNRAS.336..577R,
       author = {{Ryans}, R.~S.~I. and {Dufton}, P.~L. and {Rolleston}, W.~R.~J. and {Lennon}, D.~J. and {Keenan}, F.~P. and {Smoker}, J.~V. and {Lambert}, D.~L.},
        title = "{Macroturbulent and rotational broadening in the spectra of B-type supergiants}",
      journal = {\mnras},
         year = 2002,
        month = oct,
       volume = {336},
       number = {2},
        pages = {577-586},
          doi = {10.1046/j.1365-8711.2002.05780.x},
       adsurl = {https://ui.adsabs.harvard.edu/abs/2002MNRAS.336..577R}
}

@INPROCEEDINGS{2002RMxAC..14..111A,
       author = {{Abt}, H.~A. and {Levato}, H. and {Grosso}, M.},
        title = "{Rotational Velocities of B Stars}",
    booktitle = {Revista Mexicana de Astronomia y Astrofisica Conference Series},
         year = 2002,
       editor = {{Claria}, Juan Jose and {Garcia Lambas}, Diego and {Levato}, Hugo},
       series = {Revista Mexicana de Astronomia y Astrofisica Conference Series},
       volume = {14},
        month = jan,
        pages = {111},
       adsurl = {https://ui.adsabs.harvard.edu/abs/2002RMxAC..14..111A}
}

@ARTICLE{2000A&A...360..539K,
       author = {{Kn{\"o}dlseder}, J.},
        title = "{Cygnus OB2 - a young globular cluster in the Milky Way}",
      journal = {\aap},
         year = 2000,
        month = aug,
       volume = {360},
        pages = {539-548},
          doi = {10.48550/arXiv.astro-ph/0007442},
archivePrefix = {arXiv},
       eprint = {astro-ph/0007442},
 primaryClass = {astro-ph},
       adsurl = {https://ui.adsabs.harvard.edu/abs/2000A&A...360..539K}
}

@ARTICLE{2000A&A...355L..27H,
       author = {{H{\o}g}, E. and {Fabricius}, C. and {Makarov}, V.~V. and {Urban}, S. and {Corbin}, T. and {Wycoff}, G. and {Bastian}, U. and {Schwekendiek}, P. and {Wicenec}, A.},
        title = "{The Tycho-2 catalogue of the 2.5 million brightest stars}",
      journal = {\aap},
         year = 2000,
        month = mar,
       volume = {355},
        pages = {L27-L30},
       adsurl = {https://ui.adsabs.harvard.edu/abs/2000A&A...355L..27H}
}

@ARTICLE{2000ARA&A..38..143M,
       author = {{Maeder}, Andr{\'e} and {Meynet}, Georges},
        title = "{The Evolution of Rotating Stars}",
      journal = {\araa},
         year = 2000,
        month = jan,
       volume = {38},
        pages = {143-190},
          doi = {10.1146/annurev.astro.38.1.143},
archivePrefix = {arXiv},
       eprint = {astro-ph/0004204},
 primaryClass = {astro-ph},
       adsurl = {https://ui.adsabs.harvard.edu/abs/2000ARA&A..38..143M}
}

@ARTICLE{1999A&A...352L..31H,
       author = {{Hummel}, W. and {Szeifert}, T. and {G{\"a}ssler}, W. and {Muschielok}, B. and {Seifert}, W. and {Appenzeller}, I. and {Rupprecht}, G.},
        title = "{A spectroscopic study of Be stars in the SMC cluster NGC 330}",
      journal = {\aap},
         year = 1999,
        month = dec,
       volume = {352},
        pages = {L31-L35},
       adsurl = {https://ui.adsabs.harvard.edu/abs/1999A&A...352L..31H}
}

@ARTICLE{1999A&A...349..189S,
       author = {{Spruit}, H.~C.},
        title = "{Differential rotation and magnetic fields in stellar interiors}",
      journal = {\aap},
         year = 1999,
        month = sep,
       volume = {349},
        pages = {189-202},
          doi = {10.48550/arXiv.astro-ph/9907138},
archivePrefix = {arXiv},
       eprint = {astro-ph/9907138},
 primaryClass = {astro-ph},
       adsurl = {https://ui.adsabs.harvard.edu/abs/1999A&A...349..189S}
}

@ARTICLE{1998ApJ...499..758J,
       author = {{Johnstone}, Doug and {Hollenbach}, David and {Bally}, John},
        title = "{Photoevaporation of Disks and Clumps by Nearby Massive Stars: Application to Disk Destruction in the Orion Nebula}",
      journal = {\apj},
         year = 1998,
        month = may,
       volume = {499},
       number = {2},
        pages = {758-776},
          doi = {10.1086/305658},
       adsurl = {https://ui.adsabs.harvard.edu/abs/1998ApJ...499..758J}
}

@INPROCEEDINGS{1998APS..APR..I801M,
       author = {{Mouschovias}, Telemachos Ch.},
        title = "{Angular Momentum Transport During Star Formation}",
    booktitle = {APS April Meeting Abstracts},
         year = 1998,
       series = {APS Meeting Abstracts},
        month = apr,
          eid = {I8.01},
        pages = {I8.01},
       adsurl = {https://ui.adsabs.harvard.edu/abs/1998APS..APR..I801M}
}

@ARTICLE{1997MNRAS.284..265H,
       author = {{Howarth}, Ian D. and {Siebert}, Kaj W. and {Hussain}, Gaitee A.~J. and {Prinja}, Raman K.},
        title = "{Cross-correlation characteristics of OB stars from IUE spectroscopy}",
      journal = {\mnras},
         year = 1997,
        month = jan,
       volume = {284},
       number = {2},
        pages = {265-285},
          doi = {10.1093/mnras/284.2.265},
       adsurl = {https://ui.adsabs.harvard.edu/abs/1997MNRAS.284..265H}
}

@ARTICLE{1996ApJ...463..737P,
       author = {{Penny}, Laura R.},
        title = "{Projected Rotational Velocities of O-Type Stars}",
      journal = {\apj},
         year = 1996,
        month = jun,
       volume = {463},
        pages = {737},
          doi = {10.1086/177286},
       adsurl = {https://ui.adsabs.harvard.edu/abs/1996ApJ...463..737P}
}

@ARTICLE{1994ApJ...429..781S,
       author = {{Shu}, Frank and {Najita}, Joan and {Ostriker}, Eve and {Wilkin}, Frank and {Ruden}, Steven and {Lizano}, Susana},
        title = "{Magnetocentrifugally Driven Flows from Young Stars and Disks. I. A Generalized Model}",
      journal = {\apj},
         year = 1994,
        month = jul,
       volume = {429},
        pages = {781},
          doi = {10.1086/174363},
       adsurl = {https://ui.adsabs.harvard.edu/abs/1994ApJ...429..781S}
}

@ARTICLE{1992A&A...261..209H,
       author = {{Herrero}, A. and {Kudritzki}, R.~P. and {Vilchez}, J.~M. and {Kunze}, D. and {Butler}, K. and {Haser}, S.},
        title = "{Intrinsic parameters of galactic luminous OB stars.}",
      journal = {\aap},
         year = 1992,
        month = jul,
       volume = {261},
        pages = {209-234},
       adsurl = {https://ui.adsabs.harvard.edu/abs/1992A&A...261..209H}
}

@ARTICLE{1991AJ....101.1408M,
       author = {{Massey}, Philip and {Thompson}, A.~B.},
        title = "{Massive Stars in CYG OB2}",
      journal = {\aj},
         year = 1991,
        month = apr,
       volume = {101},
        pages = {1408},
          doi = {10.1086/115774},
       adsurl = {https://ui.adsabs.harvard.edu/abs/1991AJ....101.1408M}
}

@ARTICLE{1991ApJ...370L..39K,
       author = {{Koenigl}, Arieh},
        title = "{Disk Accretion onto Magnetic T Tauri Stars}",
      journal = {\apjl},
         year = 1991,
        month = mar,
       volume = {370},
        pages = {L39},
          doi = {10.1086/185972},
       adsurl = {https://ui.adsabs.harvard.edu/abs/1991ApJ...370L..39K}
}

@ARTICLE{1990A&A...237..137D,
       author = {{Dravins}, D. and {Lindegren}, L. and {Torkelsson}, U.},
        title = "{The rotationally broadened line profiles of Sirius.}",
      journal = {\aap},
         year = 1990,
        month = oct,
       volume = {237},
        pages = {137},
       adsurl = {https://ui.adsabs.harvard.edu/abs/1990A&A...237..137D}
}

@ARTICLE{1977ApJ...213..438C,
       author = {{Conti}, P.~S. and {Ebbets}, D.},
        title = "{Spectroscopic studies of O-type stars. VII. Rotational velocities V sin i and evidence for macroturbulent motions.}",
      journal = {\apj},
         year = 1977,
        month = apr,
       volume = {213},
        pages = {438-447},
          doi = {10.1086/155173},
       adsurl = {https://ui.adsabs.harvard.edu/abs/1977ApJ...213..438C}
}

@ARTICLE{1976PASP...88..809S,
       author = {{Smith}, M.~A. and {Gray}, D.~F.},
        title = "{Fourier analysis of spectral line profiles: a new tool for an old art.}",
      journal = {\pasp},
         year = 1976,
        month = dec,
       volume = {88},
        pages = {809-823},
          doi = {10.1086/130029},
       adsurl = {https://ui.adsabs.harvard.edu/abs/1976PASP...88..809S}
}

@ARTICLE{1975ApJS...29..137S,
       author = {{Slettebak}, A. and {Collins}, II, G.~W. and {Boyce}, P.~B. and {White}, N.~M. and {Parkinson}, T.~D.},
        title = "{A system of standard stars for rotational velocity determinations.}",
      journal = {\apjs},
         year = 1975,
        month = may,
       volume = {29},
        pages = {137-159},
          doi = {10.1086/190338},
       adsurl = {https://ui.adsabs.harvard.edu/abs/1975ApJS...29..137S}
}

@ARTICLE{1958ArM.....3..469H,
       author = {{Hodges}, J.~L.},
        title = "{The significance probability of the smirnov two-sample test}",
      journal = {Arkiv for Matematik},
         year = 1958,
        month = jan,
       volume = {3},
       number = {5},
        pages = {469-486},
          doi = {10.1007/BF02589501},
       adsurl = {https://ui.adsabs.harvard.edu/abs/1958ArM.....3..469H}
}

@ARTICLE{1933MNRAS..93..478C,
       author = {{Carroll}, J.~A.},
        title = "{The spectroscopic determination of stellar rotation and its effect on line profiles}",
      journal = {\mnras},
         year = 1933,
        month = may,
       volume = {93},
        pages = {478-507},
          doi = {10.1093/mnras/93.7.478},
       adsurl = {https://ui.adsabs.harvard.edu/abs/1933MNRAS..93..478C}
}
        }
    }

    {
    \begin{appendix}
        \onecolumn   

        \section{Initial sample of \spectype{B}{0}-type stars in Cygnus~OB2} \label{A: sample}
            {
            \begin{table*}[th!]
                \caption{
                    Identifications, J2016.0 coordinates, G and B magnitudes, along with the $\mathrm{\textit{G}_{BP} - \textit{G}_{RP}}$ color index of the initial \spectype{B}{0} sample from Cygnus~OB2. 
                } 
                \label{T: Bstars_Sample}
                \centering
                \tiny
                \begin{tabular}{ccccccccc}
                    \hline
                    \hline \\ [-1.8ex]
                    2MASS Identifier  & GAIA DR3 Identifier & Alternative Name & RA [hh mm ss] & DEC [º ' ''] & \textit{B} mag. & \textit{G} mag. & $\mathrm{\textit{G}_{BP} - \textit{G}_{RP}}$ & Source \\ \hline \\ [-1.8ex]
                    J20325904+4117589 & 2067784723027769728 & Schulte 43       & 20 32 59.03   & +41 17 58.93 & 17.46           & 14.96           & 2.27                                         & [1]    \\
                    J20325976+4114196 & 2067782837540661248 & Schulte 40       & 20 32 59.75   & +41 14 19.63 & 17.13           & 13.95           & 2.63                                         & [1]    \\
                    J20332563+4115247 & 2067783524735406336 & Schulte 67       & 20 33 25.63   & +41 15 24.75 & 17.22           & 14.24           & 2.40                                         & [1]    \\
                    J20323949+4052475 & 2067768195993175936 & [CPR2002] A31    & 20 32 39.50   & +40 52 47.38 & 15.20           & 12.07           & 2.74                                         & [4]    \\
                    J20323498+4052390 & 2067768372090380800 & [CPR2002] A33    & 20 32 34.98   & +40 52 38.87 & 15.83           & 12.50           & 2.61                                         & [7]    \\
                    J20331869+4059378 & 2067759056302388992 & [CPR2002] B15    & 20 33 18.70   & +40 59 37.79 & 15.69           & 12.41           & 2.53                                         & [6]    \\
                    J20330661+4121131 & 2067784967844456192 & ALS 15165        & 20 33 06.61   & +41 21 13.06 & 15.97           & 13.22           & 2.29                                         & [3]    \\
                    J20294666+4105083 & 2067810183593068160 & ALS 19636        & 20 29 46.67   & +41 05 08.27 & 13.74           & 11.49           & 1.93                                         & [7]    \\
                    J20305112+4120218 & 2067842756625081856 & ALS 15171        & 20 30 51.11   & +41 20 21.70 & 16.66           & 13.05           & 2.77                                         & [3]    \\
                    -                 & 2067781944187448704 & ALS 15161        & 20 33 10.40   & +41 13 05.93 & 16.00           & 15.83           & 1.49                                         & [3]    \\
                    J20334234+4111456 & 2067779779523905792 & ALS 15160        & 20 33 42.34   & +41 11 45.47 & 15.71           & 12.83           & 2.37                                         & [3]    \\
                    J20331050+4122224 & 2067785375864885888 & ALS 15132        & 20 33 10.50   & +41 22 22.35 & 14.54           & 12.15           & 2.08                                         & [5]    \\
                    J20331859+4124493 & 2067879354045188992 & ALS 15180        & 20 33 18.58   & +41 24 49.28 & 16.29           & 13.97           & 2.33                                         & [3]    \\
                    J20321312+4127243 & 2067835682818358400 & Cyg OB2 4B       & 20 32 13.12   & +41 27 24.33 & 13.08           & 11.28           & 1.62                                         & [3]    \\
                    J20321568+4046170 & 2067743976676134784 & -                & 20 32 15.68   & +40 46 16.92 & 15.75           & 12.59           & 2.51                                         & [7]    \\
                    J20330526+4143367 & 2067934394049131520 & ALS 15130        & 20 33 05.26   & +41 43 36.69 & 14.43           & 11.99           & 2.15                                         & [5]    \\
                    J20272099+4121261 & 2067691195819277056 & Schulte 25       & 20 27 21.00   & +41 21 26.06 & 14.65           & 12.07           & 2.07                                         & [6]    \\
                    J20282772+4104018 & 2067625195062093312 & -                & 20 28 27.72   & +41 04 01.73 & 14.97           & 12.12           & 2.32                                         & [7]    \\
                    J20332099+4135518 & 2067929862860202752 & ALS 15177        & 20 33 20.99   & +41 35 51.76 & 16.51           & 13.29           & 2.55                                         & [3]    \\
                    J20295701+4109538 & 2067816681882846464 & ALS 19634        & 20 29 57.01   & +41 09 53.76 & 13.04           & 11.69           & 2.02                                         & [2]    \\
                    J20331106+4110321 & 2067778783091509888 & ALS 15181        & 20 33 11.06   & +41 10 31.96 & 16.97           & 13.41           & 2.70                                         & [3]    \\
                    J20281547+4038196 & 2067421304373639552 & ALS 11369        & 20 28 15.47   & +40 38 19.74 & 10.66           & 9.51            & 1.14                                         & [6]    \\
                    J20315898+4107314 & 2067821664044745984 & GSC 03157-01377  & 20 31 58.98   & +41 07 31.31 & 16.34           & 12.97           & 2.58                                         & [6]    \\
                    J20331129+4042338 & 2067751604537938688 & Cyg OB2-22A      & 20 33 11.30   & +40 42 33.67 & 16.41           & 12.49           & 2.97                                         & [7]    \\
                    J20331740+4112387 & 2067780397999213952 & ALS 15169        & 20 33 17.40   & +41 12 38.65 & 16.34           & 13.26           & 2.50                                         & [3]    \\
                    J20335925+4105380 & 2067766276146708608 & ALS 15147        & 20 33 59.25   & +41 05 38.02 & 15.30           & 12.70           & 2.19                                         & [5]    \\
                    J20312210+4112029 & 2067826332670106112 & Cyg OB2-A30      & 20 31 22.10   & +41 12 02.82 & 15.23           & 12.30           & 2.31                                         & [6]    \\
                    J20292449+4052597 & 2067429168454304896 & -                & 20 29 24.49   & +40 52 59.75 & 14.81           & 11.85           & 2.38                                         & [7]    \\
                    J20322774+4128522 & 2067837263366481536 & ALS 11410        & 20 32 27.73   & +41 28 52.18 & 12.01           & 10.98           & 1.38                                         & [2]    \\
                    J20305552+4054541 & 2067795752502969472 & [CPR2002] A35    & 20 30 55.51   & +40 54 53.95 & 14.78           & 12.02           & 2.26                                         & [7]    \\
                    J20314341+4100021 & 2067797195612486656 & -                & 20 31 43.41   & +41 00 02.00 & 16.84           & 13.17           & 2.74                                         & [6]    \\
                    J20314605+4043246 & 2067742877164542080 & [CPR2002] A44    & 20 31 46.05   & +40 43 24.58 & 14.48           & 11.67           & 2.29                                         & [7]    \\ \hline
                \end{tabular}
                \tablefoot
                    {
                    The \textit{B} magnitudes are compiled from multiple studies (see last column for reference); for those stars with no \textit{B} data available, we estimated their magnitudes using Johnson-Cousins relationships with Gaia photometry \citep{2021A&A...649A...3R}.
                    }
                \tablebib
                    {
                    [1] \citet{1991AJ....101.1408M}, [2] \citet{2000A&A...355L..27H}, [3] \citet{2003AJ....125.2531R}, [4] \citet{2003AJ....125..984M}, [5] \citet{2007AJ....133.1092W}, [6] \citet{2012yCat.1322....0Z}, [7] \citet{2016A&A...595A...1G, 2021A&A...649A...3R, 2023A&A...674A...1G}.
                    }
            \end{table*}
            }
    
        \newpage

        \section{Updated spectral types and projected rotational velocities of the initial \spectype{B}{0}-type sample} \label{A: SpT_vsini}
            {
            \begin{table*}[th!]
                \caption{
                    New spectral types (SpT) and projected rotational velocities ($\vsini$) for our sample of initial \spectype{B}{0} stars.
                } 
                \label{T: OBstars_SpT_vsini}
                \centering
                \tiny
                \begin{tabular}{ccccccc}
                    \hline
                    \hline \\ [-1.8ex]
                    Identifier        & SpT Literature                                                 & Source & New SpT                     & $v\sin{i} \, [\kmps]$ & Line                     & Resolution \\ \hline \\ [-1.8ex]
                    J20325904+4117589 & B                                                              & [1]    & \spectype{B}{2.5}[V]        & 205                   & \specline{He}{I}[6678]   & 5000       \\
                    J20325976+4114196 & B                                                              & [1]    & \spectype{B}{2:}[II:p]      & 88                    & \specline{He}{I}[6678]   & 5000       \\
                    J20332563+4115247 & B                                                              & [1]    & \spectype{B}{2}[IV:\,nn]    & 425                   & \specline{He}{I}[6678]   & 5000       \\
                    J20323949+4052475 & \spectype{B}{0:}[V{:}]                                         & [2]    & \spectype{B}{0.2}[IV]       & 113                   & \specline{He}{I}[6678]   & 5000       \\
                    J20323498+4052390 & \spectype{B}{0.2}[IV]                                          & [2]    & \spectype{B}{0.2}[IV]       & 95                    & \specline{He}{I}[6678]   & 5000       \\
                    J20331869+4059378 & \spectype{B}{0.5}[IIIe]                                        & [2]    & \spectype{B}{0.5}[IIIe]     & 59                    & \specline{He}{I}[6678]   & 5000       \\
                    J20330661+4121131 & \spectype{B}{0.5}[V]                                           & [3]    & \spectype{B}{0.5}[V]        & 186                   & \specline{He}{I}[6678]   & 5000       \\
                    J20294666+4105083 & \spectype{B}{0.5}[V] + \spectype{B}{1}[V:]-\spectype{B}{2}[V:] & [4]    & \spectype{B}{0.5}[V\,(n)]   & 141                   & \specline{He}{I}[6678]   & 5000       \\
                    J20305112+4120218 & \spectype{B}{0}[V]                                             & [2]    & \spectype{O}{9.7}[IV]       & 36                    & \specline{He}{I}[6678]   & 5000       \\
                    ALS 15161         & \spectype{B}{0}[V]                                             & [3]    & -                           & 118                   & \specline{He}{I}         & 2165       \\
                    J20334234+4111456 & \spectype{B}{0}[V]                                             & [3]    & \spectype{O}{9.7:}[V:]      & 48                    & \specline{He}{I}[6678]   & 5000       \\
                    J20331050+4122224 & \spectype{B}{0}[V] + (\spectype{B}{3}[V]+\spectype{B}{6}[V])   & [4]    & \spectype{B}{0}[V]          & 137                   & \specline{He}{I}         & 2165       \\
                    J20331859+4124493 & \spectype{B}{0:}                                               & [3]    & \spectype{B}{0.2}[V]        & 163                   & \specline{He}{I}[6678]   & 5000       \\
                    J20321312+4127243 & \spectype{B}{0}[Vp]                                            & [1]    & \spectype{B}{0:}[III:\,(n)] & 246                   & \specline{He}{I}[6678]   & 5000       \\
                    J20321568+4046170 & \spectype{B}{0.2}[IV]                                          & [2]    & \spectype{B}{0}[V]          & 29                    & \specline{He}{I}[6678]   & 5000       \\
                    J20330526+4143367 & \spectype{B}{1}[V] + B                                         & [4]    & \spectype{B}{1}[V]          & 134                   & \specline{He}{I}         & 5000       \\
                    J20272099+4121261 & \spectype{B}{0.5}[V]                                           & [5]    & \spectype{B}{0.2:}[IV:]     & 194                   & \specline{He}{I}[6678]   & 5000       \\
                    J20282772+4104018 & \spectype{B}{0.5}[V]                                           & [2]    & \spectype{B}{0.5}[V]        & 160                   & \specline{He}{I}[6678]   & 5000       \\
                    J20332099+4135518 & \spectype{B}{0}[III-IV]                                        & [3]    & \spectype{B}{0}[V]          & 15                    & \specline{He}{I}[6678]   & 5000       \\
                    J20295701+4109538 & \spectype{B}{0}[V]                                             & [2]    & \spectype{O}{9.5}[IV]       & 67                    & \specline{O}{III}[5592]  & 13600      \\
                    J20331106+4110321 & \spectype{B}{0}[V]                                             & [3]    & \spectype{B}{0.7:}[V:]      & 78                    & \specline{He}{I}[6678]   & 5000       \\
                    J20281547+4038196 & \spectype{B}{0}[V:]                                            & [2]    & \spectype{O}{9.7}[IV]       & 60                    & \specline{He}{I}[6678]   & 5000       \\
                    J20315898+4107314 & \spectype{B}{1}[V]                                             & [5]    & \spectype{B}{1}[V]          & 235                   & \specline{He}{I}[6678]   & 5000       \\
                    J20331129+4042338 & \spectype{B}{0:}[III:]                                         & [2]    & \spectype{O}{9.7}[III]      & 233                   & \specline{He}{I}[6678]   & 5000       \\
                    J20331740+4112387 & \spectype{B}{0:}[V]                                            & [3]    & \spectype{B}{0.5}[V\,(n)]   & 355                   & \specline{He}{I}[6678]   & 5000       \\
                    J20335925+4105380 & \spectype{B}{0}[V]                                             & [3]    & \spectype{B}{0}[V]          & 141                   & \specline{He}{I}[6678]   & 5000       \\
                    J20312210+4112029 & \spectype{B}{0}[III]                                           & [2]    & \spectype{B}{0.2}[V\,(n)]   & 198                   & \specline{He}{I}[6678]   & 5000       \\
                    J20292449+4052597 & \spectype{B}{0.2}[IV]                                          & [2]    & \spectype{B}{0.2}[IV]       & 137                   & \specline{Si}{III}[4552] & 7500       \\
                    J20322774+4128522 & \spectype{B}{0}[Ib]                                            & [3]    & \spectype{B}{0}[Ib]         & 21                    & \specline{He}{I}[6678]   & 5000       \\
                    J20305552+4054541 & \spectype{B}{0}[Ib]                                            & [2]    & \spectype{O}{9.7}[III]      & 186                   & \specline{He}{I}         & 4688       \\
                    J20314341+4100021 & \spectype{B}{1}[V]                                             & [5]    & \spectype{B}{1}[V\,(n)]     & 92                    & \specline{He}{I}[6678]   & 5000       \\
                    J20314605+4043246 & \spectype{B}{0.5}[IV]                                          & [2]    & \spectype{B}{0.5}[IV]       & 59                    & \specline{Si}{III}[4552] & 7500       \\ \hline \\ [-1.8ex]
                \end{tabular}
                \tablefoot
                    {
                    The second and third columns present the literature-based spectral classification and reference sources, respectively. The fourth column details the updated SpT obtained in this work. The fifth column gives the projected rotational velocity from \texttt{iacob-broad}. The sixth and seventh columns list the diagnostic line and the spectral resolution of the spectrum used in the computation.
                    }
                \tablebib
                    {
                    \citet{1991AJ....101.1408M}, [2] \citet{2012A&A...543A.101C}, [3] \citet{2007ApJ...664.1102K}, [4] \citet{ 2016ApJS..224....4M}, [5] \citet{2018A&A...612A..50B}.
                    }
            \end{table*}
            }

        \newpage        

        \section{Revising the spectral type of 2MASS~J20295701+4109538} \label{A: SpT_class}
            {
            \begin{figure}[h!]
                \centering
                \resizebox{\hsize}{!}{\includegraphics[scale=1]{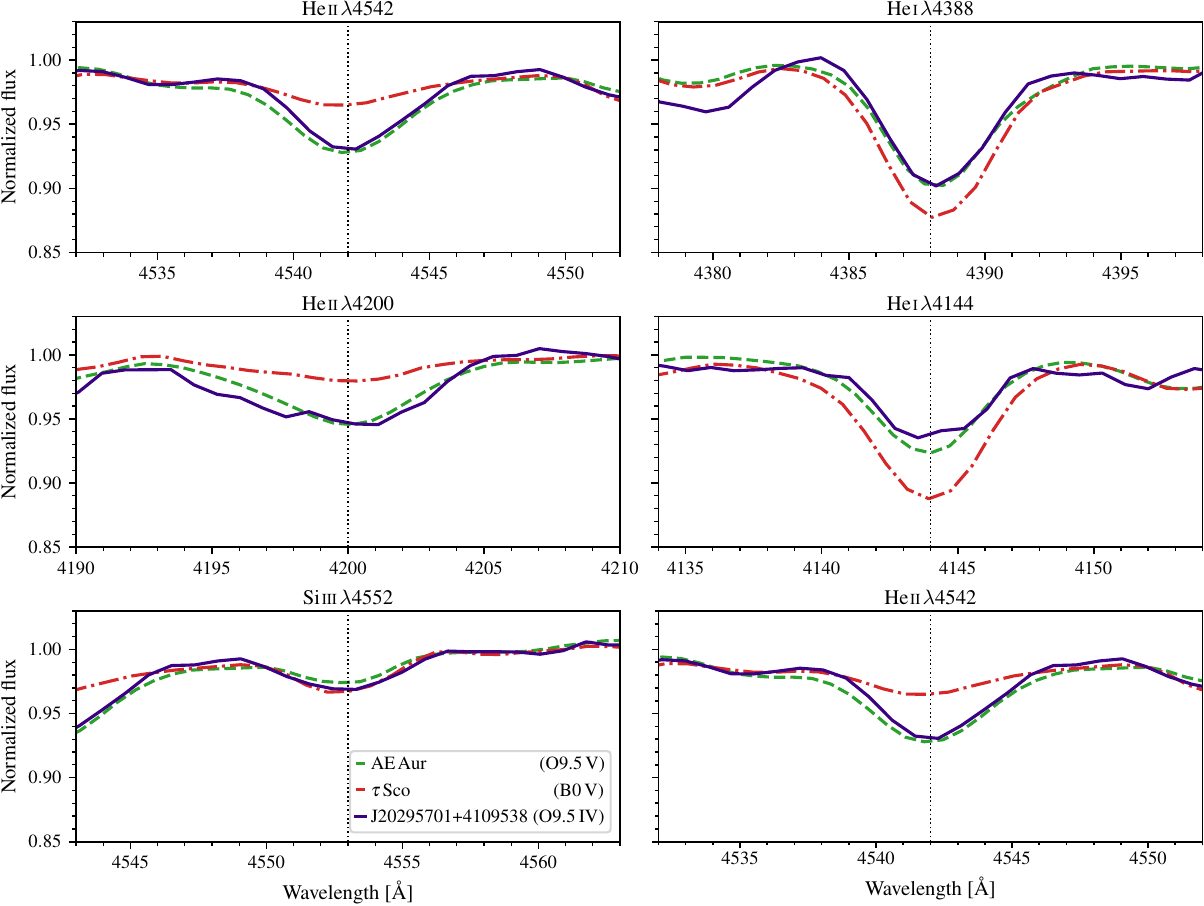}}
                \caption
                    {
                        Reclassification of 2MASS~J20295701+4109538 from \spectype{B}{0} to \spectype{O}{9.5}. The target spectrum, obtained with OSIRIS+@GTC at $\mathrm{R \!\sim\! 2500}$, is shown in \textit{dark blue}. Each panel displays the diagnostic lines used to distinguish O- and early B-type stars, following the criteria of \citet{2011ApJS..193...24S}. For comparison, the O- and B-type spectra from the GOSSS catalog \citep{2016ApJS..224....4M} are included: the \spectype{O}{9.5}[V] standard AE~Aur (in green) and the \spectype{B}{0}[V] standard $\tau$~Sco (in red).
                    }
                \label{Fig: Spec_classification_J20295701+4109538}
            \end{figure}
            
            2MASS~J20295701+4109538 was previously classified as \spectype{B}{0}[V] by \citet{2012A&A...543A.101C}. However, a detailed inspection of its optical spectrum reveals features more consistent with a late O-type classification. As shown in Fig.~\ref{Fig: Spec_classification_J20295701+4109538}, the helium and silicon diagnostic lines provide clear evidence supporting this revision.

            The \specline{He}{II} absorption lines at $4542\,\AA$ and $4200\,\AA$ are clearly present, although weaker than the \specline{He}{I} lines at $4388\,\AA$ and $4144\,\AA$. This relative line strength pattern is characteristic of late-O and early-B stars, according to \citet{2011ApJS..193...24S}. However, a \spectype{B}{0} star would show substantially weaker \specline{He}{II} lines compared to \specline{He}{I} (see Table~\ref{T: LineRatios_SpecClass}). Moreover, the \specline{Si}{III}[4552] line appears weak with respect to \specline{He}{II}[4542], indicating an O-type rather than early-B classification.

            We performed the spectral classification by comparing the observed spectrum of 2MASS~J20295701+4109538 (OSIRIS+@GTC, $\mathrm{R \!\sim\! 2500}$) with standard spectra using the \texttt{MGB} code. This tool enables a direct comparison of diagnostic line ratios to identify the best-matching standard star, ensuring an accurate determination of the spectral type. This analysis utilizes O-type standards from the GOSSS catalog \citep{2016ApJS..224....4M} and a dedicated collection of \spectype{B}{0} standard spectra. As shown in Fig.~\ref{Fig: Spec_classification_J20295701+4109538}, the observed spectrum of 2MASS~J20295701+4109538 closely matches the \spectype{O}{9.5}[V] standard AE~Aur, while differing markedly from the \spectype{B}{0}[V] standard $\tau$~Sco (plotted in \textit{dark blue}, green and red, respectively). The difference is especially clear in the strength of the \specline{He}{II} features and in the \specline{Si}{III}[4552] line. Together, these diagnostics confirm that 2MASS~J20295701+4109538 should be reclassified as \spectype{O}{9.5}. 
            
            This misclassification by \citet{2012A&A...543A.101C} was likely caused by the relatively low spectral resolution and limited signal-to-noise ratio of the data. Their spectra, with a resolving power of $\mathrm{R \!\sim\! 1000}$, make it challenging to detect subtle variations in key diagnostic lines. This is particularly important when distinguishing between late-\spectype{O}{9} and \spectype{B}{0} spectral types: differences in line strengths and profiles are minimal and can be easily masked by noise or instrumental broadening. In these cases, even small inaccuracies in line measurement can lead to misclassification.

            \begin{table}[h]
                \caption{
                    Classification criteria for spectral types at the \spectype{O}{9.5}-\spectype{B}{0} range \citep{2011ApJS..193...24S}. 
                } 
                \label{T: LineRatios_SpecClass}
                \centering
                \small
                \begin{tabular}{ccc}
                \hline
                \hline \\ [-1.8ex]
                Spectral & \ratio{He}{II}[4542]{He}{I}[4388] & \ratio{Si}{III}[4552]{He}{II}[4542] \\
                type     & and                               &                                     \\
                         & \ratio{He}{II}[4200]{He}{I}[4144] &                                     \\ \hline \\ [-1.8ex]
                O9       & $=$                               & $\ll$                               \\
                O9.5     & $\leq$                            & $<$                                 \\
                O9.7     & $<$                               & $\leq$ to $\geq$                    \\
                B0       & $\ll$                             & $\gg$                               \\ \hline
                \end{tabular}
            \end{table}

            This example highlights the advantages of using the interactive features of \texttt{MGB}. With this tool users can closely examine line profiles and ratios, and compare multiple diagnostic lines simultaneously. When combined with high-quality data -- spectra with sufficient signal-to-noise and resolution -- this approach significantly improves the precision of spectral classification, critical in the transition between late-O and early B-type stars.
            }

        \section{Comparative analysis of projected rotational velocity distributions via the Anderson-Darling k-sample test} \label{A: AD_test}
            {
            To check if the rotational velocities of Cygnus OB2 statistically follow the same distribution as those in reference studies, we applied the Anderson-Darling k-sample test \citep[AD, see][]{ADtest, Scholz+87}. The AD test was intentionally chosen over the commonly used Kolmogorov-Smirnov two-sided test \citep[KS, ][]{1958ArM.....3..469H}: while the KS test is more sensitive to differences at the center of the distribution, the AD test emphasizes the tails -- a critical advantage given our focus on the high-velocity region. Additionally, the AD test shows greater sensitivity than the KS test when dealing with small samples, making it particularly well-suited for our Cygnus~OB2 distribution. 
    
            \begin{table}[th]
                \caption{
                    Test statistics and p-values from the Anderson–Darling k-sample test.
                } 
                \label{T: Ktest}
                \centering
                \tiny
                \begin{tabular}{cccc}
                    \hline
                    \hline \\ [-1.8ex]
                    \multicolumn{2}{c}{$v \sin{i}$ distribution comparison}                   & AD statistic & p-value           \\ \hline \\ [-1.8ex]
                    \multirow{2}{*}{\citet{2022AA...665A.150H}} & Complete                    & 10.65        & $1 \cdot 10^{-4}$ \\
                                                                & $\mathrm{M\leq\!32\,\Msol}$ & 7.08         & $7 \cdot 10^{-4}$ \\
                    \citet{2013AA...560A..29R}                  & Complete                    & 3.00         & $2 \cdot 10^{-2}$ \\
                    \citet{2025AA...695A.248B}                  & Complete                    & 4.47         & $6 \cdot 10^{-3}$ \\
                    \citet{2019AA...626A..50D}                  & Complete                    & 1.94         & $5 \cdot 10^{-2}$ \\ \hline
                \end{tabular}
                \tablefoot
                    {
                    The projected rotational velocity distributions of O-type stars (first column) are compared to that of the Cygnus~OB2 association, which includes our new O stars combined with measurements from \citet{2020AA...642A.168B}. From top to bottom, the comparison samples correspond to the O population in the Galactic region \citep{2022AA...665A.150H}, 30~Doradus \citep{2013AA...560A..29R}, Carina~OB1 \citep{2025AA...695A.248B}, and NGC~346 \citep{2019AA...626A..50D}.
                    }
            \end{table}
            
            The results of the AD test, summarized in Table~\ref{T: Ktest}, show statistically significant differences between the $\vsini$ distributions of Cygnus~OB2 and other reference environments. The strongest contrast is found with the Galactic sample \citep{2022AA...665A.150H}, both for the full dataset and for stars with $\mathrm{M\!<\!32\,\Msol}$. These cases yield the highest test statistics and lowest p-values, strongly rejecting the null hypothesis of a shared parent distribution. Similarly, the 30~Doradus \citep{2013AA...560A..29R} and Carina~OB1 \citep{2025AA...695A.248B} samples show statistical differences with p-values below 0.02 and 0.01, respectively. In contrast, the NGC~346 distribution \citep{2019AA...626A..50D} yields a marginal p-value of 0.05, suggesting only weak evidence against the null hypothesis. Overall, these AD results indicate that the Cygnus~OB2 rotational velocity distribution is statistically distinct from the other studied massive star environments.
            }

        \section{Statistical inference of projected rotational velocity distributions via bootstrap analysis} \label{A: Bootstrap}
            {
            In Sect.~\ref{S: Cygnus OB2 rotational velocity distribution from its O-type population}, we present the distributions of projected rotational velocities ($\vsini$) for O-type stars across several Galactic and extragalactic regions. Fig.~\ref{Fig: vsini_Distribution_2} compares the $\vsini$ distribution of Cygnus~OB2 (shown in black) with four reference populations: the Galactic sample \citep[top left,][]{2022AA...665A.150H}, the 30~Doradus population \citep[top right,][]{2013AA...560A..29R}, Carina~OB1  \citep[bottom left,][]{2025AA...695A.248B}, and NGC~346 \citep[bottom right,][]{2019AA...626A..50D}. In each case, the upper subplot displays the normalized histogram, while the lower subplot shows the corresponding cumulative distribution function. 

            As previously discussed, significant differences are observed between stellar populations, especially in the high-velocity region ($\vsini\!\gtrsim\!200\,\kmps$). Some populations exhibit a higher fraction of rapid rotators, as the Milky Way and 30~Doradus (see Table~\ref{T: N_samples}). These variations may reflect differences in evolutionary state, properties related to the molecular cloud (e.g., multiplicity fraction or initial rotational distribution), or dynamical mechanisms such as runaway ejections. While alternative explanations can be considered, we consider them less pausible; these include biased $\vsini$ arising from spin-axis alignment and reduced rotation due to magnetic braking (see Sect.~\ref{S: Discussion} for a detailed discussion).

            To verify that the observed differences in the high-velocity region are not artifacts of binning or differences in sample size, we perform an intensive bootstrap analysis \citep{Kushary01052000, 10.1214/16-SS113}. This non-parametric method enables statistically robust comparisons: it reduces the impact of small-number statistics and provides reliable uncertainties for each velocity bin, without assuming any underlying distribution. For each population, we randomly select subsamples of $40$ stars -- with replacement -- from the original dataset; with this subsample size we ensure an unbiased and consistent comparison across all regions (see Table~\ref{T: N_samples}). Each bootstrap iteration generates a synthetic sample, with its $\vsini$ values organized into fixed-width bins of $20\,\kmps$. This procedure is repeated $10^4$ times to achieve high statistical precision, and final histograms are derived by averaging the binned counts of all iterations. The corresponding standard deviation provides a robust, data-driven estimate of the uncertainty in each bin.

            \begin{table}[h]
                \caption{
                    Number of O-type stars with measured projected rotational velocities in each stellar sample.
                } 
                \label{T: N_samples}
                \centering
                \tiny
                \setlength{\tabcolsep}{3.7pt} 
                \begin{tabular}{ccclcc}
                    \hline
                    \hline \\ [-1.8ex]
                    \multicolumn{2}{c}{\multirow{2}{*}{$\vsini$ distribution}}              & \multicolumn{2}{c}{All $\vsini$}   & \multicolumn{2}{c}{$\vsini>\!200\,\kmps$} \\
                    \multicolumn{2}{c}{}                                                      & \multicolumn{2}{c}{$\mathrm{N_{i}}$} & $\mathrm{N_{i}}$ & $\mathrm{N_i/N_{total}}$ \\ \hline \\ [-1.8ex]
                    \multirow{2}{*}{Cygnus~OB2}                 & Complete                    & \multicolumn{2}{c}{69}               & 9                & 0.13                     \\
                                                                & $\mathrm{M\leq\!32\,\Msol}$ & \multicolumn{2}{c}{44}               & 6                & 0.09                     \\
                    \multirow{2}{*}{\citet{2022AA...665A.150H}} & Complete                    & \multicolumn{2}{c}{285}              & 51               & 0.18                     \\
                                                                & $\mathrm{M\leq\!32\,\Msol}$ & \multicolumn{2}{c}{84}               & 27               & 0.09                     \\
                    \citet{2013AA...560A..29R}                  & Complete                    & \multicolumn{2}{c}{216}              & 51               & 0.24                     \\
                    \citet{2025AA...695A.248B}                  & Complete                    & \multicolumn{2}{c}{37}               & 7                & 0.19                     \\
                    \citet{2019AA...626A..50D}                  & Complete                    & \multicolumn{2}{c}{47}               & 9                & 0.19                     \\ \hline
                \end{tabular}
                \tablefoot
                    {
                    The first column lists the regions included in this work: from top to bottom, Cygnus~OB2 \citep[combining our dataset with measurements from][]{2020AA...642A.168B}, the Galactic region \citep{2022AA...665A.150H}, 30~Doradus \citep{2013AA...560A..29R}, Carina~OB1 \citep{2025AA...695A.248B}, and NGC~346 \citep{2019AA...626A..50D}.
                    }
            \end{table}

            The resulting $\vsini$ distributions for O-type stars are presented in Fig.~\ref{Fig: vsini_Distribution_4}, following the same format as Fig.~\ref{Fig: vsini_Distribution_2}. Bootstrapped uncertainties, shown as error bars, provide direct estimates of statistical errors in the normalized histograms and cumulative distributions. Crucially, the bootstrapped distributions closely match the original histograms, reinforcing the statistical robustness of the observed features. In Cygnus~OB2, this agreement confirms that the lack of fast rotators is not a consequence of binning artifacts or statistical fluctuations. Instead, the bootstrap analysis reveals a significant and persistent deficit of fast-rotating stars, suggesting that the absence of a high-velocity tail is intrinsic to the sample. While this result is statistically robust, it may still be affected by observational biases, particularly due to the limited spatial coverage of the Cygnus~OB2 sample.

            \begin{figure}[]
                \centering
                \resizebox{\hsize}{!}{\includegraphics[scale=1]{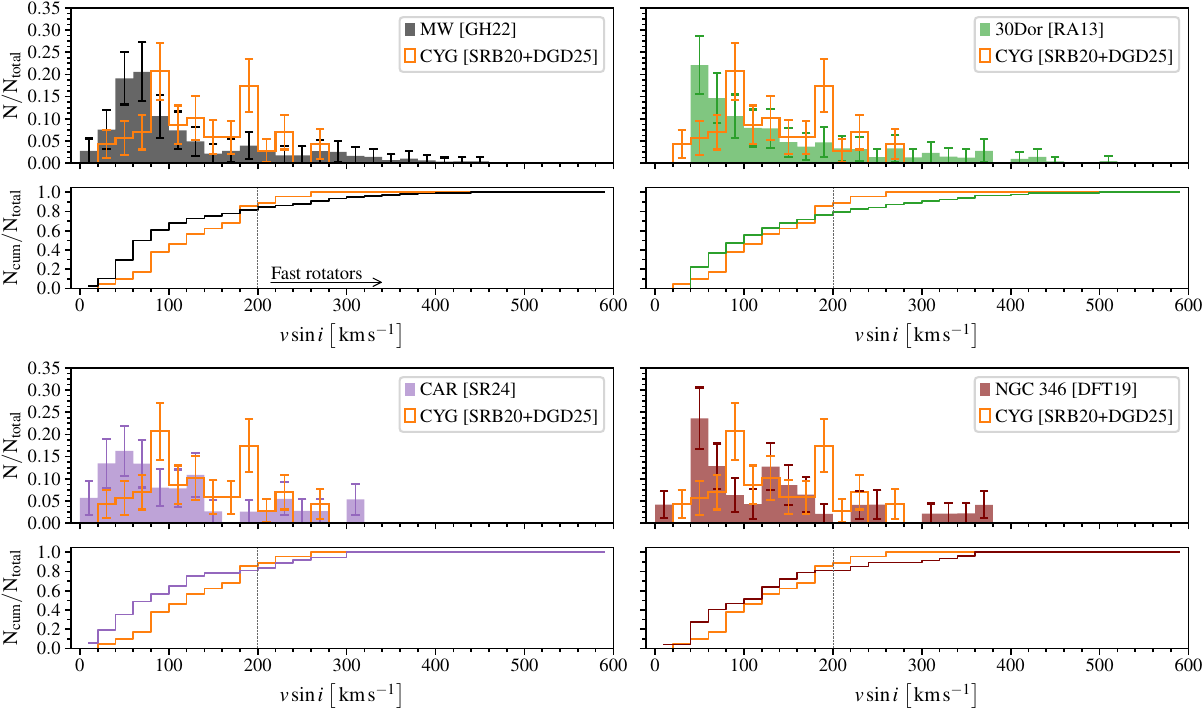}}
                \caption
                    {
                         Bootstrapped distribution of projected rotational velocities for O-type stars across different regions within the local Universe. The histograms in black, green, purple, and brown show the $\vsini$ measurements for Galactic stars in the solar neighborhood \citep{2022AA...665A.150H}, 30~Doradus \citep{2013AA...560A..29R}, Carina~OB1 \citep{2025AA...695A.248B}, and NGC~346 \citep{2019AA...626A..50D}, respectively. The orange histogram represents the rotational velocity distribution for Cygnus~OB2, incorporating O-type stars from \citet{2020AA...642A.168B} alongside those recently reclassified from the \spectype{B}{0} sample. The error bars reflect the standard deviation in each velocity bin obtained from the bootstrap analysis.
                    }
                \label{Fig: vsini_Distribution_4}
            \end{figure}
            }
            
    \end{appendix}
    }

\end{document}